\documentclass[11pt,a4paper]{article}
\pdfoutput=1
\usepackage{jinstpub}

\usepackage{booktabs} % for much better looking tables
\usepackage{tabularx}
\usepackage{caption}
\usepackage{array} % for better arrays (eg matrices) in maths
\usepackage{paralist} % very flexible & customisable lists
\usepackage{verbatim} 
\usepackage{subfig} 
\usepackage{siunitx} % for easier number formatting
\usepackage{longtable} % multipage tables
\usepackage{upgreek} % needed for old arXiv hack

\usepackage{url}
\makeatletter
\def\url@leostyle{%
  \@ifundefined{selectfont}{\def\UrlFont{\sf}}{\def\UrlFont{\small\ttfamily}}}
\makeatother
\title{The IceCube Neutrino Observatory:\\Instrumentation and Online Systems}

\collaboration{The IceCube Collaboration}
\author[b]{M.~G.~Aartsen,}
\author[ba]{M.~Ackermann,}
\author[p]{J.~Adams,}
\author[l]{J.~A.~Aguilar,}
\author[ad]{M.~Ahlers,}
\author[aq]{M.~Ahrens,}
\author[x]{D.~Altmann,}
\author[ag]{K.~Andeen,}
\author[aw]{T.~Anderson,}
\author[l]{I.~Ansseau,}
% ...
\affiliation[b]{Department of Physics, University of Adelaide, Adelaide, 5005, Australia}
% ...
\affiliation[l]{Universit\'e Libre de Bruxelles, Science Faculty CP230, B-1050 Brussels, Belgium}
% ...
\affiliation[p]{Dept.~of Physics and Astronomy, University of Canterbury, Private Bag 4800, Christchurch, New Zealand}
% ...
\affiliation[x]{Erlangen Centre for Astroparticle Physics, Friedrich-Alexander-Universit\"at Erlangen-N\"urnberg, D-91058 Erlangen, Germany}
% ...
\affiliation[ad]{Dept.~of Physics and Wisconsin IceCube Particle
Astrophysics Center, University of Wisconsin, Madison, WI 53706, USA}
% ...
\affiliation[ag]{Department of Physics, Marquette University, Milwaukee, WI, 53201, USA}
% ...
\affiliation[aq]{Oskar Klein Centre and Dept.~of Physics, Stockholm University, SE-10691 Stockholm, Sweden}
% ...
\affiliation[aw]{Dept.~of Physics, Pennsylvania State University, University Park, PA 16802, USA}
% ...
\affiliation[ba]{DESY, D-15735 Zeuthen, Germany}
% ...

\emailAdd{jkelley@icecube.wisc.edu}
\emailAdd{drwilliams3@ua.edu}

\abstract{The IceCube Neutrino Observatory is a cubic-kilometer-scale
  high-energy neutrino detector built into the ice at the South Pole.
  Construction of IceCube, the largest neutrino detector built to date, was
  completed in 2011 and enabled the discovery of high-energy
  astrophysical neutrinos.  We describe here the design, production, and
  calibration of the IceCube digital optical module (DOM), the cable
  systems, computing hardware, and our methodology for
  drilling and deployment. We also describe the online triggering and
  data filtering systems that select candidate neutrino and cosmic ray
  events for analysis. Due to a rigorous pre-deployment protocol, 98.4\% of
  the DOMs in the deep ice are operating and collecting data. IceCube routinely
  achieves a detector uptime of 99\% by emphasizing software stability and
  monitoring.  Detector operations have been stable since
  construction was completed, and the detector is expected to operate at
  least until the end of the next decade.} 


\arxivnumber{1612.05093}

\begin{document}
\maketitle
\clearpage

%auto-ignore
\section{Introduction}
\label{sec:intro}

In the six decades following the 
discovery of the neutrino by Cowan and Reines \cite{reines1960detection}, detectors have been realized
that have explored the properties and sources of neutrinos. Early developments in the field included 
radiochemical detection of MeV-scale neutrinos from
nuclear fusion in the Sun \cite{Homestake} and the first detections of
higher energy
atmospheric neutrinos in deep underground telescopes
\cite{Achar,Witwatersrand}. Advanced
atmospheric and solar neutrino observatories provided definitive
evidence of neutrino mass and constrained neutrino mixing
parameters \cite{SK,SNO}.  The first neutrinos to be detected from outside
the solar system were from Supernova 1987A
\cite{SK1987A,IMB1987A,BUST1987A}. The search for neutrinos emanating from astrophysical
processes led to the development of a new generation of large-scale neutrino detectors. 

Atmospheric neutrinos have energies in the GeV--TeV range and result from the interactions of cosmic ray
particles with the atmosphere. Cosmic rays in turn are
believed to be accelerated in astrophysical objects such as supernova
remnants, active galactic nuclei, and gamma ray bursts. These
astrophysical acceleration sites should also produce neutrinos up to 
PeV-scale energies. Neutrinos, because of their weak interaction and
electrically neutral character, are useful probes of high-energy phenomena
in the Universe. Unlike photons, detecting their origins in astrophysical 
acceleration sites would unambiguously indicate hadronic acceleration and
provide identification of the origins of cosmic rays. Neutrinos arrive 
undeflected and unscattered upon detection and thus point back to their
sources, providing
a clear view of the physics deep within shrouded and compact sources. At
the most extreme energies, they are the only particles that can reach 
us from sources at cosmological distances. Notwithstanding the neutrinos seen from SN1987A, a remarkable but
extremely rare phenomenon, the detection of neutrinos originating in
astrophysical processes outside our solar system requires detector facilities of
extreme dimensions to detect the faint fluxes of these weakly-interacting
particles. 

The DUMAND experiment \cite{DUMAND} pioneered the idea of using large-area
photomultiplier tubes in the deep ocean to detect high energy
neutrinos. Although DUMAND was never realized, experience in operating
detectors remotely in a harsh environment led to the successful
efforts of Baikal \cite{Baikal}, AMANDA \cite{AMANDA:detector} and
ANTARES \cite{ANTARES}. These experiments provided key measurements of the
high-energy atmospheric neutrino spectrum, constrained optimistic models of
astrophysical neutrino production, and demonstrated the feasibility of the
technique. However, the detection of astrophysical neutrinos proved
elusive, suggesting that a kilometer-scale array was required to
achieve the necessary sensitivity \cite{Halzen:2002pg,Learned:2000sw,Gaisser:1994yf}.  

The IceCube Neutrino Observatory is a cubic-kilometer neutrino detector
built into the ice at the South Pole. Construction was completed
on December 18, 2010, and commissioning was completed in 2011.  Its primary scientific objective has been the discovery of
astrophysical neutrinos, which was achieved in 2013 \cite{IC3:evidence}, and the 
identification and characterization of their sources.  Other science
objectives include indirect detection of dark matter, searches for other exotic particles,
studies of neutrino oscillation physics, and detection of the neutrino burst
from a Galactic core-collapse supernova \cite{halzen_klein_review}.  A multi-messenger collaboration
with optical, X-ray, gamma-ray, radio, and gravitational wave observatories
provides multiple windows onto the potential neutrino sources.  A key to
the success of these initiatives is the reliability and performance of the
IceCube instrumentation, as well as the flexibility and stability of the
online software systems.  

\subsection{A Functional Description of the IceCube Observatory}

In order to detect neutrinos, IceCube exploits the fact that charged
particles resulting from neutrino interactions move through the
ice faster than the phase velocity of light in ice, and therefore emit Cherenkov photons. An enormous detection volume
is required due to the small interaction cross-section of neutrinos
and the extremely low fluxes expected at Earth from astrophysical
objects. The ice cap at the South Pole is about three kilometers thick and is
a suitable
operational site, since it not only offers a large quantity of interaction
material but also a medium with excellent optical qualities.  With a
Cherenkov photon yield of $\mathcal{O}(\num{E5})$ visible photons per
\SI{}{\giga\electronvolt} of secondary particle shower energy, the long
optical attenuation length of South Pole ice, and large-area PMTs, it is possible to instrument cubic kilometers of
ice with a rather sparse spacing of detectors. IceCube is located at Amundsen-Scott
South Pole Station, which offers the logistical support
required for the construction and operation of the observatory.

The basic detection unit in IceCube is the
digital optical module (DOM), covered in detail in section~\ref{sec:dom}.
Encapsulated in a glass pressure sphere 
to withstand the extreme pressure in the deep ice, the main components of a DOM
are a \SI{10}{''} PMT, embedded high-voltage generation, a
light-emitting diode (LED) Flasher 
Board, and a Main Board containing the analog and digital processing circuitry
for PMT pulses.  Detector calibration
is described in section~\ref{sec:dom_calibration}.  Digitized, timestamped
PMT signals are sent from the DOMs to a
central computing facility at the surface via a cable system described in
section~\ref{sec:cable}.  Aspects of detector deployment and ice drilling are
covered in section~\ref{sec:drill-deploy}.  An overview of the data flow including
DOM readout, event triggering, processing and filtering are presented in
section~\ref{sect:online}, along with data handling, monitoring, and operational performance of
the observatory.

\begin{figure}[!ht]
 \centering
 \includegraphics[width=\textwidth]{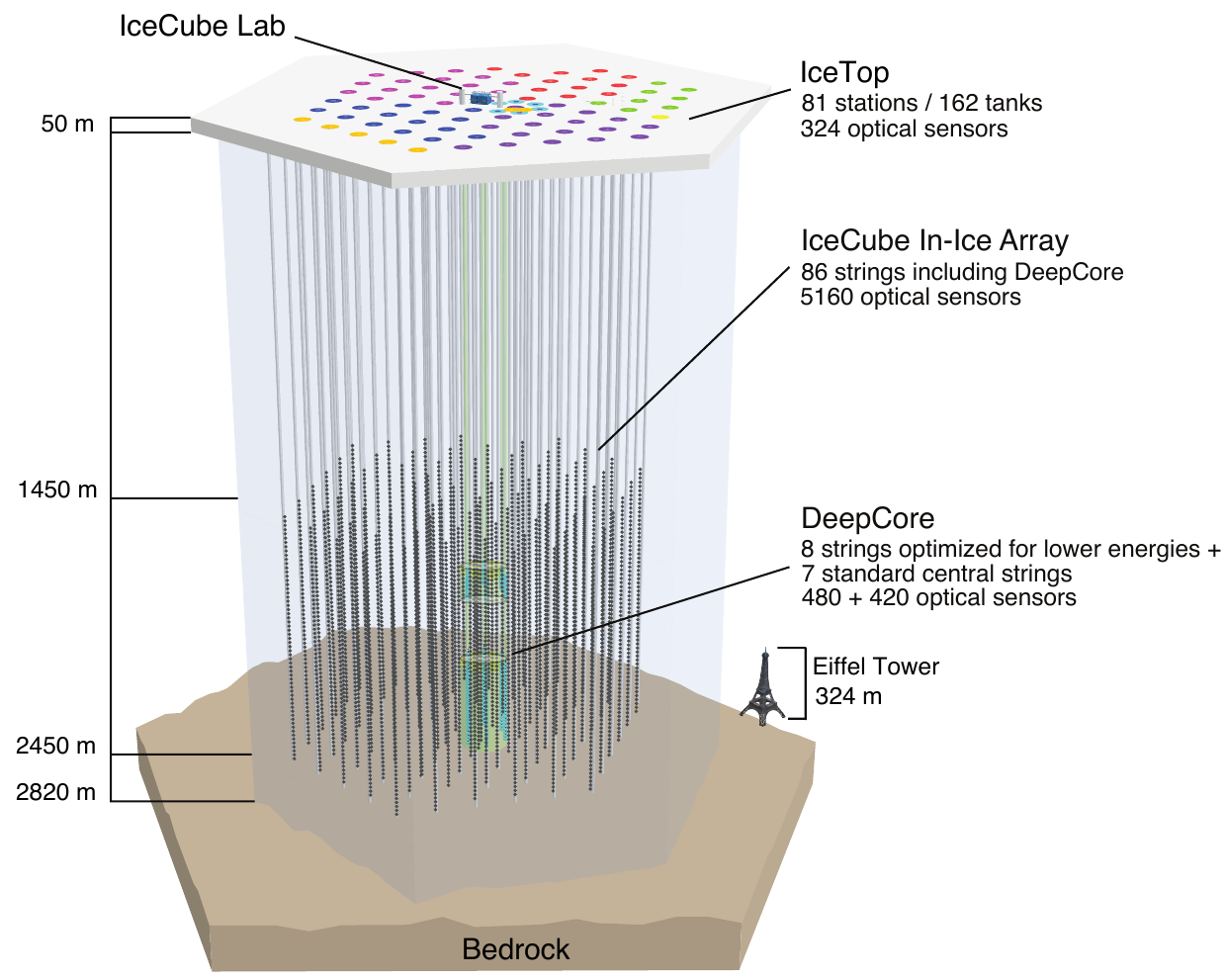}
 \caption{The IceCube Neutrino Observatory with the in-ice array, its sub-array DeepCore, and
   the cosmic-ray air shower array IceTop. The different string/station colors
   represent different deployment seasons.}
 \label{fig:array}
\end{figure}

The IceCube Neutrino Observatory consists of a subsurface
``in-ice'' array of DOMs, including the more densely instrumented
DeepCore sub-array, and the IceTop surface array.  The entire detector
uses the same DOM design and associated surface readout. A schematic layout
of the observatory is shown in figure~\ref{fig:array}. 

\subsubsection{IceCube In-Ice Array}

In order to detect the Cherenkov photons emitted by charged particles
traversing the ice, \num{5160} DOMs are deployed between \SI{1450}{\meter}
and \SI{2450} {\meter} below the surface of the ice on \num{86} vertical
strings. Each string consists of \num{60} DOMs deployed along a
single cable containing twisted copper-wire pairs. The 
primary in-ice array consists of \num{78} strings with a vertical
separation of the DOMs of \SI{17}{\meter}.  The strings are
deployed within a hexagonal footprint on a triangular grid with
\SI{125}{\meter} horizontal spacing, 
instrumenting a volume of one cubic kilometer of ice.  This design was chosen in
order to meet the primary science requirement of detecting astrophysical
neutrinos in the energy range of $\mathcal{O}(\SI{}{\tera\electronvolt})$--
$\mathcal{O}(\SI{}{PeV})$.  

Two different event topologies form the standard signatures of neutrinos in
IceCube.  Track-like events originate from a charged-current interaction of
a high-energy muon neutrino with a nucleus, producing a hadronic shower at
the vertex and an outgoing muon that emits Cherenkov light in a cone along its
track.  The angular resolution for muon tracks and hence the incident
neutrino direction is typically $\SI{0.6}{\degree}$, confirmed by analysis
of the shadows of the Moon and Sun
in cosmic rays \cite{IC3:moon,IC3:sun}.  Muons with energies above a
critical energy, about $\SI{1}{\tera\electronvolt}$ in ice,
predominantly lose energy by radiative processes that exhibit a stochastic behavior with large fluctuations.  This results in
large variability in the amount of energy deposited 
for different muons of the same energy.  A second class of events are
electromagnetic or hadronic showers from interactions of all neutrino
flavors, resulting in a more spherical light generation in the detector.
Since the total light output of such a shower is directly proportional to its energy, and
the showers are often well-contained in the detector, the neutrino energy
reconstruction for such events is much more precise than for track-like
events. The average \emph{deposited} energy resolution for both event types
is about 15\% \cite{IC3:ereco}.

\subsubsection{DeepCore}

A subset of in-ice DOMs is deployed deeper than \SI{1750}{\meter} with a denser instrumented volume and
correspondingly lower energy threshold. This
sub-array, DeepCore \cite{ICECUBE:DC}, consists of eight specialized and
closely-spaced strings of sensors in the center of the array, along with
the seven central standard IceCube strings. The inter-string spacing
in DeepCore varies from \SI{41}{\meter} to \SI{105}{\meter}, with an
average spacing of \SI{72}{\meter}.

The eight specialized DeepCore strings have a DOM-to-DOM spacing of
\SI{7}{\meter} for the bottom 50 DOMs, deployed at depths of
\SI{2100}{\meter} to \SI{2450}{\meter}.  The remaining 10 DOMs are
deployed at depths shallower than \SI{2000}{\meter} with a spacing of
\SI{10}{\meter} to form a veto cap, allowing better rejection of downgoing
atmospheric muons.  Depths from \SI{2000}{\meter} to \SI{2100}{\meter}
are not instrumented, as the optical scattering and absorption is
significantly increased in this region of the ice (the ``dust layer''~\cite{Aartsen:2013rt}).

Six of the specialized DeepCore strings are fully instrumented with
DOMs using PMTs with 35\% higher quantum efficiency than the
standard IceCube modules. The remaining two specialized strings are
equipped with a mixture of standard and higher quantum efficiency DOMs. The denser geometry and increased efficiency
result in a lower energy threshold of about
\SI{10}{\giga\electronvolt}, compared to about
\SI{100}{\giga\electronvolt} for most IceCube analyses. The DeepCore design
is optimized for the detection of neutrinos with energies
from \SI{10}{\giga\electronvolt} to \SI{100}{\giga\electronvolt},
increasing IceCube's ability to detect atmospheric neutrino
oscillations, neutrinos from WIMP dark matter annihilation, and Galactic
supernovae~\cite{ICECUBE:DC}. 

\subsubsection{IceTop}

The cosmic ray air shower array IceTop \cite{ICECUBE:IceTop}, located
on the surface of the ice at 2835~m above sea level, consists
of \num{162} ice-filled tanks, instrumented with PMTs that detect Cherenkov
radiation and arranged in \num{81} stations on the
surface, using approximately the same grid on which the in-ice
array is deployed. A denser infill array is formed by the eight
stations in the center of IceTop, corresponding to the denser
inter-string spacing
in DeepCore. Each tank is filled with ice to a height of \SI{0.90}{\meter}.  The two tanks at each surface station are separated from
each other by \SI{10}{\meter}. Each tank contains two standard IceCube
DOMs, one ``high-gain'' DOM operated at a PMT gain of $5 \times 10^{6}$, and one
``low-gain'' DOM operated at a gain of $10^{5}$, to increase the dynamic
range for air shower detection.  Cosmic-ray-initiated air showers are typically
spread over a number of stations. The light generated in the tanks by the
shower particles (electrons, photons, muons and hadrons) is a measure of
the energy deposition of these particles in the tanks. IceTop is sensitive to
primary cosmic rays in the energy range of \SI{}{PeV} to \SI{}{EeV}
with an energy resolution of 25\% at \SI{2}{PeV}, improving to 12\% above \SI{10}{PeV} \cite{IT:measurement}. For the infill
array, the energy threshold is lowered to about \SI{100}{TeV}. The energy
range of IceCube as a cosmic-ray detector fully covers the ``knee'' region
of the spectrum and extends to the energies where a transition from
Galactic cosmic rays to a population of extra-galactic 
particles may occur. The IceTop array has additionally been used to study
high-$p_T$ muons, PeV gamma rays, and transient events, such as the
radiation effects of solar flares. It also serves as a partial veto for the
detection of downward-going neutrinos with IceCube.   

\subsubsection{The IceCube Laboratory}

\begin{figure}[!ht]
 \centering
 \includegraphics[width=0.6\textwidth]{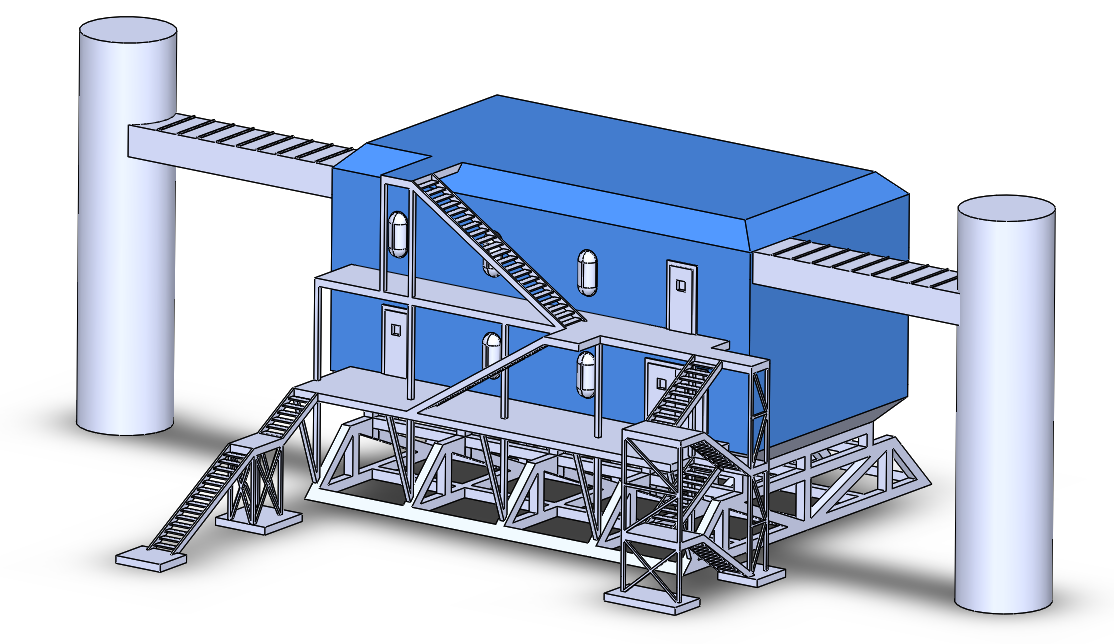}
 \caption{The IceCube Laboratory (ICL) is the central operations building
   for the experiment.  Two cable towers connect the array's surface cables
   to readout computers on the second floor.}
 \label{fig:icl}
\end{figure}

The IceCube Laboratory (ICL), located at the surface in the center
of the array, is the central operations building for the experiment
(figure~\ref{fig:icl}). Surface cables from the array are routed up two cable
towers on either side of the structure and into a server room on the second
floor (section~\ref{sec:cable}). The server room houses the South Pole System
(section~\ref{sect:sps}), including all data acquisition and online filtering
computers. Power is supplied by the South Pole Station generators.

The building is temperature-controlled to a target of $18\ ^{\circ}$C; continuous
airflow through the server room is important to avoid overheating.  The second
floor is shielded against electromagnetic interference with a metal
mesh, and the building shield is connected to the shields of each surface
cable as they enter from the cable tower bridges.  Because of the low
humidity, strict protocols to minimize electrostatic discharge when working
on ICL equipment are necessary.

%auto-ignore

% additional definitions
\newcommand{\degC}[1]{$\unit[#1]{^\circ{C}}$}
% definition to produce a "less than or similar to" symbol
\def\lsim{\mathrel{\rlap{\raise 0.2ex\hbox{$\,<\,$}}{\lower 0.9ex\hbox{$\,\sim\,$}}}}
% definition to produce a "greater than or similar to" symbol
\def\gsim{\mathrel{\rlap{\raise 0.2ex\hbox{$\,>\,$}}{\lower 0.9ex\hbox{$\,\sim\,$}}}}

\section{\label{sec:dom}The Digital Optical Module}

\subsection{\label{sec:dom_functional}A Functional Description of the DOM}

\begin{figure}[!h]
  \captionsetup[subfigure]{labelformat=empty}
  \centering
  \subfloat[]{\includegraphics[width=0.5\textwidth]{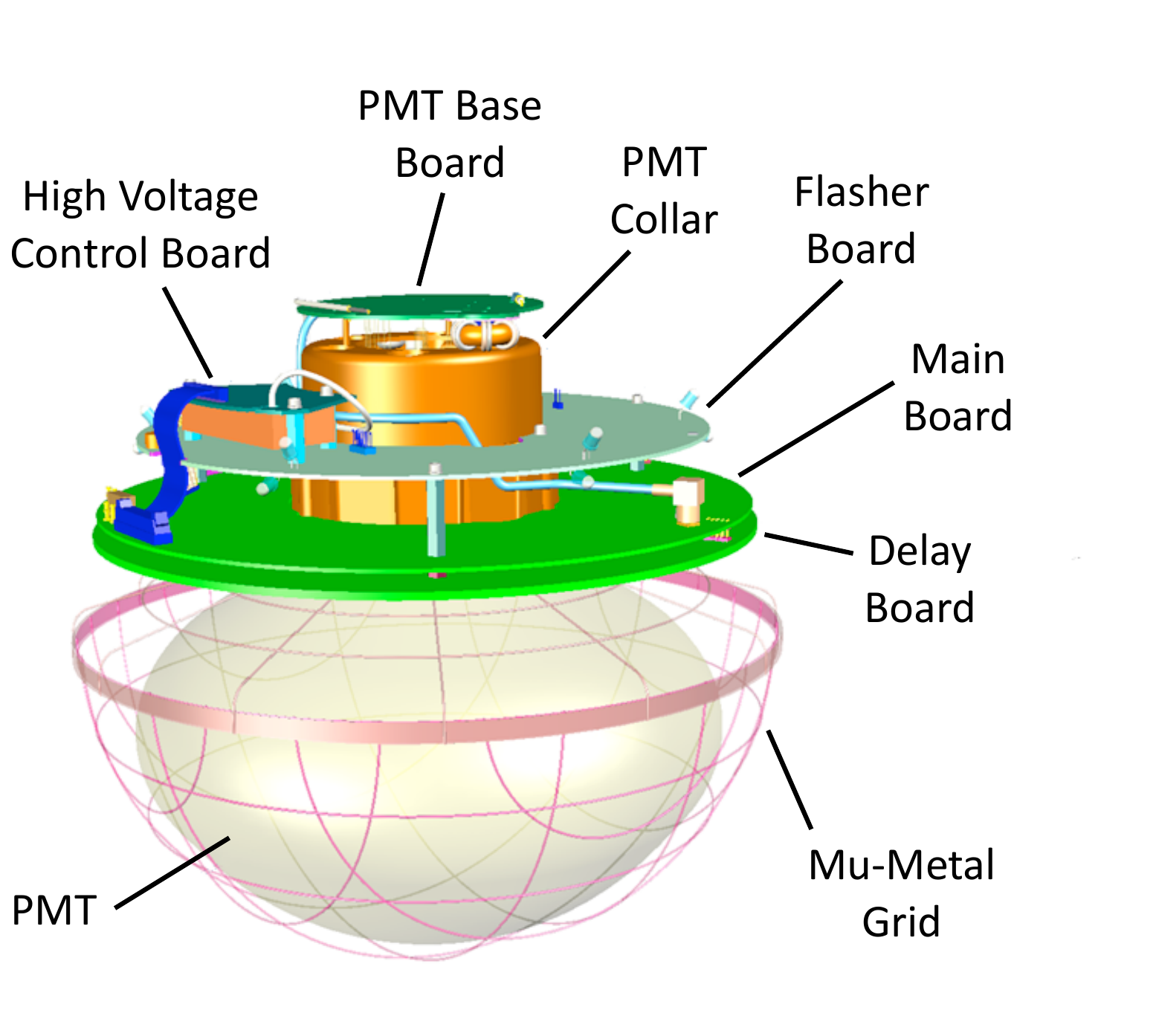}}
  \subfloat[]{\includegraphics[width=0.5\textwidth]{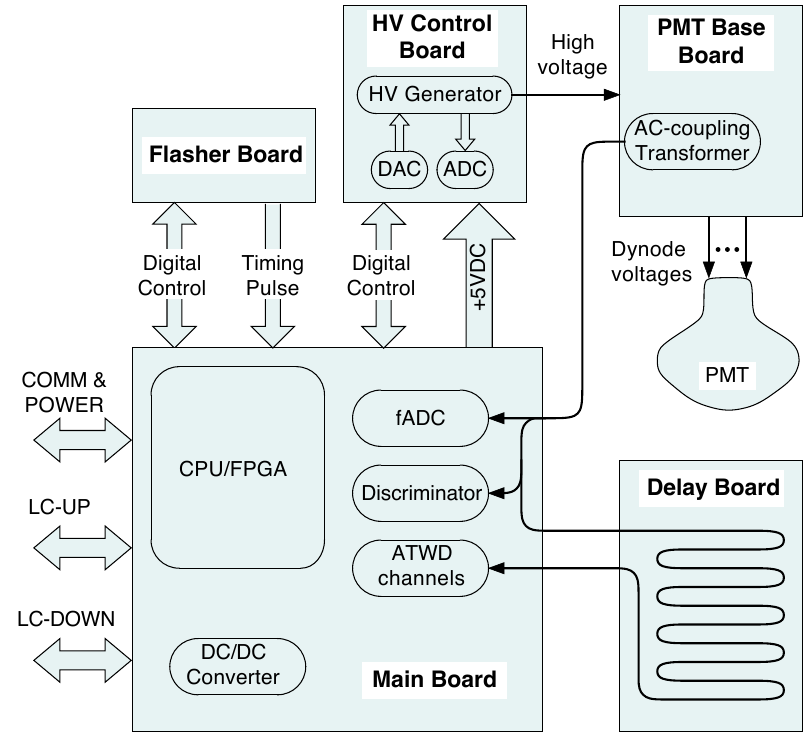}}
  \caption{Components of the DOM, showing mechanical layout (left) and functional connections (right).}
  \label{fig:domcomponents}
\end{figure}

The DOM is the fundamental light sensor and data acquisition unit for IceCube.
It consists of a spherical glass housing 
containing a downward-facing \SI{10}{''}-diameter PMT~\cite{ICECUBE:PMT}
and associated circuit boards that allow near-autonomous operation (figure~\ref{fig:domcomponents}).
Data acquisition, control, calibration, communication, and low-voltage power conversion 
are integrated in one annular circuit board (Main Board) that fits around the neck of the PMT~\cite{ICECUBE:DAQ}. 
Separate circuit boards generate PMT high voltage, interface to the PMT pins,
delay PMT signals, and generate calibration light flashes that can reach other DOMs.
Key requirements for the DOM include
the precise recording of a wide variety of PMT pulse widths and amplitudes
with nanosecond time resolution, robustness in 
a challenging deployment environment, and long-term reliability.

%============================================================

%============================================================

The PMT detects signals from particles interacting in the ice, typically ranging over
energies from \qty{10}{GeV} to \qty{10}{PeV} and distances up to \qty{500}{m}
away.  At a gain of $10^7$ (section~\ref{sec:hv}), corresponding PMT waveforms can have amplitudes from \qty{1}{mV} up
to and beyond the linearity limit of the PMT ($\sim$\qty{2}{V}) and widths
from \qty{12}{ns} up to around \qty{1500}{ns}.  In order to accommodate such a variety
of signals, the DOM includes multiple digitizers with overlapping dynamic
range and different sampling speeds
(section~\ref{sec:mainboard}).  Each DOM independently
detects individual photons, starting a recording of
the PMT waveform that also includes photons arriving up to
\qty{6.4}{\micro\second} later (a ``hit'').  The hit time is saved along with the
waveform shape, allowing the determination of the times of arriving photons
relative to this reference.  The DOM accumulates such hit
data for a period of about \qty{1}{s} before sending the data up as a block.
However, if data readout is interrupted, the DOM can store
$\mathcal{O}(\SI{10}{\second})$ of data before overflowing local memory
(16~MB of SDRAM), depending on hit rate.  Separately, the PMT hit rate is recorded by
the DOM in \qty{1.6384}{ms} intervals, as a collective increase of all rates
could signify detection of many low energy neutrinos in case of a Galactic
supernova event (section~\ref{sect:SNDAQ}) \cite{IC3:supernova}.

DOMs transmit their data to computers in the ICL
over a twisted wire pair that also provides power (section~\ref{sec:cable}).
Wire pairs are bundled to form the vertical in-ice cables and the horizontal surface
cables.  Each wire pair is shared between two DOMs, with data transfers
initiated by a surface computer.  Separately, dedicated local coincidence
(LC) wiring to neighbor DOMs above and below allows quick recognition of neighboring
coincident hits, where
nearest or next-to-nearest neighbors are hit within a common time
window. The time window is configurable and is set to $\pm1~\mu\mathrm{s}$
for both in-ice and IceTop DOMs. The signals are forwarded from one DOM to the next
through the dedicated wiring.  The span of the forwarding is
software-configurable and is currently set to two for in-ice DOMs,
i.e. a DOM signals its neighbor and next-to-nearest neighbor DOMs in
both up and down directions along the string. The local coincidence
connections for IceTop, which allow coincidences between the two tanks in a
station, are described in ref.~\cite{ICECUBE:IceTop}. Local coincidence
hits (``HLC'' hits) often have complex PMT waveforms
indicating multiple photons detected in each DOM and are therefore saved
in full detail; otherwise, the DOM saves abbreviated information appropriate
to single photon detection (section~\ref{sect:online:payloads}).

The DOM is capable of interpreting commands from the surface that specify
tasks for configuration, data-taking and transmission, monitoring or
self-calibration.  Self-calibration functions establish PMT and amplifier
gains as well as sampling speed (section~\ref{sec:domcal}).  The RAPCal
system (section~\ref{sect:dom:rapcal}) is implemented for tracking each
local DOM clock's offset from universal time, allowing PMT pulses that were
independently recorded in many DOMs to be built into events by surface
computers.

\subsection{\label{sec:dom_components}Components}

\subsubsection{\label{sec:sphere}Glass Sphere and Harness}

The glass sphere housing has an outer diameter of \SI{13}{''} and thickness
of \SI{0.5}{''}.
The spheres protect the inside electronics and PMT against long-term applied pressure of 
\qty{250}{bar} (\qty{2.6}{km}-equivalent water depth)
as well as temporary overpressure up to \qty{690}{bar} during the refreezing of melted ice in the drill hole.
The housings were produced by Benthos (Falmouth, Massachusetts), based on a design for deep-sea
environments but using borosilicate glass from Kopp Glass
with very low potassium and other radioactive trace elements that would contribute to the dark noise
count rate (section~\ref{sect:darknoise}).  
Optical transmission was measured in representative glass samples as 93\% at \qty{400}{nm},
decreasing to 50\% at \qty{340}{nm} and 10\% at \qty{315}{nm} (normal
incidence). Fresnel reflection is not included in the quoted
transmission, since the effect of Fresnel reflection is small in ice,
where the refractive index is better matched to the glass.

\begin{figure}
\vspace{3pt}
\centering
\begin{tabular}{c@{\hspace{0.5in}}c}
\includegraphics[width=0.4\textwidth,clip=true]{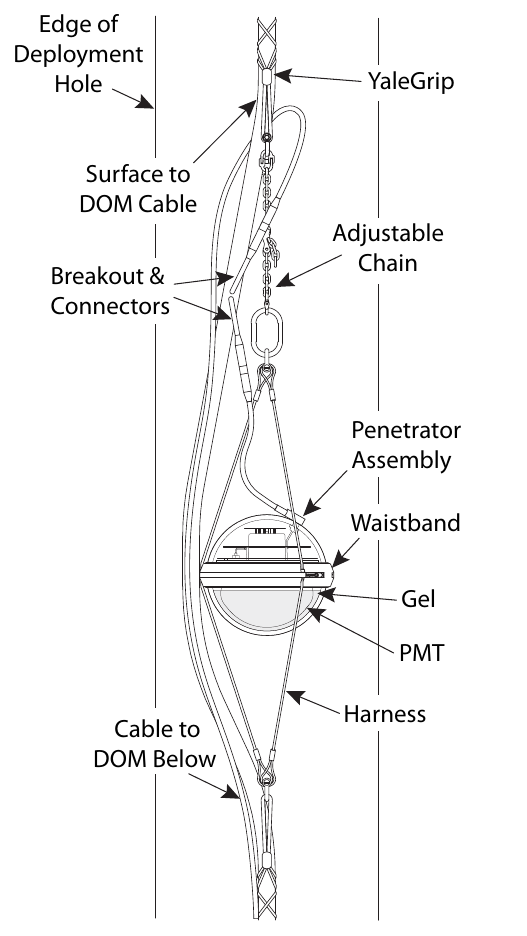} & \
\includegraphics[width=0.45\textwidth,clip=true]{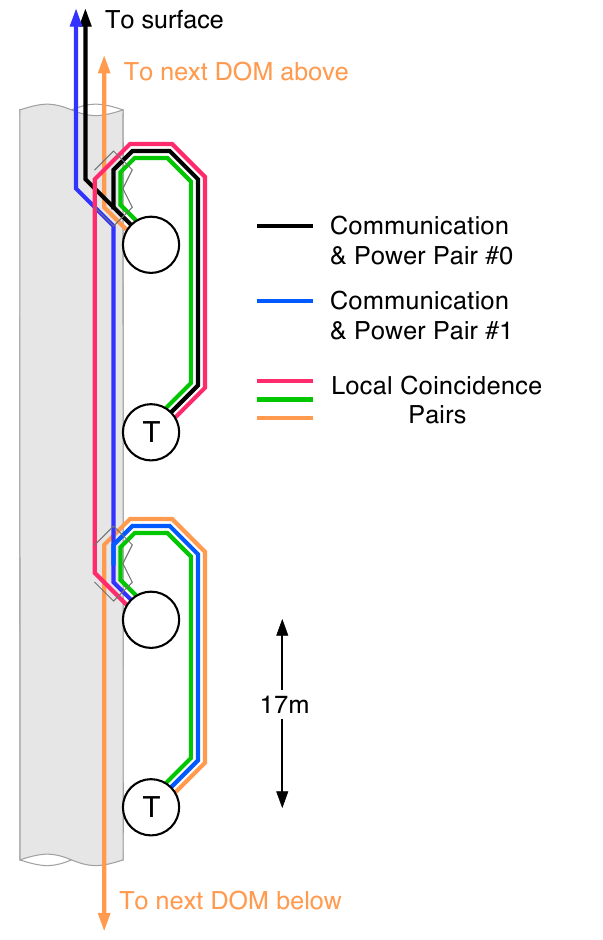} \\
\end{tabular}
\caption{
(Left) DOM as deployed on main in-ice cable, showing cable breakout to the penetrator
assembly and the mechanical support system.  (Right) Schematic of cable connections for a set
of four DOMs serviced by two wire pairs from the surface that carry power and
communications.  The ``T'' labels indicate where electrical termination (\qty{140}{\ohm}) is
installed in one of two DOMs that share such a wire pair.  Other wire pairs are used for
bidirectional signaling between neighboring DOMs, in order to check for in-time coincident
detections.
}
\label{fig:domcable}
\end{figure}

All spheres were tested up to \qty{690}{bar} hydrostatic pressure by the manufacturer.
Each was delivered as two hemispheres that mate precisely at the equator
and were sealed during assembly (section~\ref{sec:dom_prodtest}).  The DOM
is held by an aluminum waistband with rubber gaskets against 
the glass above and below the equator seam. 
Figure~\ref{fig:domcable} shows how the DOM is attached to the main in-ice cable via a harness
of steel rope and a chain that carries the weight load around the DOM.
The main cable bends around the DOM, and the DOM axis stays vertically aligned with the string.

%============================================================
%============================================================

\subsubsection{\label{sec:penetrator}Cable Penetrator, Cable and Connector}

A penetrator assembly brings three wire pairs out through a \qty{16.3}{mm} hole in
the DOM glass sphere.  The wires are routed inside a customized cable, shown in figure~\ref{fig:domcable},
and terminate at a pressure-tight, waterproof connector that mates with a similar connector
that continues each pair into the main cable.  One wire pair carries power and a
bidirectional digital communications stream, connecting ultimately
with a computer in the
IceCube Laboratory building (section~\ref{sec:cable}).
The other wires lead to neighboring DOMs directly above and below,
carrying LC digital pulses that signify time-correlated hits in nearby DOMs (section~\ref{sec:mainboard}).

DOMs were produced in two versions, in which the communications wire pair was either electrically
terminated (\qty{140}{\ohm}) or unterminated inside the DOM.  The
terminated DOM is deployed \qty{17}{m} below the unterminated one (\qty{7}{m}
or \qty{10}{m} in DeepCore strings) and therefore includes a correspondingly 
long penetrator assembly cable (figure~\ref{fig:domcable}).

The entire penetrator assembly was designed and produced by SEACON Brantner \& Associates (El Cajon,
California).  The part that seals against the DOM glass is a
hollow steel bolt that is secured inside the DOM by a nut and spring
washer and compresses a fluorosilicone O-ring against the outside surface.
The steel part includes additional sealing around the wires that pass
through it.  Outside the DOM, a plastic shell is molded around the steel
and onto the cable jacket.  External mechanical features like the
penetrator are subject to large stresses during deployment and the
refreezing process; a right-angle bend outside the DOM was included for
robustness, based on previous experience deploying AMANDA modules.

\subsubsection{\label{sec:pmt}PMT, Gel and Magnetic Shield}

DOMs use the \SI{10}{''}-diameter Hamamatsu R7081-02 PMT, 
or the corresponding high-quantum-efficiency (HQE) version, Hamamatsu R7081-02MOD, for DeepCore strings.
The PMT properties have been measured and described in ref.~\cite{ICECUBE:PMT}.
The PMT is specified by Hamamatsu for the wavelength range
\qty{300}{nm}--\qty{650}{nm}, with peak quantum efficiency around 25\% (34\%
for HQE) near \qty{390}{nm}.  It features a box-and-line dynode chain with 10 stages,
with in-ice DOMs (both standard and HQE) operated at a gain of $10^7$ (section~\ref{sec:domcal}).

The PMT bulb faces downwards in the bottom glass hemisphere, secured in high-strength 
silicone gel to a depth sufficient to surround the photocathode area.  
The gel provides both good optical coupling and mechanical support for the
whole assembly of PMT and circuit boards. The gel thickness between the PMT
envelope and glass sphere is approximately \qty{1}{cm}.   
Originally the gel was supplied from General Electric as RTV6136-D1,
and later as a similar formulation from Quantum Silicones (Virginia, USA).  
It is optically clear with transmission of 97\% at \qty{400}{nm}, 91\% at \qty{340}{nm}, and 65\% at \qty{300}{nm}
(normal incidence).  The refractive index is 1.41, yielding less than 0.1\% reflection as light
passes from the sphere glass into the gel and then into the PMT envelope.
The characteristics of the cured gel are specified to remain stable in the
temperature range $-70^\circ$C to $45^\circ$C.  Visual inspection of
non-deployed DOMs reveals no indication of cracks (``crazing'') after more than 10
years, and studies of the long-term optical efficiency of deployed DOMs
reveal no measurable aging effects (section~\ref{sec:optical_stability}).  

To reduce effects of the ambient South Pole magnetic field (\qty{550}{mG}, $17^\circ$
from vertical) on the PMT collection efficiency, a mu-metal cage surrounds the PMT bulb up to
the neck.  It was constructed as a wire mesh with typical wire spacing \qty{66}{mm} and
wire diameter \qty{1}{mm}, blocking about 4\% of the incident light,
and delivered by ITEP Moscow.
Without such a shield, this PMT exhibits 5--10\% lower
collection efficiency, poorer single photoelectron resolution, and gain variations of 20\% depending on 
azimuthal orientation, for a South Pole magnetic field strength and orientation~\cite{calvo}.
With the shield in place, the interior magnetic field is 2.8 times
smaller than the external field, pointing mostly along the axis and therefore reducing efficiency by
less than 2\% for this type of PMT.

Other interior DOM components are held in place by attachment to the PMT, mostly via screws into
a molded plastic collar glued around the PMT neck.  The PMT Base Board is
soldered directly to the PMT pins.

\subsubsection{\label{sec:hv}High Voltage Supply and Divider}

The PMT high voltage subsystem consists of a resistive 
voltage divider circuit (PMT Base Board) directly
solder-mounted on the PMT and a separate High Voltage Control Board. 
The High Voltage Control Board includes a DAC and ADC for setting and reading out the PMT high voltage,
connected to the Main Board with a digital interface, as well as the high
voltage generator itself.

The high voltage generator is a custom encapsulated module (\#9730A) designed by
EMCO High Voltage (California).  The maximum high voltage is
\qty{2047}{volts}, specified for up to $30\,{\rm \upmu A}$ current.  The
voltage setting, in steps of 0.5~V, is controlled by the DAC
output, and the actual voltage is monitored via a high-impedance divider and the ADC.  The output ripple
is less than \qty{1}{mV}, and stability is better than \qty{1}{V} RMS.  Power
consumption of the high voltage supply is less than \qty{300}{mW} at full
load. 

The generator output is carried to the PMT Base Board \cite{ICECUBE:PMT} via a high voltage
coaxial cable.  The voltage divider, designed for low power consumption,
presents a total resistive load of \qty{130}{\mega\ohm}. 
The PMT is operated with cathode at ground potential, so the anode signal output is AC-coupled using 
a 1:1 bifilar-wound toroid transformer mounted on the Base Board; this
toroid was modified once during DOM production in order to reduce
distortion of high-amplitude signals (section~\ref{sec:waveformcal}).
The transformer secondary is then wired to the Main Board analog input with a coaxial cable.
The single photoelectron (SPE) output waveform has been described in ref.~\cite{ICECUBE:PMT}.  
With a \qty{100}{\ohm} load connected to the transformer, and operating
at standard PMT gain of $10^7$, the SPE 
peak voltage before front-end amplification is approximately \qty{8}{mV}
with a FWHM of 7--8 ns.  Several effects combine to increase
the FWHM of digitized SPE waveforms to $\sim$\qty{13}{ns} (peak $\sim$\qty{5}{mV}).

\subsubsection{\label{sec:mainboard}Main Board and Delay Board}

The Main Board, designed at Lawrence Berkeley National Laboratory,
has been described in detail in \cite{ICECUBE:DAQ}.   
Essentially an embedded single-board data-acquisition computer, the Main
Board interfaces to other boards as shown in figure~\ref{fig:domcomponents} and
provides many key functions of the DOM, including:

\begin{enumerate}
\item{Control of all the devices inside the DOM, including the high voltage power supply for the PMT, 
the flasher board, and various sensors (pressure, temperature, power supply voltage monitor). 
Also supplies necessary DC power to the subsystems.}
\item{Digitization of the PMT waveforms, using a custom integrated circuit (ATWD: Analog
  Transient Waveform Digitizer~\cite{ICECUBE:DAQ}) and a continuously sampling fast ADC (fADC).}
\item{Providing computational functions, including PMT gain calibration, compressing 
digitized waveforms, temporarily storing the data, and creating time-stamped data packets.}
\item{Communicating with the data acquisition (DAQ) system on the surface.}
\item{Exchanging timing pulses with the surface DAQ to calibrate the internal DOM clock. }
\item{Exchanging LC pulses with the adjacent DOMs.}
\item{Uniquely identifying each DOM by providing a Main Board ID generated from a 
component device identifier.}
\item{Providing an onboard adjustable low-intensity optical source
    for direct calibration of PMT gain and timing, and hosting a
    daughter board with an adjustable high-intensity optical source for inter-module calibration.}
\end{enumerate}

\begin{figure}[h]
 \centering
 \includegraphics[width=0.9\textwidth]{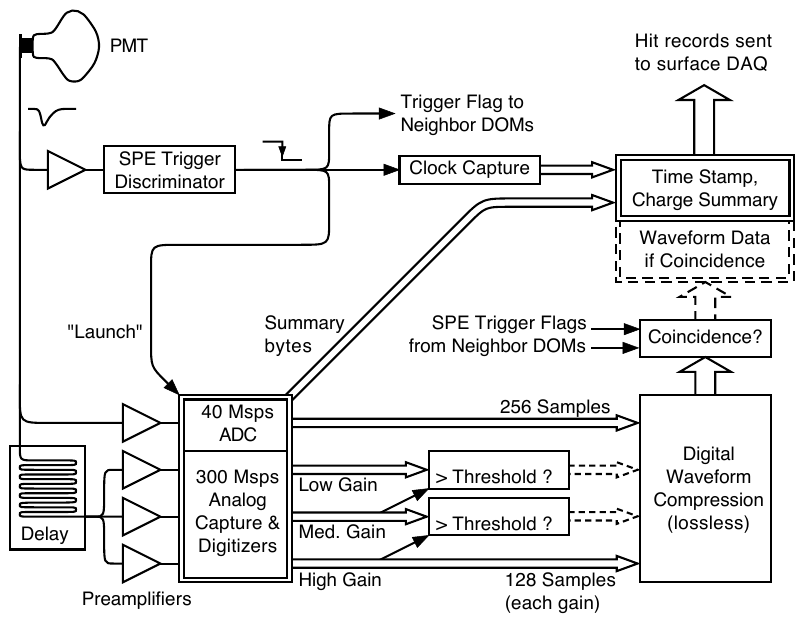}
 \caption{Data flow diagram for recording and processing of PMT waveforms in the DOM to form 
 "Hit Records" that are sent to the surface DAQ computers.  As shown by dashes, full waveform data are only included
 when neighbor DOMs report time-coincident signals above the SPE discriminator threshold.  Additionally,
 data from low-gain channels are omitted for waveforms that are within range of higher-gain channels.}
 \label{fig:domdataflow}
\end{figure}

The data flow starting from the PMT is shown in figure~\ref{fig:domdataflow}.
PMT waveforms are amplified and compared to a discriminator threshold.  Two
discriminators are available, the SPE discriminator that is used for in-ice DOMs
and typically set to a voltage threshold corresponding to \qty{0.25}{PE},
and an MPE discriminator used for the larger-amplitude signals in IceTop.
A discriminator crossing begins a ``launch'' of the high-speed waveform
capture and digitization. Each DOM is equipped with two ATWD chips,
and each chip is provided with three different amplifier
gains with nominal values of 16, 2, and 0.25 in order to completely cover the 
dynamic range of the PMT output (up to 150~mA, or 7.5~V, when saturated).  A
fourth ATWD channel on each chip is used for calibration inputs and is not normally read out.
The ATWD chips are configured to sample the input voltage at \qty{300}{Msps}
and operate by analog storage of waveform samples in switched capacitor arrays of depth 128,
followed by a 10-bit digitization \cite{atwd}.  In order to record the waveform starting from before the discriminator
threshold crossing, the signal is first routed through the Delay Board.  Here a total delay of about
\qty{75}{ns} is accomplished by an approximately \qty{10}{m}-long, \qty{0.25}{mm}-wide
serpentine copper trace embedded in the dielectric and sandwiched between
ground planes.  The highest-gain channel is used
for most pulses, and lower-gain recordings are also retained as needed when pulses reach 75\% of the range 
of a higher-gain channel, in order to avoid any loss of information due to
digital saturation.  The mean amplifier gains of all deployed DOMs for high-,
medium-, and low-gain channels are $15.7\pm0.6$, $1.79\pm0.06$, and $0.21\pm0.01$ respectively.

\begin{figure}[h]
 \centering
 \includegraphics[width=0.9\textwidth]{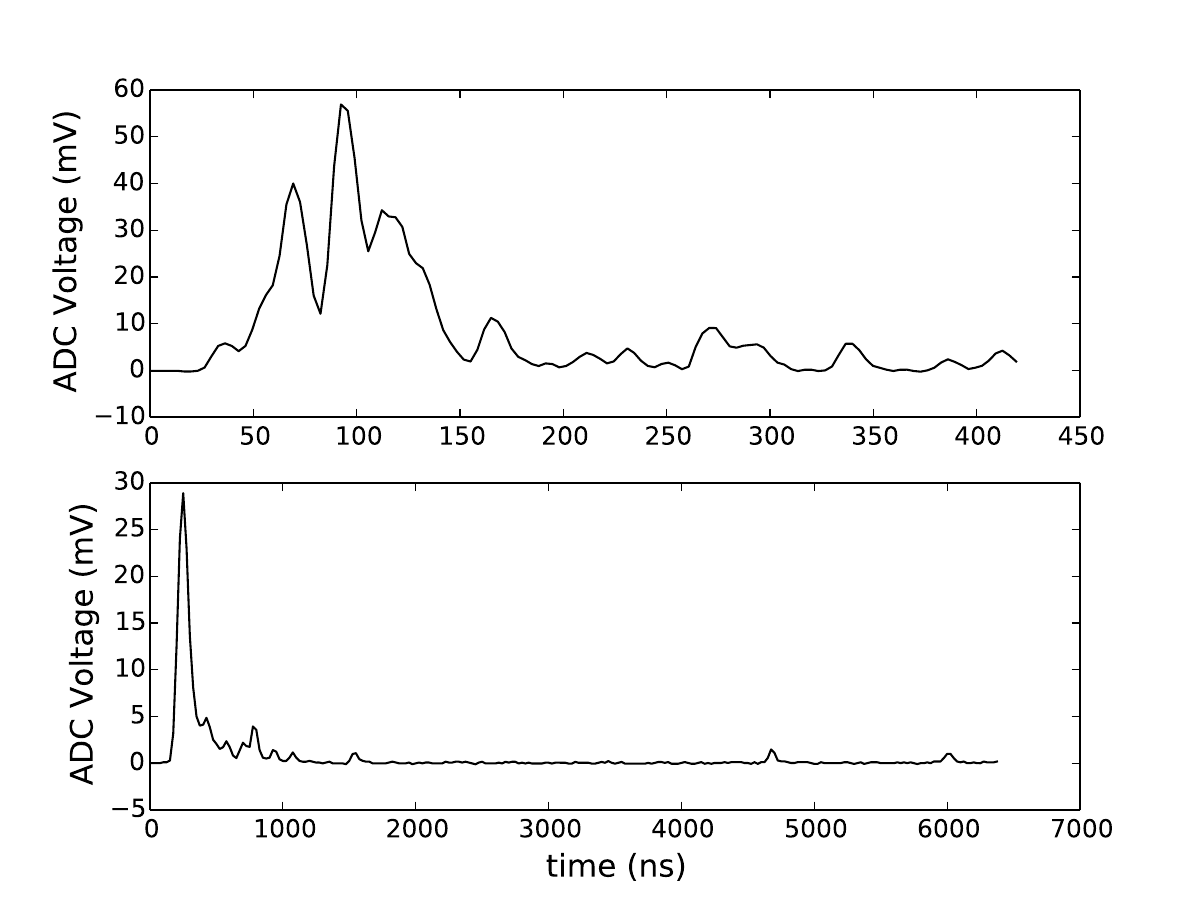}
 \caption{The same signal sampled in the ATWD (top) and the fADC (bottom):
   the ATWD recording duration is 427~ns whereas the fADC recording
   duration is 6.4~$\mu$s. Energy reconstruction in IceCube uses
   the charge and time recorded in the waveform~\cite{IC3:ereco}.}
 \label{fig:mpe_waveform}
\end{figure}

The ATWD recording duration is \qty{427}{ns}.  This is sufficient for
reconstructing light produced within tens of meters of
a DOM, but photons from farther away may arrive over a broader time
interval due to the optical scattering of the ice.  Such distant signals are
also lower in amplitude, and the information is captured in the 10-bit \qty{40}{Msps} fADC.
The fADC samples continuously, and the FPGA is programmed to save an
interval of \qty{6.4}{\micro\second} after the launch. Its amplifier provides a
dynamic range comparable to the high-gain ATWD channel, but has extra pulse shaping to accommodate the lower
sampling speed. An example of a digitized waveform with multiple pulses is shown in
figure~\ref{fig:mpe_waveform}.

Every digitizer launch results in a ``hit'' record.  Hits are
transferred from the FPGA to SDRAM lookback memory (LBM) via Direct Memory Access
(DMA), and the Main Board CPU bundles them and sends them on request to the surface
computers.   The amount of
information included in a hit depends on whether a signal was also detected in one of the neighboring DOMs.
In case of an isolated signal (no coincidence), only a time stamp and brief charge summary are sent, and
the digitization process is aborted.  Conversely, when a nearest or next-to-nearest neighbor DOM 
also signals a launch within $\pm$\qty{1}{\micro\second} (local coincidence), the full waveform is compressed
and included in the hit record.  The LC signaling operates via digital pulse codes sent on
the extra wire pairs described in section~\ref{sec:penetrator}.

\begin{figure}[]
 \centering
 \includegraphics[width=1.0\textwidth]{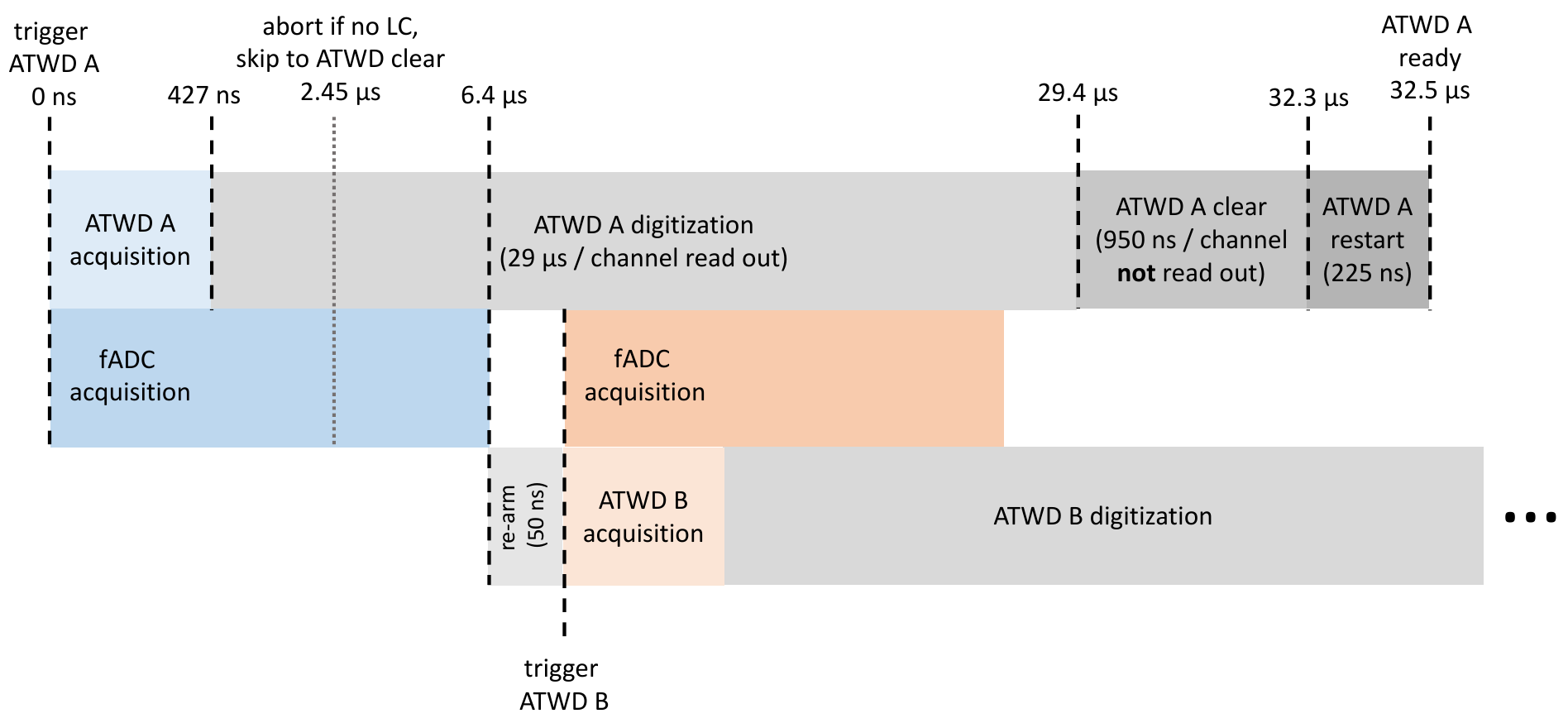}
 \caption{Timing of ATWD and fADC acquisition and associated deadtime, for
   back-to-back HLC hits with one ATWD gain channel of four digitized and read out.  The
   horizontal (time) axis is not to scale.}
 \label{fig:atwd_timing}
\end{figure}

As explained in ref.~\cite{ICECUBE:DAQ}, two sets of ATWD chips are
operated alternately in order to reduce deadtime; the second ATWD is
available to launch during the digitization step of the first,
after a re-arm delay of $50~\mathrm{ns}$.  Significant deadtime
only occurs after two back-to-back launches and depends on how many
ATWD channels are digitized, and whether the initial hit had an LC
condition.  Since the full waveform is not needed in the absence of LC, the
digitization can be aborted early, and the ATWD channels can be cleared and
reset.  The timing sequence for back-to-back hits is shown in
figure~\ref{fig:atwd_timing}.

The total accumulated deadtime for each individual DOM is measured by counting
discriminator crossings when both ATWDs and the fADC are not acquiring
data. This deadtime varies seasonally based on the atmospheric muon
flux~\cite{ICECUBE:IceTop}.  The median fractional deadtime during a
high-rate period for in-ice DOMs 
is $6.6\times10^{-5}$, for IceTop low-gain DOMs is $7.2\times 10^{-6}$, and
for IceTop high-gain DOMs is $3.2 \times 10^{-3}$.

\subsubsection{\label{sec:flasher}Flasher Board}

Each DOM contains an LED Flasher Board, which is used to generate
light \emph{in situ} for a
variety of calibration purposes~\cite{IC3:SC,Aartsen:2013rt}, including: 

\begin{enumerate}
\item Verifying the timing response of the DOMs throughout the analysis
  software chain.
\item Measuring the position of the deployed DOMs in ice.
\item Measuring the optical properties of the ice.
\item Verifying the performance of shower reconstruction algorithms
  in measuring position, direction, and energy.
\end{enumerate}

The standard Flasher Board is
included in every DOM except the ``color DOMs''
described below. It is an annular board fitted with 12 LEDs (ETG-5UV405-30)
specified with output wavelength $405\pm5$ nm.  Laboratory
measurements with sample DOMs yield a peak at
\qty{399}{nm} and spectral width \qty{14}{nm} (FWHM) when measured at
$-20^{\circ}$~C, where the peak wavelength is shifted by 
\qty{-1}{nm} compared to room temperature~\cite{Aartsen:2013rt}.
The LEDs are arranged in six pairs, evenly spaced around the board
with a 60$^{\circ}$ separation between adjacent pairs. One LED in each pair
is pointed downward at an angle of 10.7$^{\circ}$; after refraction through the DOM glass and into
the ice, the LED
emits light horizontally into the ice. The other LED is tilted upward
at an angle of 51.6$^{\circ}$; after refraction the tilted LED
emits light upward at an angle 
of 48$^{\circ}$, close to the Cherenkov angle in ice. The angular
emission profile of the flasher LEDs was measured in the lab by
rotating a PMT connected to an
optical fiber pickup around the DOM; the readout angle was recorded
using the resistance of a potentiometer at the rotation axis.
The angular emission profile of each LED has a FWHM of
30$^{\circ}$ in air and is modeled as a Gaussian emission profile
with $\sigma = 13^{\circ}$. After refraction through the DOM glass and into
the ice, the emission profile is modified to $\sigma = 9.7^{\circ}$ in the polar direction
and $9.8^{\circ}$ in the azimuthal direction for the tilted LEDs, and $\sigma=9.2^{\circ}$ in the polar direction
and $10.1^{\circ}$ in the azimuthal direction for the horizontal LEDs.
About 10\% of the light is emitted outside the Gaussian beam, modeled by
a secondary profile proportional to $(1+\cos{\alpha})$, where $\alpha$ is the angle
away from the LED axis.

\begin{figure}[h]
 \centering
 \includegraphics[width=0.8\textwidth]{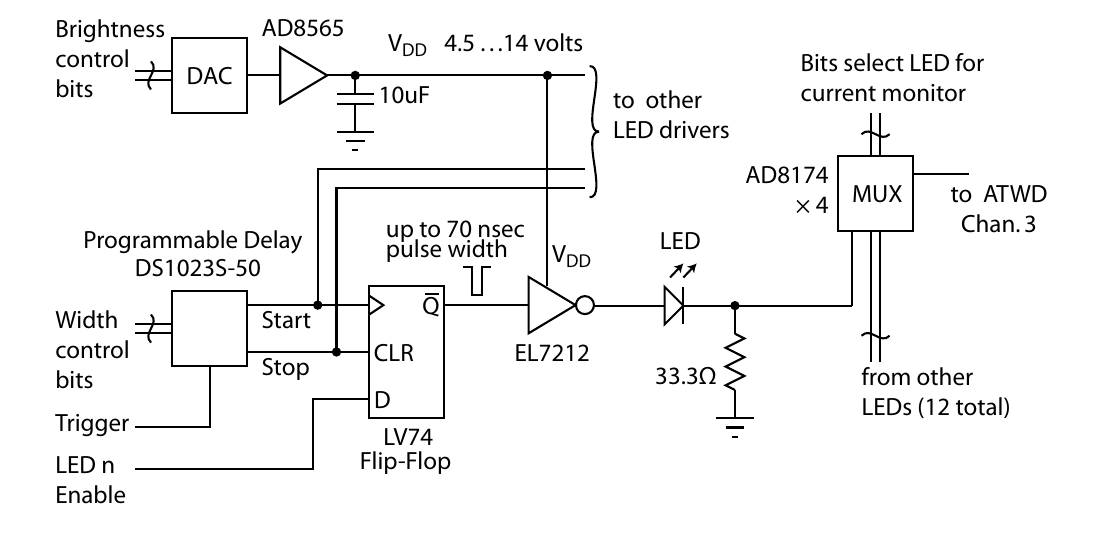}
 \caption{LED flasher circuit diagram for one of twelve LEDs, including current pulse monitor (simplified).}
 \label{fig:flasherdiagram}
\end{figure}

\begin{figure}[h]
 \centering
 \includegraphics[width=0.6\textwidth]{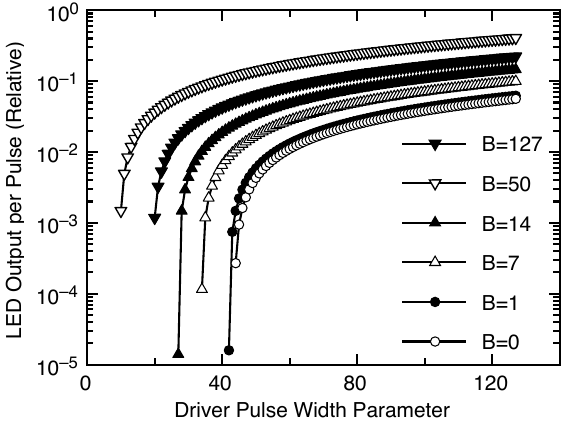}
 \caption{Light output from flasher LED pulses (relative to maximum), depending
on brightness parameter (B) and width configuration parameters.  Additional dynamic range is available
by enabling from 1 to 12 individual LEDs per DOM.}
 \label{fig:flasheroutput}
\end{figure}

The LEDs are controlled via individual high-speed MOSFET drivers; the
flasher circuit diagram is shown in Fig.~\ref{fig:flasherdiagram}. The LEDs can be turned on individually or in any
combination of the 12, by setting bits in a configuration parameter.
The photon output of each LED depends on the width and
amplitude of the driving current pulse, which are controlled as common
values for all enabled LEDs in each DOM (figure~\ref{fig:flasheroutput}).  
The pulse width parameter controls the width up to a maximum of \qty{70}{ns}; 
for sufficiently short driving current pulses the light output narrows to \qty{6}{ns} (FWHM) with
10\% afterglow decaying over 15--20 ns. The brightness parameter (0--127) controls the driving voltage between
$4.5$ and \qty{15}{V}, which yields a peak current up to
\qty{300}{mA} through the LED and current-limiting resistor.
By varying brightness and width settings as well as the number of LEDs enabled, DOMs can generate flashes
from $10^6$ to $1.4\times10^{11}$ photons, similar to the total light from
neutrino interaction showers between \qty{7}{GeV} and \qty{1}{PeV} energy.
The low end of this dynamic range requires fine tuning of driving
parameters in order to operate LEDs very close to threshold.

The LED current waveforms are recorded in an auxiliary ATWD channel, supplying
a rising edge time that also establishes the onset of the optical pulse (after a known
turn-on delay).
The repetition rate is programmable up to \qty{610}{Hz}.
Although flashers can be
operated in multiple DOMs in the same run, the DAQ does not support
time-synchronized flashing of LEDs on different DOMs, so coincident flasher
events happen only by chance. 

Sixteen ``color DOMs'' (cDOMs) are fitted with multi-wavelength
Flasher Boards; 8 are deployed on String~79 in the center of IceCube, and 8
are deployed on String~14 on the edge of IceCube.  Each
cDOM includes LEDs with nominal wavelengths of 505~nm, 450~nm, 370~nm,
and 340~nm. The LEDs are arranged in six pairs as on the
standard flasher board, three pairs of 370~nm and 450~nm and three
pairs of 340~nm and 505~nm, but all LEDs point outward horizontally. 
The properties of the LEDs on the standard DOMs and the cDOMs are
given in table~\ref{table:cdom_properties}. Differences between the nominal and
measured wavelengths are within expectations based on normal LED
variation from the manufacturer.

\begin{table}
\caption{Properties of the standard IceCube flasher LED (tilted (t)
  and horizontal (h)) and the cDOM LEDs, including wavelength $\lambda$,
  emission FWHW $\sigma$ in air, DOM polar
  angular emission FWHM in ice $\sigma_{\theta}$, and DOM azimuthal angular emission
  FWHM in ice $\sigma_{\phi}$.}
\begin{tabularx}{\linewidth}{lXXXXXX}
\toprule
 LED& nominal $\lambda$ (nm) & measured $\lambda$ (nm) & $\sigma$ air ($^{\circ}$) &
 $\sigma_{\theta}$ ($^{\circ}$) & $\sigma_{\phi}$ ($^{\circ}$)\\
\midrule
ETG-5UV405-30 & 405 & 399 & 30.0 & 9.7 (t)& 9.8 (t) \\
 &  &  &  & 9.2 (h)& 10.1 (h)\\
UVTOP335-FW-TO39 & 340 & 338 & 51.0 & 36.1 & 42.9 \\
%\hline
NS370L\_5RFS & 370 & 371 & 55.2 & 39.1 & 42.9 \\
%\hline
LED450-01 & 450 & 447 &	6.8 & 4.8 &	5.3 \\
%\hline
B5-433-B505 & 505 & 494 & 6.4 &	4.5 & 4.9 \\
\bottomrule
\end{tabularx}
\label{table:cdom_properties}
\end{table}

\subsection{\label{sec:dom_prodtest}Production and Testing}

Approximately 5800 DOMs were built and tested, with
approximately 5500 delivered to the South Pole. The DOMs satisfied
stringent requirements, needed to ensure reliable operation in the deep ice
for at least 20 years. As hot-water drilling was the principal 
driver of the deployment timeline, the DOM production schedule was
structured to supply DOMs as needed and to avoid any inventory shortfall.
The production was implemented in a 3-stage approach. Stage 1 was
the production of the initial 400 DOMs at three sites: one in the
United States at the University of Wisconsin--Madison's Physical
Sciences Lab (PSL, Stoughton, Wisconsin) and two
in Europe (DESY, Zeuthen, Germany, and Stockholm University,
Sweden). DOM testing was performed at PSL, DESY, and Uppsala University,
Sweden. This
quantity of DOMs was sufficient to verify production readiness, supply
instrumentation for the first year drilling plan, and validate the design after a deployment
season.  During Stage 2, material and supplies were procured, and another
1000 DOMs were produced and tested. Finally, Stage 3 involved procurements,
integration, and testing of the remaining DOMs.

DOM production utilized a formalized process to track each DOM through to
the end of the testing process, with each step recorded in a DOM Process
Traveler.  Technicians were trained and certified to perform DOM
integration and test tasks, and each site had separate quality control
personnel. Commercial materials were confirmed to be fully tested by the
suppliers, and regular visits were made to key vendors.  Measurement
equipment was calibrated and records maintained that verified
traceability to a reference standard.  DOM integration took place in
an electrostatic discharge (ESD)-, temperature-, and humidity-controlled environment.  The introduction
of these manufacturing protocols based on electronics industry best
practices enabled each production site to work independently yet
produce DOMs that performed identically.

DOM integration started with the attachment of the PMT collar
to the PMT.  The collar provides a mounting point for the electronic boards inside
a DOM. The PMT was then mounted into a special jig for precise
placement inside the bottom glass hemisphere.  In parallel, the mu-metal
shield was placed inside the bottom hemisphere, and the
gel was mixed and poured into the same hemisphere. The gel was then
degassed by placing under a partial vacuum, in order to avoid bubbles and
crazing in the gel. 
After degassing, the PMT was placed in the gel, and the gel was allowed to
cure for 48 hours.  After curing, the PMT Base Board was soldered onto the
leads of the 
PMT.  Separately, the PC Board Stack was assembled by attaching the Delay
Board, Main Board, Flasher Board, and High Voltage Control Board together.
The Board Stack was then mounted onto the PMT collar, the penetrator assembly
was mounted into the top hemisphere, and the two halves of the sphere were
joined.  With the entire assembly under a bell jar, the spheres were
evacuated and backfilled with dry nitrogen, 
a butyl rubber sealant applied around the seam, and the seam covered with
wide plastic tape. The interior gas pressure was reduced to 0.5 bar (at
room temperature) so that the seal remains tight even at the assumed minimum
ambient South Pole air pressure of 0.64 bar.

As the DOMs are not serviceable after deployment, an extensive testing
protocol (Final Acceptance Testing, or FAT) including temperature-cycling
and cold-soaking ensured that bad modules and early component failures were
identified and repaired before shipping.  This testing at production sites
was performed in Dark Freezer Labs (DFLs), light-tight walk-in 
enclosures capable of sustaining temperatures down to $-55^\circ$C.  Main Board
and DOM functionality was tested by self-diagnostic software installed on
each module.  Other tests included gain calibration, dark noise monitoring,
LC validation, and booting after cold-soaking.  Optical sensitivity, time resolution,
and linearity were measured using external light sources fed into the DFLs
via optical fibers and diffused over the DOM lower hemispheres at each
testing station. Sensitivity was measured primarily with a
monochromator-selected white lamp, time resolution was measured with a
pulsed diode laser (405~nm), and linearity was measured with a bright
pulsed LED source (405~nm).

\begin{figure}[!h]
 \centering
 \includegraphics[width=0.6\textwidth]{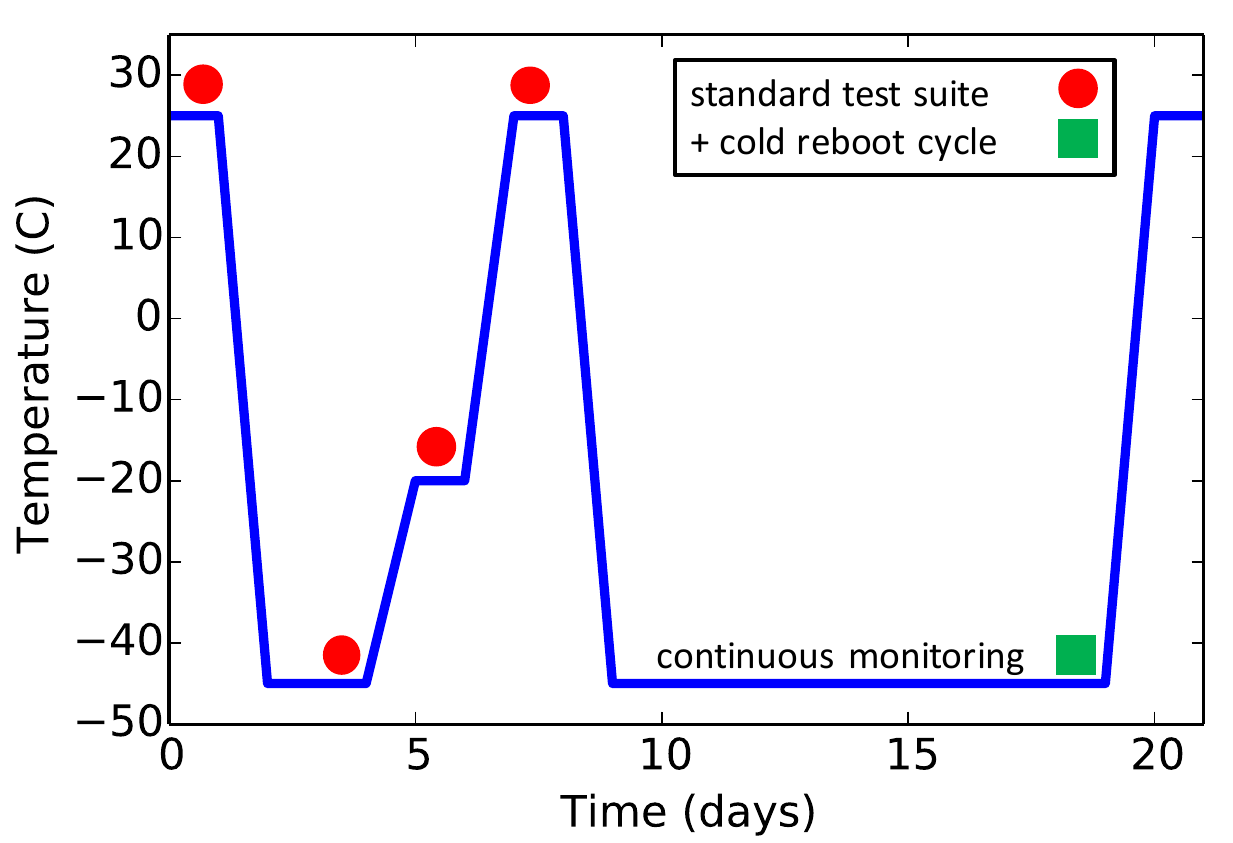}
 \caption{Final Acceptance Testing (FAT) temperature profile, including
   DOM testing steps performed at each stage.}
 \label{fig:fat_cycle}
\end{figure}

A typical FAT temperature and testing cycle is shown in
figure~\ref{fig:fat_cycle}. The initial pass rate of DOMs during FAT was
92\%.  The primary causes of failures were elevated noise rates detected during the
low-temperature soak, functional issues on the Main Board or Flasher Board,
and PMT gain instability.  The majority of failing DOMs were retested and
subsequently passed, while DOMs with serious issues were repaired if possible and
retested prior to shipment. After the successful completion of FAT, DOM
deployment harnesses were attached (figure~\ref{fig:domcable}) and the
DOMs individually packed for shipment. 

All DOMs were also re-tested at the South Pole before final deployment, to
screen out any modules damaged during transit.  The self-diagnostic
program and dark noise measurements were performed with the DOMs still in
their packing boxes, covered with additional light-tight material.  Of the
approximately 5500 DOMs shipped to South Pole, about 30 (0.5\%) were
returned to the U.S.~after failing on-ice testing.    

\subsection{\label{sec:reliability}Reliability}

As of 2016, 5397 of the 5484 deployed DOMs ($98.4\%$) are operating in
data-taking mode in the data acquisition system; the remaining 87~DOMs
have failed (table
\ref{tab:dom_failures}).  We classify DOM 
failures into two broad categories: failures during deployment and
freeze-in, and failures during subsequent operation.  The majority of the
failures (55) occurred before post-deployment commissioning; we hypothesize
that these are primarily attributable to cable failures, water leaks,
or freeze-in damage. 32 DOMs have failed after commissioning, and
we include in this count modules on a wire pair taken out of service when
the partner DOM on the same pair failed.  No particular pattern in the
failures is observed, other than they are typically during non-standard
operation or an exceptional event: a power outage, calibration run, or
flash filesystem upgrade.  The most recent two DOMs failed on May 23, 2013,
losing communications after a power outage.  Diagnosis of DOM failures
beyond identifying electrical shorts is challenging.

A total of 171~DOMs have developed issues that affect their data-taking
configuration but are still usable.  For example, DOMs with a single functional
ATWD chip have a higher deadtime but are otherwise good.  The LC settings
of functional DOMs adjacent on a string to 
dead DOMs must also be modified. In most cases, local coincidence is
disabled in the direction of a dead neighbor DOM, but in a few cases a
malfunctioning DOM can still be configured to forward the local
coincidence signals along the string even if it is not used in
data-taking. These are enumerated in table \ref{tab:dom_failures}.  

\begin{table}[h]
  \centering
  \caption{Number of DOM failures during deployment/freeze-in and after
    commissioning during detector operation, as well as DOMs with various
    issues causing them to be operated in a
    non-standard data-taking mode.  The majority of DOMs with non-standard
    LC settings function normally but have a neighbor with an issue.}  
  \label{tab:dom_failures}
  \begin{tabular}{rc|rc}
    \toprule
    DOM failures & $N$ & DOMs in non-standard mode & $N$\\
    \hline    
    deployment / freeze-in & 55 &single functional ATWD & 12\\
    post-commissioning & 32  & reduced PMT gain & 1 \\
   & & non-standard LC & 158 \\
 \bf{total} & \bf{87} & \bf{total}& \bf{171}\\
    \bottomrule 
  \end{tabular}
\end{table}

\begin{figure}[!h]
 \centering
 \includegraphics[width=0.95\textwidth]{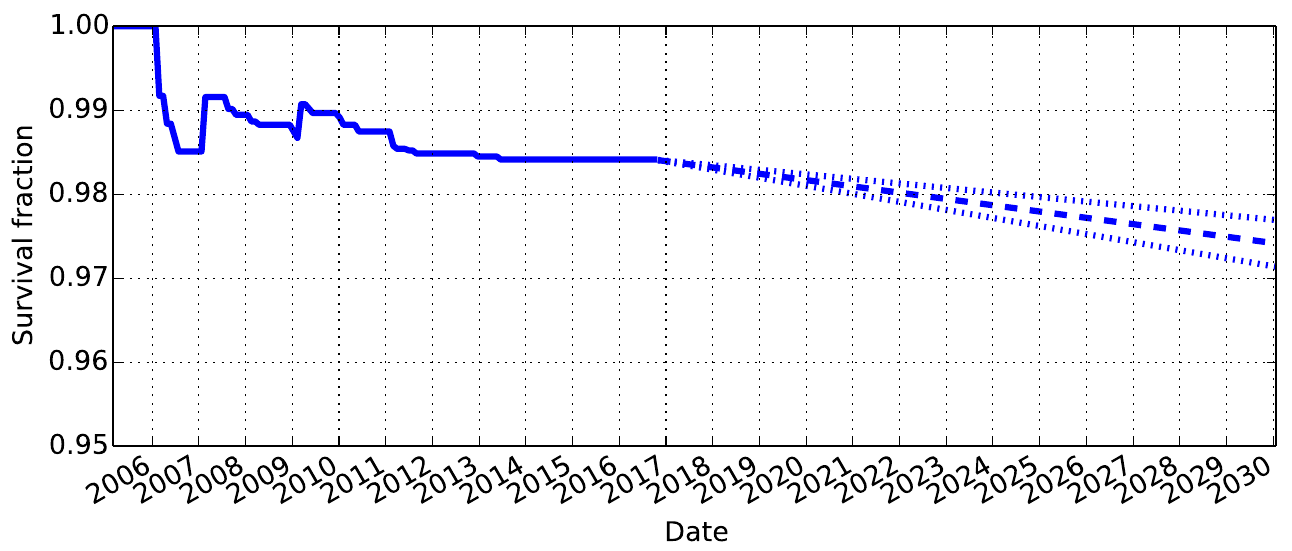}
 \caption{Actual and predicted fraction of surviving DOMs versus time, based on an assumed
 constant post-freeze-in failure rate.  The dotted lines indicate the
 central and 95\% CL estimates.  Increases before 2011 are due
 to deployments of new strings.} 
 \label{fig:dom_survival}
\end{figure}

We can estimate the surviving fraction of DOMs 25 years after the original
deployment, assuming a constant, random failure rate after freeze-in.
Specifically, we calculate the Wilson score binomial confidence interval \cite{Wilson_Score} of
survival probability using the post-commissioning failure rate of DOMs.
The estimated survival fraction as a function of 
time is shown in figure~\ref{fig:dom_survival}.  Currently we estimate the
mean failure rate to be $4.1\pm1.2~\mathrm{yr}^{-1}$, resulting in a
survival fraction in 2030 of $97.4\pm0.3\%$.  While this simplified 
model does not account for an increase in failure rate due to component aging, the
recent observed failure rate since detector completion of $1.7~\mathrm{yr}^{-1}$ is
significantly lower than the mean predicted rate, since the failure rate
during construction was higher.  We attribute
this to infant mortality and/or to improved operational protocols that
minimize the number of DOM power cycles.  DOMs are not regularly
power-cycled during data-taking but only when required to resolve an
intermittent problem.

%auto-ignore
\section{\label{sec:dom_calibration}Calibration and Stability}

Regular calibration of each individual DOM allows translation of the
recorded signals into a measurement of incident Cherenkov light at a
particular time, the basis of particle event reconstructions.  The
primary DOM calibration routine is DOMCal (section~\ref{sec:domcal}), a
calibration procedure run yearly on in-ice DOMs and monthly on IceTop
DOMs that provides the constants used to convert DOM waveforms into
physical units (section~\ref{sec:waveformcal}).  Global time calibration across the array of DOMs is 
provided by the RAPCal procedure (section~\ref{sect:dom:rapcal}) that runs during
data-taking.  The relative and absolute optical efficiency of the DOMs is
determined both by laboratory and \emph{in situ} measurements
(section~\ref{sec:domeff}). The stability of these calibrations is also
relevant, as not every calibration parameter can be tracked continuously during
data-taking (sections~\ref{sec:baselines} to \ref{sec:optical_stability}).
Understanding the statistical properties and time-dependence of the background ``dark
noise'' is also important for low-energy neutrino analyses, supernova
searches, and detector simulation (section~\ref{sect:darknoise}).

The determination of the optical properties of the ice is crucial to
the understanding of IceCube data but is beyond the scope of this work.
Further details can be found in refs.~\cite{Aartsen:2013rt,IC3:spice_lea}.

\subsection{\label{sec:domcal}DOMCal}

The calibration of the PMT waveforms recorded by the DOMs, i.e. translation
of digitized samples into voltage and time, as well as accounting for the
gain of the PMT itself, is achieved via DOM-by-DOM calibration constants
determined by the DOMs themselves. The primary reference inputs to
the DOM calibration software (DOMCal) are a precision electronic
pulser circuit providing known charges; a circuit providing a reference DC
bias voltage level; the 20 MHz oscillator on the Main
Board, used as the timing reference; and single photoelectrons, either from
background ``dark noise'' photons or a low-luminosity LED on the Main
Board.  The Main Board oscillator is manufactured by Vectron
International (model C2560A-0009 TCXO) and has a certified stability
of roughly $1 \times 10^{-11}$ for a sample interval of 5 seconds~\cite{ICECUBE:DAQ}.
The Main Board LED is used to illuminate the PMT on the same DOM, as opposed
to the LED flashers described in 
section~\ref{sec:flasher} which stimulate other DOMs. This LED produces
zero to a few tens of photoelectrons at the PMT~\cite{ICECUBE:DAQ}. Analysis and fitting of the
calibration results is done by the DOMCal software, with the results being
transmitted from each DOM to the surface as an XML file of fit parameters.

Because the operating conditions for
the in-ice DOMs are so stable, DOMCal is only run once a year on the full
detector. IceTop DOMs are calibrated once per month in order to track
temperature-dependent changes.  Calibration is
typically performed on half the detector at a time, the other half
remaining in data-taking mode in case of a transient event (such as a
supernova) during calibration. The total calibration run length is a
little under three hours. Because the calibration procedure produces
light, these runs are excluded from normal analyses.

First, the discriminator circuits used to launch the DOM are calibrated
using the electronic pulser circuit, which is capacitively coupled into the
analog front-end before the Delay Board~\cite{ICECUBE:DAQ}.  This
calibration is later refined using actual 
PMT waveforms, once the PMT gain is known.  Next, the ATWD voltage levels
are calibrated by sweeping the input DC bias voltage and recording the
corresponding waveforms at each DC level.  Because of the slight variations in the ATWD circuits, this
calibration provides an individual linear relationship between ATWD counts
and voltage for each time sample and gain channel.

The average ATWD offset at zero input voltage, or baseline, is needed for
charge integration and can in principle be determined using the previous
ATWD calibration. However, in
practice, these baselines are extremely sensitive to the operating
condition of the DOMs, and since data-taking conditions cannot be exactly
replicated while running DOMCal, the baselines used during data-taking are
determined instead by using averaged forced triggers taken during a normal
run (section~\ref{sec:baselines}).  DOMCal can still use baselines that it measures
for numerical charge integration during the calibration procedure.

The highest-gain channels of each ATWD are calibrated using the electronic
pulser circuit, and then the gains of the other ATWD channels and the fADC
are determined by varying the pulser output and comparing the charges
simultaneously recorded in multiple gain channels.  This relative
calibration is later refined using PMT waveforms stimulated by the Main
Board LED.

The tunable ATWD sampling speed is calibrated by digitizing the Main Board oscillator
waveform and recording the number of clock cycles as a function of ATWD
speed setting; the relationship between configuration setting and sampling
speed is fit with a quadratic polynomial.  The fADC sampling speed is
slaved to the 20 MHz oscillator, which is used as a timing reference, and
so is not separately calibrated.  The relative timing of 
the ATWD and fADC waveforms is determined using the electronic pulser circuit;
non-linear fitting of the digitized waveforms to known pulse templates is
required in order to determine the fADC time offset to the required accuracy.
The transit time of the PMT and Delay Board as a function of PMT high
voltage is determined by calculating the delay between the digitized
current pulse through the Main Board LED and the observed light pulse in
the ATWDs.

\begin{figure}[!h]
 \centering
 \includegraphics[width=0.6\textwidth]{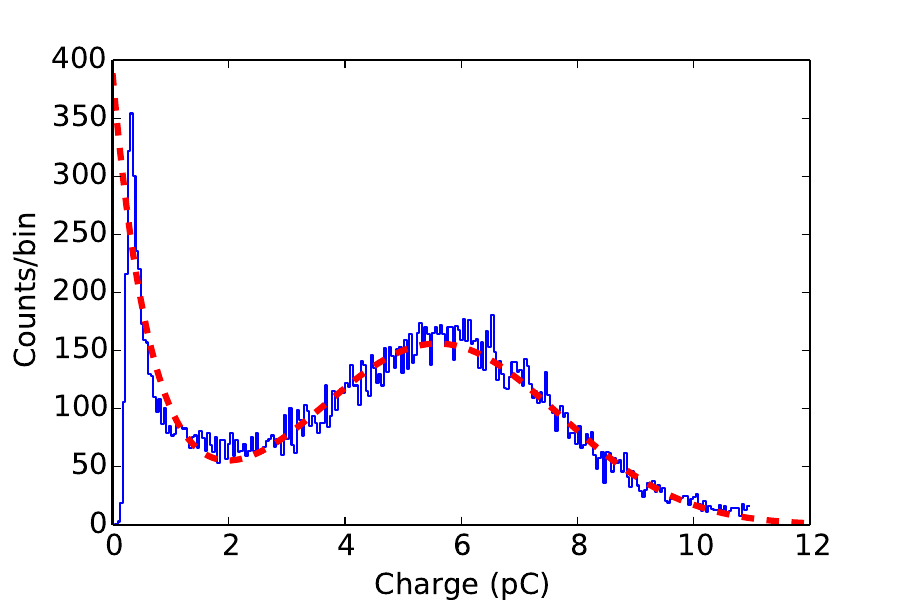}
 \caption{Sample SPE charge spectrum at a gain of $10^{7.5}$ as recorded by DOMCal and fit
   \textit{in situ} with the sum of an exponential and a Gaussian function.}
 \label{fig:domcal_hvfit}
\end{figure}

\begin{figure}[!h]
 \centering
 \includegraphics[width=0.6\textwidth]{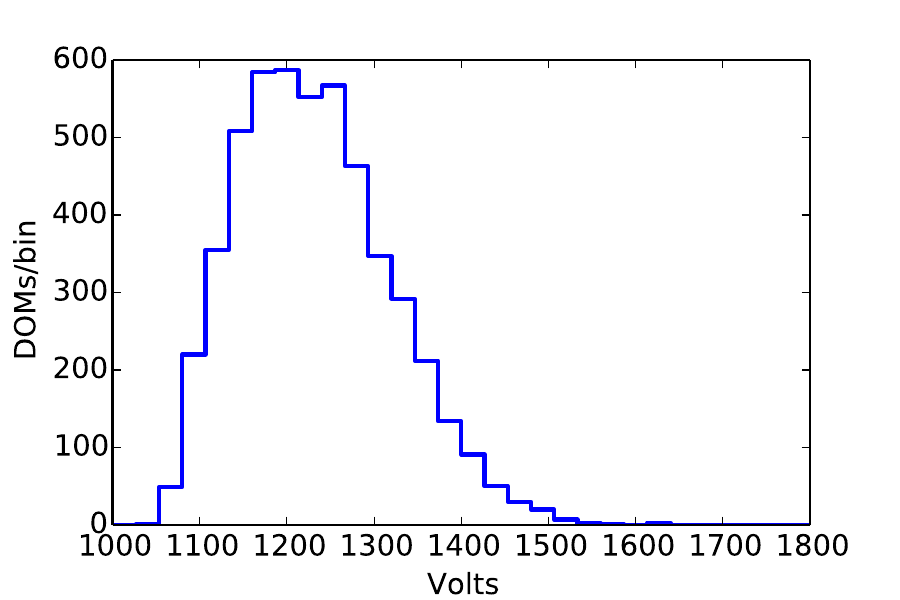}
 \caption{Distribution of PMT high voltages at $10^7$ gain for in-ice DOMs.}
 \label{fig:domcal_hv_settings}
\end{figure}

The PMT gain as a function of high voltage is calibrated using background
``dark noise'' photons --- the charge $e$ prior to amplification is quantized
and known.  At each voltage level, a histogram of many observed charges is
recorded, where the charge is determined by numerical integration of the waveform.
Each histogram is fit with the sum of an exponential and a Gaussian
function (figure~\ref{fig:domcal_hvfit}).  The peak of the Gaussian component is used
to 
determine the amplification of the PMT at each voltage, and a linear fit
of $\log_{10}(\mathrm{gain})$ versus $\log_{10}(\mathrm{voltage})$ allows
the high voltage of each PMT to be tuned to the desired operating point ($10^7$
for in-ice DOMs; figure~\ref{fig:domcal_hv_settings}).  Small 
corrections ($3-5\%$) to the gain of each DOM are determined using charge
distributions recorded by the Processing and Filtering system
(section~\ref{sect:online:filter}) during normal data-taking.  These
corrections are used to eliminate a systematic difference in the charge as
determined by DOMCal's numerical integration and the waveform pulse
unfolding used in data processing. 

\subsection{\label{sec:waveformcal}Waveform Calibration}

The deposited charge is the basis for all energy measurements in IceCube. The
calibration constants needed are 1) the pedestal, the value the digitizer
reads when the input voltage is zero, and 2) the gain, the input voltage change
required to increase the readout value by one count.  An ATWD waveform
consists of readouts from 128 effectively independent 
digitizers, while an fADC waveform consists of successive outputs of a 
pipelined digitizer (256 samples). Two calibration constants per digitizer are needed to turn each of these
raw readout values (an integral number of ADC counts) into a
voltage, from which the deposited charge is calculated.

The baseline is the mean of the pedestals of each sample in a waveform, while the pedestal pattern is
the deviation of each sample's pedestal from the common baseline.  The fADC
has no additional pedestal pattern, but for the ATWD, it is important to distinguish
between the pedestals of individual samples 
and the common baseline of the entire waveform.  In order to remove the
sample-dependent offset, the DOM subtracts the pedestal pattern from
the ATWD waveform before sending it to the surface.  Thereafter, both ATWD
and fADC waveforms can be calibrated by first subtracting the common
baseline from each sample, then multiplying by the gain. Correct charge
measurement and energy reconstruction depends on correct measurement
of the baseline, as discussed in section~\ref{sec:baselines}.

The pedestal pattern is computed by the DAQ software at the beginning of each 
run by averaging 200 forced-readout waveforms.  Accidental single photoelectrons in
the individual waveforms are averaged out, but a single coincident bright event can
result in an offset in the pedestal average.  In order to avoid such
light contamination, a second average is computed, and the
autocorrelation coefficient of the difference between the two pedestal
averages is computed at two different lags.  This autocorrelation test detects averages in which a
single waveform contains at least a 15 PE pulse (approximately 0.1 PE in the
average).  If light contamination is detected, the procedure is repeated.
The shift between the baseline of the pairs is also calculated to verify
that the baseline is stable.  This procedure ensures that fewer than 1 DOM
in 1000 runs will contain a contaminated baseline.

\begin{figure}[!h]
  \captionsetup[subfigure]{labelformat=empty}
  \centering
  \subfloat[]{\includegraphics[width=0.5\textwidth]{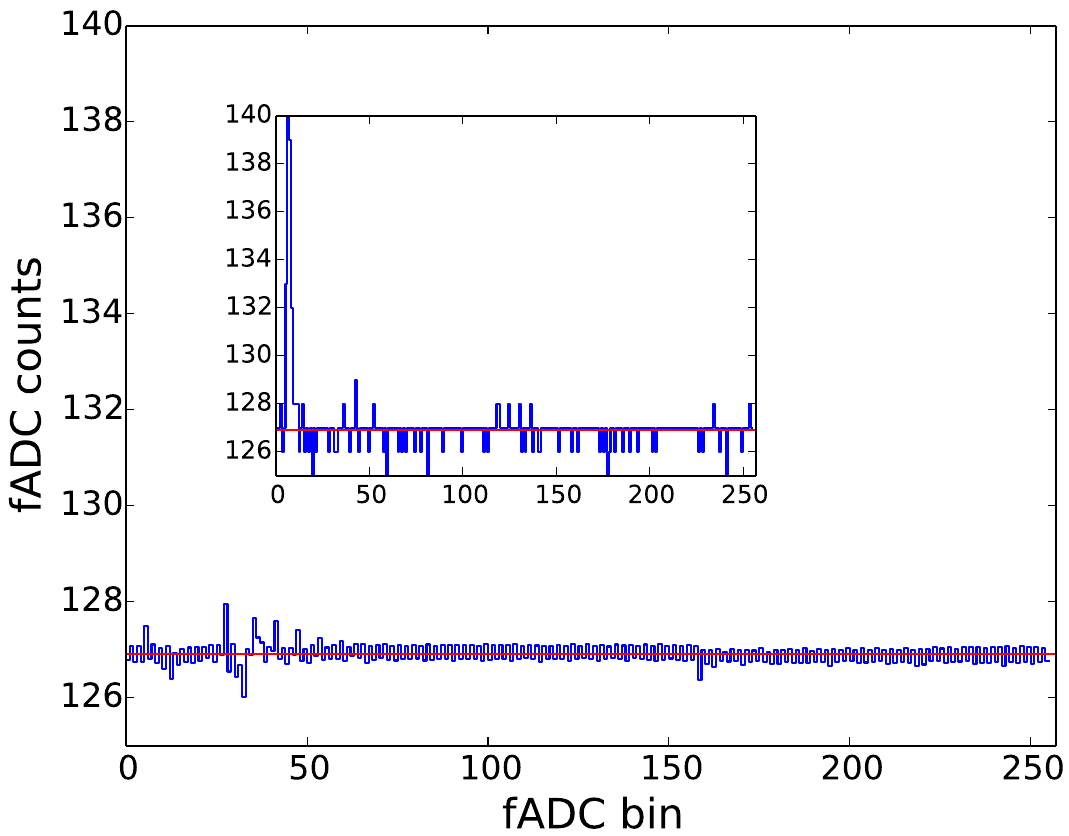}}
  \subfloat[]{\includegraphics[width=0.5\textwidth]{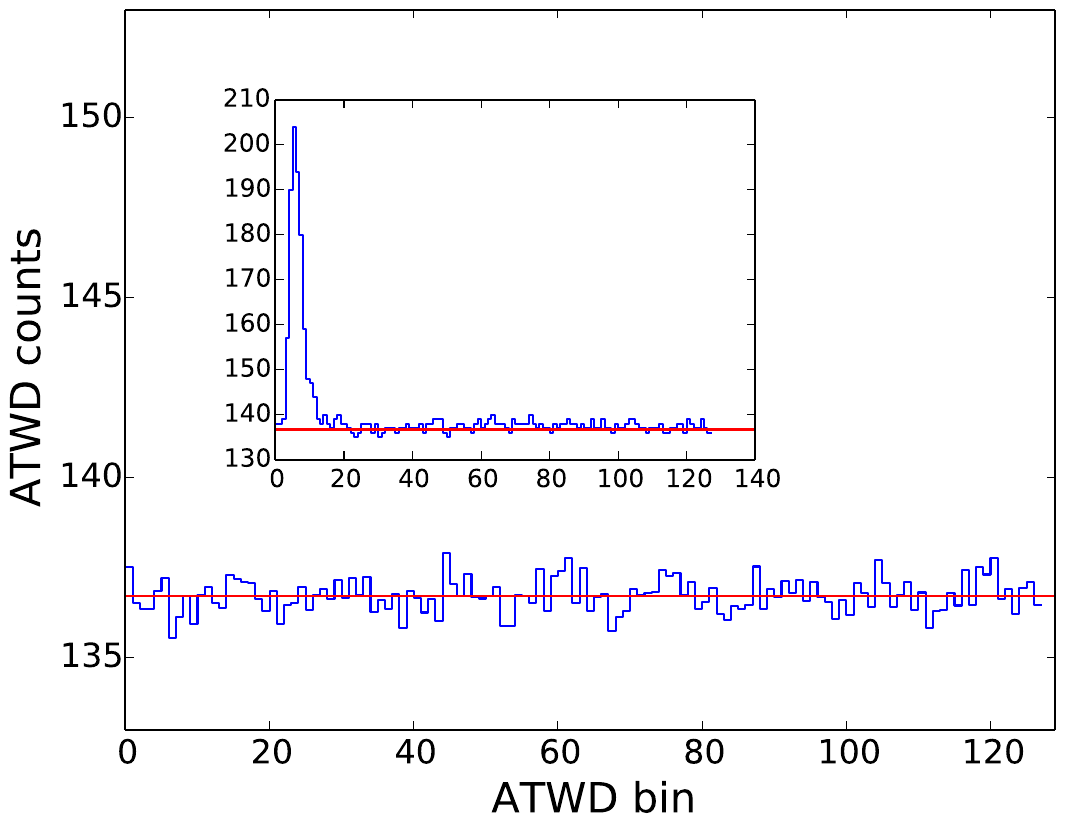}}
  \caption{Averaged beacon waveforms from the fADC (left) and ATWD (right) of
    an IceCube DOM. The waveforms are an average of approximately $1000$ beacon
   launches. The baseline, which is the mean value of the
    beacon waveform, is shown as a horizontal red line. Typical raw SPE
    waveforms are inset.}
  \label{fig:raw_baselines}
\end{figure}

The baseline is set to about 10\% of the maximum value of the
digitizer counts in
order to capture signals that go below the baseline. Since 2012, the
baseline value for each DOM is configured (section~\ref{sect:online:daqdomconfig}) in order to ensure 
stability. The baseline value differs for each digitizer channel in
each DOM, ranging from 112 to 161 counts in the fADC and 109 to 172
counts in the ATWD. The baselines for each digitizer channel in each DOM are measured with
beacon hits, forced triggers that are collected during normal data
acquisition at a rate of 1.2~Hz per DOM
in in-ice DOMs and 4.8~Hz in IceTop DOMs. Beacon waveforms
from the fADC and ATWD of a typical DOM are shown in figure~\ref{fig:raw_baselines}.

The digitizer gain is measured by DOMCal (section~\ref{sec:domcal}), a single
value for the fADC waveform and a sample-dependent value for the ATWD
waveform. The calibrated waveform voltage is then

\begin{equation}
  \mathrm{voltage} = \mathrm{(counts - baseline) \times gain}
\end{equation}

The waveform start time is then corrected for the transit time,
the average time it takes a pulse to propagate through the entire
PMT and Delay Board. The transit time correction $t_{\mathrm{transit}}$ is
dependent on the PMT high voltage $V$:

\begin{equation}
  t_{\mathrm{transit}} = \frac{m}{\sqrt{V}} + b
\end{equation}

\noindent where $m$ and $b$ are determined by DOMCal. The typical value of $m$
is 2000~ns$\cdot \sqrt{\mathrm{V}}$, and the typical value of $b$ is
80~ns, which includes the 75~ns delay line of the Main Board. The
typical transit time is therefore 130~ns.  Waveform start times from the
second ATWD chip and the fADC are further 
corrected for the delay $\Delta t$ with respect to the first ATWD
chip as determined by DOMCal, so the total start time correction is
$t_{\mathrm{transit}} + \Delta t$. 

Finally, the waveforms are corrected for the effects of droop from the
transformer that couples the Main Board to the PMT output. The toroid
coupling effectively acts as a high-pass filter on the PMT output
that makes the tails of the waveforms ``droop'' and even
undershoot the baseline. This effect is temperature-dependent and is larger
at lower temperatures. The droop correction inverts the effect of the high-pass
filter and eliminates the undershoot in the waveform tails. This is
done by calculating the expected reaction voltage from the toroid at
each sample time, and adding the reaction voltage to the calibrated waveform
to compensate. The reaction voltage decays 
according to a temperature-dependent model of the transformer behavior. 
When a readout contains consecutive launches from the same
DOM, the reaction voltage at the end of the last launch is used to
correct for the residual droop in the following launch.  DOMs
use two types of toroid transformers: the ``old toroid'' with a short
time constant that was used in early DOM production, and a ``new
toroid'' with a longer time constant that produces less
distortion. Of 5484~DOMs deployed in IceCube, 1204 are the old
toroid type, and 4280 are the new toroid type. The full correction is
modeled with two time constants, 
where the DOM's transient response $\tilde{\delta}(t)$ to an input
signal $\delta(t)$ is given by

\begin{equation}
\tilde{\delta}(t) = \delta (t) - N((1 - f) e^{-t/\tau_1} +f\ e^{-t/\tau_2})
\label{eq:droop}
\end{equation}

\noindent where the first time constant $\tau_1$ is given by 

\begin{equation}
  \tau_1(T) = A + \frac{B}{1 + e^{(-T/C)}}\ .
  \label{eq:tau1}
\end{equation}

\noindent In eqs.~\ref{eq:droop} and \ref{eq:tau1}, $T$ is the
temperature, and the constants $A$, $B$, $C$, $N$ and $f$ were
determined empirically with a dedicated analysis. For the old toroids, the
second time constant $\tau_2 =0.75\tau_1$. For the new toroids, $\tau_2$ is 500~ns.

\subsection{\label{sect:dom:rapcal}RAPCal}

\begin{figure}[!h]
 \centering
 \includegraphics[width=0.6\textwidth]{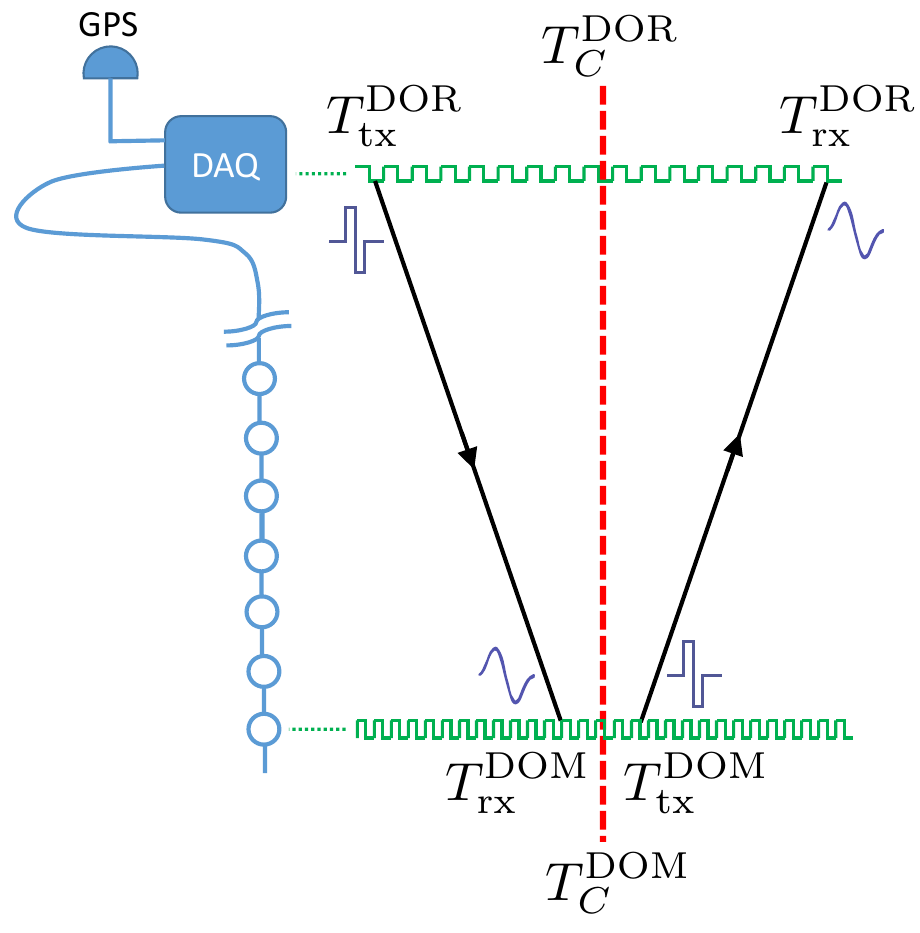}
 \caption{Diagram of the RAPCal time calibration scheme (not to scale).
   Each transmitted and received pair of signals is
   timestamped $(T_{\mathrm{tx}},T_{\mathrm{rx}})$ in the local clock domain, and by the symmetry of the
   situation the midpoints $T_C$ between transmitted and received timestamps are
   synchronous.} 
 \label{fig:rapcal_symmetry}
\end{figure}

The Reciprocal Active Pulsing Calibration (RAPCal) is a
continuously-running procedure for translating hit timestamps based on the 
individual free-running DOM clocks to the synchronized surface clocks in the
ICL, in order to establish a standard time base for the array to
$O(\mathrm{ns})$ accuracy.  Subsequently, the ICL-based hit timestamps can
then be translated to UTC.  The scheme used for the time transfer involves
transmission of a pulse to and from each DOM and is shown in
figure~\ref{fig:rapcal_symmetry}.  The base implementation is described in
\cite{ICECUBE:DAQ}; we describe here the details of the 
algorithm and validation of the DOM relative time calibration. 

While data transmission is paused, a bipolar pulse is generated and sent to
each DOM over the power/communications wire pair.  After 
this pulse has been received, a bipolar pulse --- having the same shape
as the one generated on the surface --- is generated in the DOM and sent
back to the surface.  The pulses are timestamped using the local
clock count at each transmission and reception point.  The symmetry of the
pulses sent in each direction and symmetry of the down- and up- cable
transmission enables the time transfer from each free-running DOM clock to the
surface clock and hence to UTC, without prior knowledge of the cable length
or its dispersive properties.

\begin{figure}[h]
 \centering
 \includegraphics[width=0.6\textwidth]{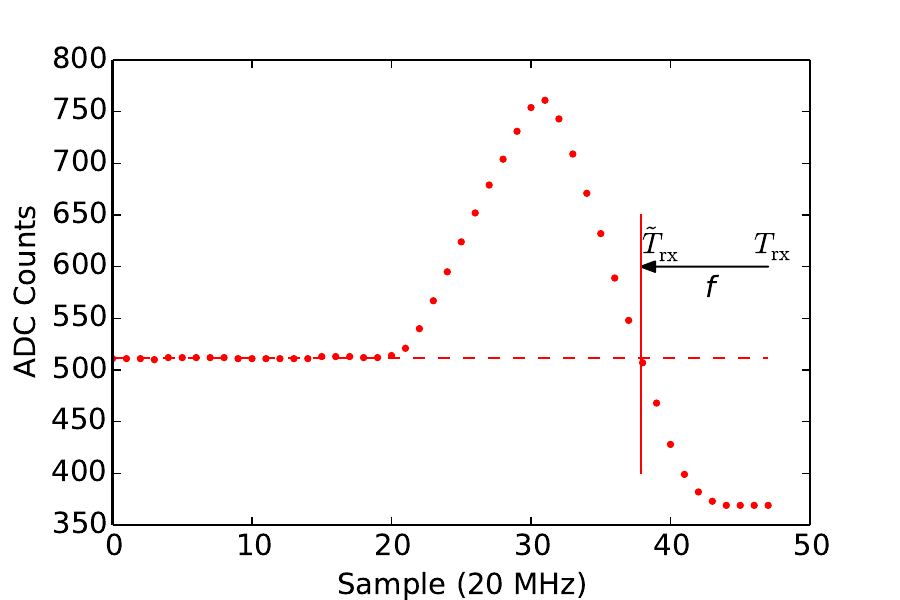}
 \caption{Digitized RAPCal pulse as received by a DOM, after cable dispersion.  The
   zero-crossing of the baseline-subtracted pulse is used as a fine-delay
   correction $f$ to the received timestamp $T_{\mathrm{rx}}$.}
 \label{fig:rapcal_zero_crossing}
\end{figure}

The received dispersed bipolar pulses are digitized by a 20 MHz communications ADC and 
timestamped using the local clock count.  The received pulse timestamp
corresponds to a fixed delay after the start of the final pulse recovery.  A RAPCal pulse sequence to and
from each DOM therefore results in a series of four timestamps 
$T_{\mathrm{tx}}^{\mathrm{DOR}}$, $T_{\mathrm{rx}}^{\mathrm{DOM}}$, 
$T_{\mathrm{tx}}^{\mathrm{DOM}}$,  and $T_{\mathrm{rx}}^{\mathrm{DOR}}$
(figure~\ref{fig:rapcal_symmetry}).  These timestamps initially count periods
of the DOM's 40 MHz clock and of the 20 MHz clock of the surface
electronics (DOR, section~\ref{sect:online:master_clock}); they are
translated into common units for the calculations described below.

Next, in order to improve the precision beyond the sampling speed of the
communications ADC, a ``fine delay'' correction $f$ to $T_{\mathrm{rx}}^{\mathrm{DOM}}$ and
$T_{\mathrm{rx}}^{\mathrm{DOR}}$ is calculated by interpolating 
to find the negative-going baseline crossing, with the baseline voltage
calculated using the initial samples of the waveform (figure~\ref{fig:rapcal_zero_crossing}):

\begin{equation}
  \tilde{T}_{\mathrm{rx}} = T_{\mathrm{rx}} - f\ .
\end{equation}

\noindent The midpoints $T_C^{\mathrm{DOR,DOM}}$ as shown in
figure~\ref{fig:rapcal_symmetry} are then determined, where

\begin{equation}
  T_C =  \frac{T_{\mathrm{tx}} + \tilde{T}_{\mathrm{rx}}}{2}\ .
\end{equation}

\noindent $T_C^{\mathrm{DOM}}$ and $T_C^{\mathrm{DOR}}$ then identify a
single point in time across the two clock domains.

The next stage of the process is to translate an arbitrary DOM hit timestamp
$t$ to UTC.  In typical operating conditions, the RAPCal procedure
is repeated once per second for each DOM.  We use the two nearest RAPCal
results before and after $t$ to derive a linear relationship

\begin{equation}
  \mathrm{UTC}(t) = (1+\epsilon)(t - T_C^{\mathrm{DOM}}) +
  T_C^{\mathrm{DOR}} + \Delta\ .
\end{equation}

\noindent The slope $(1+\epsilon)$ accounts for drifts in the DOM
clocks and is calculated from the $T_C$ pairs of the two neighboring RAPCal
results: 

\begin{equation}
  1+\epsilon = \frac{T_{C,2}^{\mathrm{DOR}} -
    T_{C,1}^{\mathrm{DOR}}}{T_{C,2}^{\mathrm{DOM}} -
    T_{C,1}^{\mathrm{DOM}}}\ .
\end{equation}

\noindent The median magnitude of $\epsilon$ is $1.34 \times 10^{-6}$.
Finally, because the timestamps $T^{\mathrm{DOR}}$ count the sub-second
time offset into the current UTC second, the UTC time offset $\Delta$ of the previous
1-second boundary, provided by the master clock, is added.  Details on the
master clock system are provided in section~\ref{sect:online:master_clock}.

The stability and repeatability of the calibration is monitored by
tracking the cable delay from RAPCal, determined by

\begin{equation}
  T_{\mathrm{cable}} = \frac{1}{2} \left( ( \tilde{T}_{\mathrm{rx}}^{\mathrm{DOR}} -
  T_{\mathrm{tx}}^{\mathrm{DOR}} ) - (1+\epsilon)(T_{\mathrm{tx}}^{\mathrm{DOM}} -
  \tilde{T}_{\mathrm{rx}}^{\mathrm{DOM}} )\right) \ .
\end{equation}

\begin{figure}[!h]
 \centering
 \includegraphics[width=0.6\textwidth]{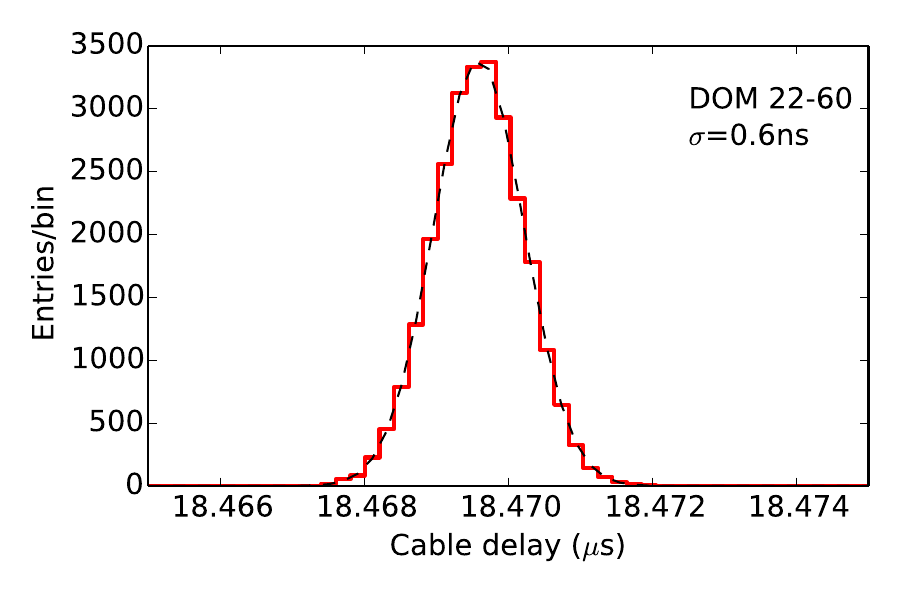}
 \caption{Distribution of one-way cable delays from multiple RAPCal
   measurements on DOM 22-60 (bottom of String 22), shown with a Gaussian fit.}
 \label{fig:rapcal_cable_len}
\end{figure}

\noindent A representative distribution of $T_{\mathrm{cable}}$ from one DOM over an 8-hour
data-taking run is shown in figure~\ref{fig:rapcal_cable_len}, with a
standard deviation of 0.6 ns.  Individual RAPCal measurements in which $T_{\mathrm{cable}}$
deviates by more than 10 ns from a moving average are discarded.

\begin{figure}[!h]
  \captionsetup[subfigure]{labelformat=empty}
  \centering
  \subfloat[]{\includegraphics[width=0.495\textwidth]{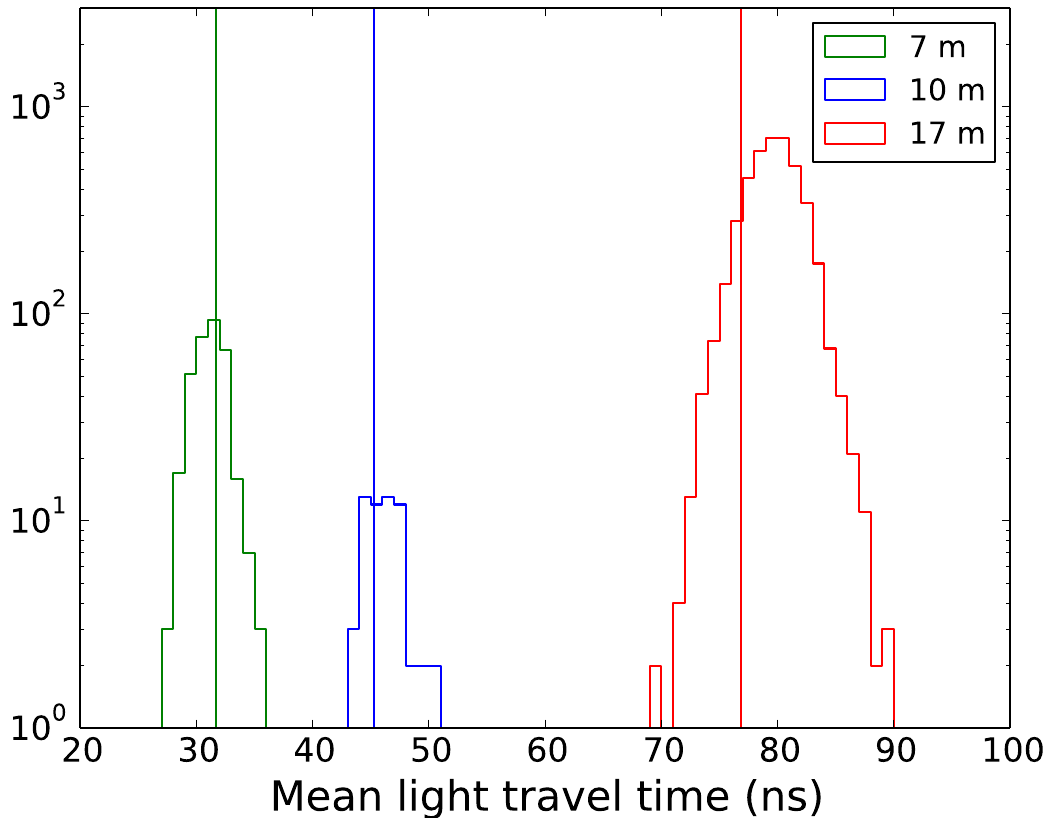}}
  \subfloat[]{\includegraphics[width=0.5\textwidth]{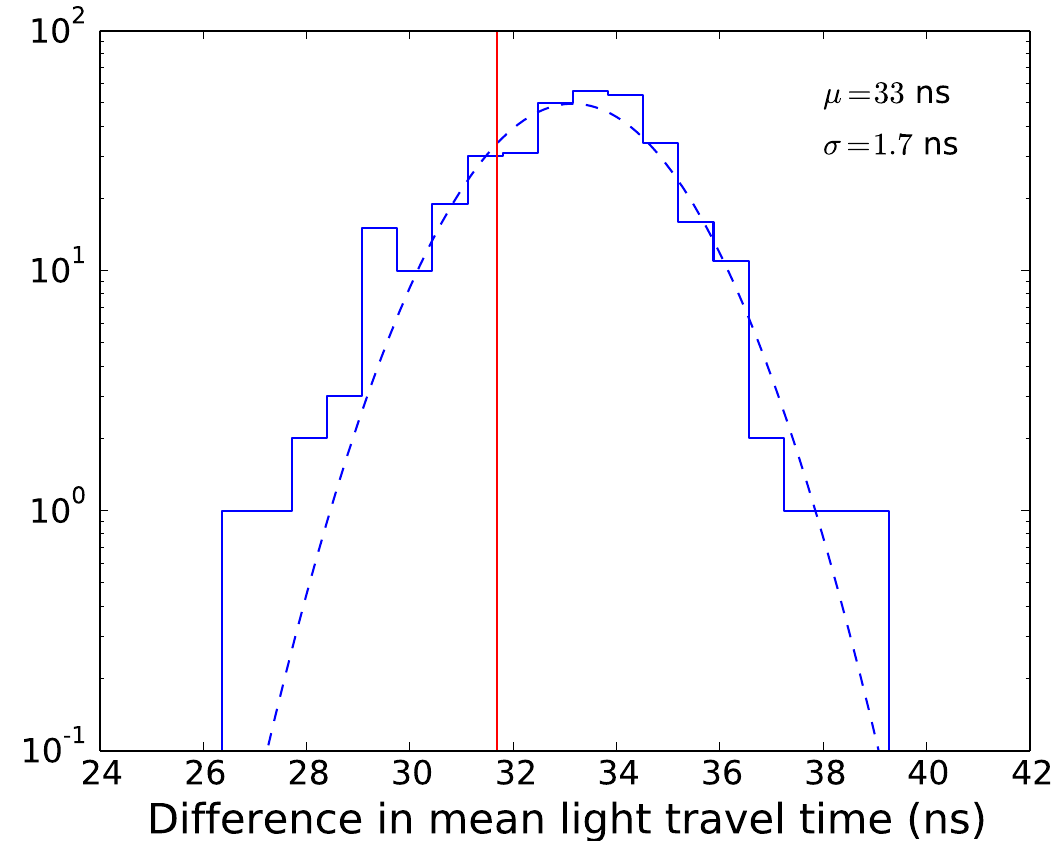}}
  \caption{Left: Time from flasher emission to detection at the DOM above for 17~m vertical spacing
    (red), 10~m vertical spacing (blue) and 7~m vertical spacing
    (green). The expected light travel times in ice for each distance are marked with
    vertical lines. Right: arrival time difference (blue) between the two
    DOMs above the flasher for DOMs with 7~m vertical spacing.  The
    vertical red line denotes expected light
    travel time in ice, and the mean of a Gaussian fit to the distribution
    (dashed blue line) is 33~ns, due to
    the effects of scattering in ice. The width of the distribution is 1.7~ns.}
  \label{fig:flashertiming}
\end{figure}

The time calibration through the entire data acquisition and software
processing chain is verified using the LED flashers. During
commissioning, performed on each string after deployment, all 12~LEDs on
each DOM are flashed simultaneously at 
maximum brightness, and the arrival times of the first photons at the DOM above the flashing DOM
are recorded. Given the vertical spacing of 17~m on a standard IceCube
string and the group index of
refraction of  1.356 in ice at 400~nm \cite{price_woschnagg_ice}, the expected light travel time
from the flashing DOM to the DOM above is 77~ns. In DeepCore, the DOM
vertical spacings are 10~m and 7~m, corresponding to light travel
times of 45~ns and 32~ns respectively. The mean light travel
time to the DOM above for all flashing DOMs in ice is shown in
figure~\ref{fig:flashertiming}. The mean arrival time agrees with the
expectation for the DeepCore DOMs. For the standard DOMs, the observed
light travel time is about 3~ns longer than the expected light travel
time, due to the effects of scattering in the ice over the longer
distance. The accuracy of the photon arrival time with respect to the
arrival time of any other photon is measured using
the difference between arrival times at the two DOMs above the
flasher, which eliminates any uncertainty in the flasher source
time. Figure~\ref{fig:flashertiming} shows the distribution of the
difference in arrival times at the two DOMs above the flasher for the
DOMs with 7~m spacing, as the DOMs with larger spacing are more
affected by scattering in ice. The width of the distribution is
1.7~ns, so the measured timing accuracy is 1.7/$\sqrt{2}$~ns $=$ 1.2~ns. Muons are also used to
verify the time calibration~\cite{ICECUBE:DAQ}, including the absolute time difference
between the IceTop surface stations and the in-ice DOMs~\cite{IC3:perf}.

\subsection{\label{sec:domeff}DOM Optical Efficiency}

\begin{figure}[!h]
 \centering
 \includegraphics[width=0.55\textwidth]{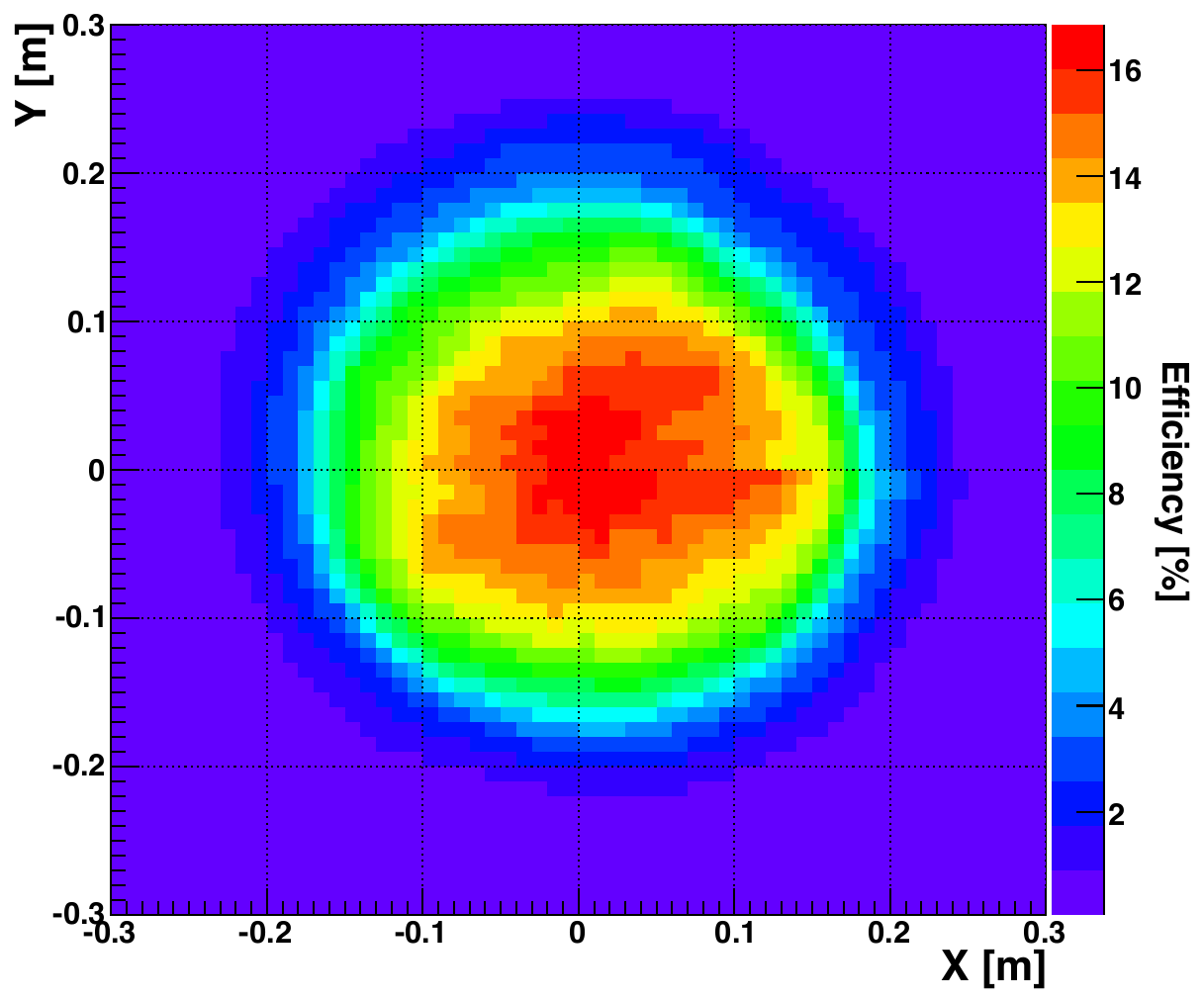}
 \caption{Position dependence of the response of a DOM, normalized to
   the absolute efficiency at 365~nm. The x-y coordinates measure
   distance from the center of the DOM.}
 \label{fig:goldendom}
\end{figure}

A baseline value for the photon detection efficiency has been established
by combining PMT measurements at 337~nm and a separate model of wavelength-
and angle-dependent effects.  Absolute sensitivity measurements were
performed on 13 IceCube PMTs, using single photons from a primary 337~nm laser beam of calibrated
intensity~\cite{ICECUBE:PMT}. The results agree well with independent
Hamamatsu measurements of sensitivity in the range 270--730~nm, which
were then normalized to the 337~nm measurement.  The resulting quantum
efficiency at the center of the PMT at 390~nm is 25\%.  A full
simulation model of the DOM includes the wavelength dependence of the PMT
response, optical absorption in the DOM glass and gel, discriminator
threshold effects, and photocathode non-uniformity.  The angular dependence is
dominated by the shape of the photocathode and its response variation away
from the center, which was measured for 135 bare
PMTs~\cite{ICECUBE:PMT}. The efficiency was also measured for 16 fully integrated
DOMs, including glass and gel; the response of an example DOM is shown
in figure~\ref{fig:goldendom}. The efficiency measurement used the
337~nm nitrogen laser, as well as LEDs at 365~nm, 470~nm, 520~nm, and
572~nm due to the low transparency of the glass at short wavelengths.

The efficiency model as determined from the laboratory measurements was supplemented with \textit{in
  situ} measurements using Cherenkov light from muons in the ice. The
efficiency measured \textit{in situ} includes effects of the cable
shadowing the DOM and the refrozen ``hole ice'' in the vicinity of the
DOM. In one
study, low-energy muons (median energy 82~GeV) with well-understood light
emission were selected to illuminate the downward-facing PMTs in the ice as
directly as possible. The number of photons detected at different
distances from these muons was then compared to predictions of the
simulation model~\cite{IC3:ereco}.  Based on this and other \textit{in
  situ} analyses, the central value for the efficiency was adjusted upward
by 10\% in the simulation, compared to the baseline.  For physics analyses,
the normalization of the absolute DOM efficiency is generally included as a
nuisance parameter with prior uncertainty of 10\% that includes other uncertainties related to
generation and propagation of the light. Additional laboratory measurements
on assembled DOMs are in progress, including wavelength- and angle-dependent
effects, and are expected to reduce uncertainties~\cite{ICECUBE:DOMEFF}.

The absolute calibration at 337~nm was performed at room temperature on a
small subset of IceCube PMTs. The relative efficiency of all assembled DOMs
was separately measured as part of production testing
(section~\ref{sec:dom_prodtest}), using a 405~nm pulsed laser and a system of
fibers and diffusers to illuminate DOMs within 50$^{\circ}$ of the
PMT axis.  Using this system, the relative efficiency of DeepCore DOMs
(high quantum efficiency type) was measured to be higher by a factor
of 1.39
compared to standard IceCube DOMs, agreeing well with the
manufacturer-specified value of 1.40. Further \textit{in situ} studies
using muons yielded a factor of 1.35~\cite{ICECUBE:DC}, an
effective value including the Cherenkov spectrum folded with the different
wavelength sensitivity curves of the two types of PMTs.  The production
testing system also established that the efficiency change from room
temperature to $-40\ ^{\circ}\mathrm{C}$ is less than 1\% when gain is maintained at
the design value of $10^7$.

\subsection{\label{sec:baselines}Baseline Voltage Stability}

\begin{figure}[!h]
 \centering
 \includegraphics[width=0.6\textwidth]{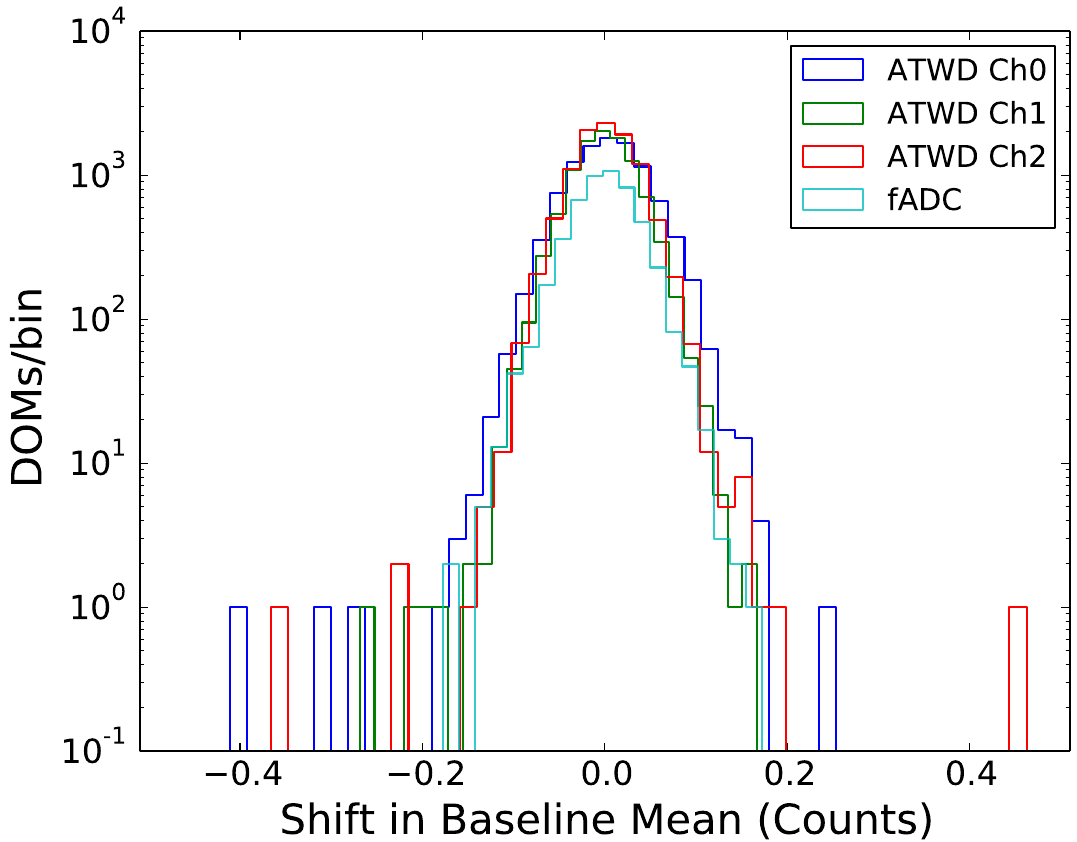}
 \caption{Distribution of shifts in baseline values in all ATWD
   channels and the fADC for all DOMs from May 2015 to April
   2016. The DOM configuration was unchanged during this period.}
 \label{fig:baseline_stability_2015}
\end{figure}

The beacon hits from which the digitizer baselines are derived
(section~\ref{sec:waveformcal}) are 
monitored continuously throughout the year. The average values of the
beacon baselines are very stable, with shifts of no more than
0.2~counts per year, corresponding to 0.018~mV in the fADC and
0.025~mV in the high-gain ATWD channel. The baseline shifts from May
2015 to April 2016 are shown in figure~\ref{fig:baseline_stability_2015}. 

\begin{figure}[!h]
 \centering
 \includegraphics[width=0.8\textwidth]{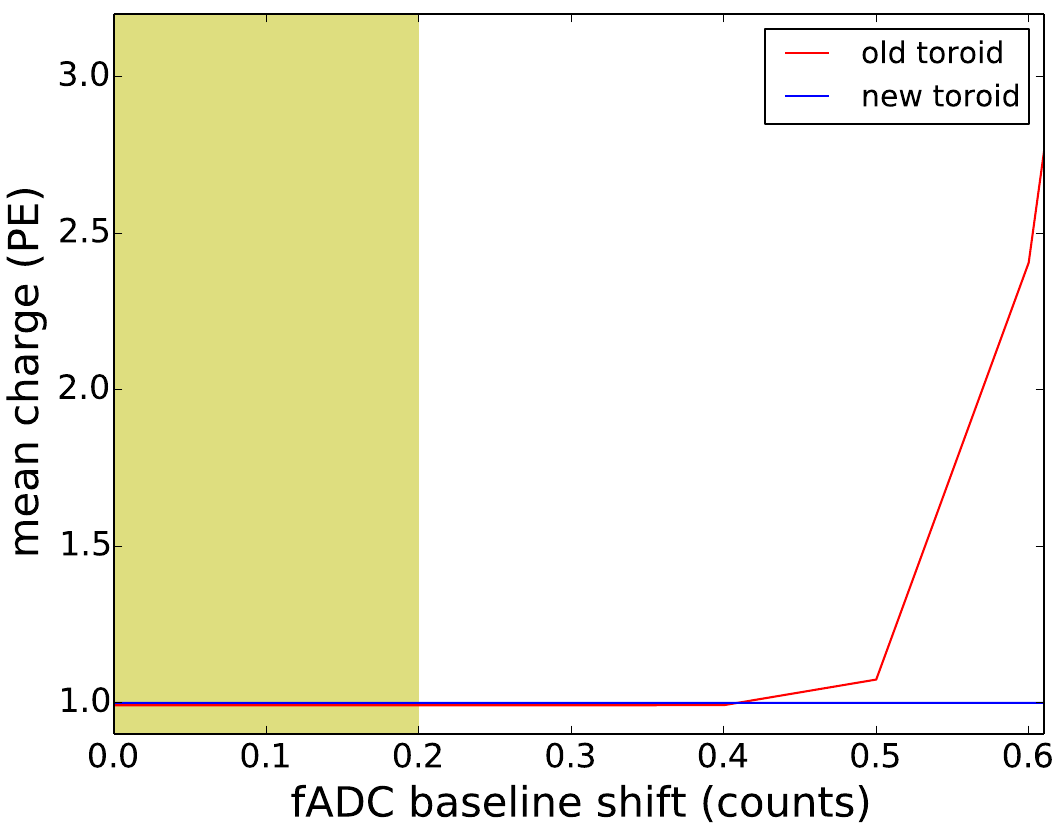}
 \caption{Reconstructed charge from a simulated single photoelectron
   deposit as a function of fADC baseline shift, for both old- and
   new-toroid DOMs. The shaded region indicates the observed range of 
   baseline variation from the nominal value in ice; no observable
   distortion in the charge spectrum is seen at these values.}
 \label{fig:charge_fadcshift}
\end{figure}

Every year when the detector
is recalibrated, adjustments in the SPE discriminator threshold can
cause shifts of up to 0.6~counts (0.54~mV) in the fADC
baselines, for reasons not completely understood. ATWD baselines are
unaffected by the SPE discriminator 
setting. Correcting recorded waveforms for the effect of transformer
coupling as described in section~\ref{sec:waveformcal} has
the side effect that small DC offset errors are converted into an
apparent PMT current that increases linearly in time over the course
of a few microseconds. The effect is stronger for DOMs with old-type toroid
transformers, where a baseline error of 0.6~counts can 
turn an actual deposited charge of 1~PE into a measured charge of over
2~PE with an unphysical time structure. The observed distortion in a
simulated single photoelectron charge due to fADC baseline shifts is
shown in figure~\ref{fig:charge_fadcshift}.  To avoid this problem, whenever the
discriminator thresholds are changed, the fADC baselines are re-measured,
and the values used for calibration are updated. As long as 
the discriminator thresholds are unchanged, the baselines are stable
to within 0.2~counts
as shown above, and no charge distortion is seen at that level of
baseline stability.

The baselines can be sensitive to radio-frequency interference (RFI). In
2009, RF signals from the COBRA meteor radar transmitter broadcasting at
46.3~MHz \cite{meteor_radar}
appeared as sinusoidal or spiky modulations in the waveform
baselines.  To mitigate this effect, one DOM was operated without PMT high
voltage and used to monitor the RF broadcasts and tag periods of RFI in
order to avoid data contamination.  Also
in 2009, RFI from DOM 68-42 affected nearby DOMs after it failed during DOM
calibration and appeared to begin sparking, resulting in sinusoidal distortions
in the baselines of neighboring DOMs. The meteor radar
transmitter is no longer operating, and investigations into RFI
from more recent radar activities at South Pole (SuperDARN
\cite{superdarn}) have not revealed any measurable interference, either in DOM
baselines or in RAPCal time stability. 

\subsection{\label{sec:gain_stability}Gain Stability}

The gain stability of the DOM, or the stability of the amplified charge
from converted photons, depends on a number of factors including stability
of the PMT high voltage, Main Board channel amplifiers, and the digitizers.
We can examine these subsystems using both historical calibration results
and by tracking the SPE charge during data-taking.

\begin{figure}[!h]
  \captionsetup[subfigure]{labelformat=empty}
  \centering
  \subfloat[]{\includegraphics[width=0.5\textwidth]{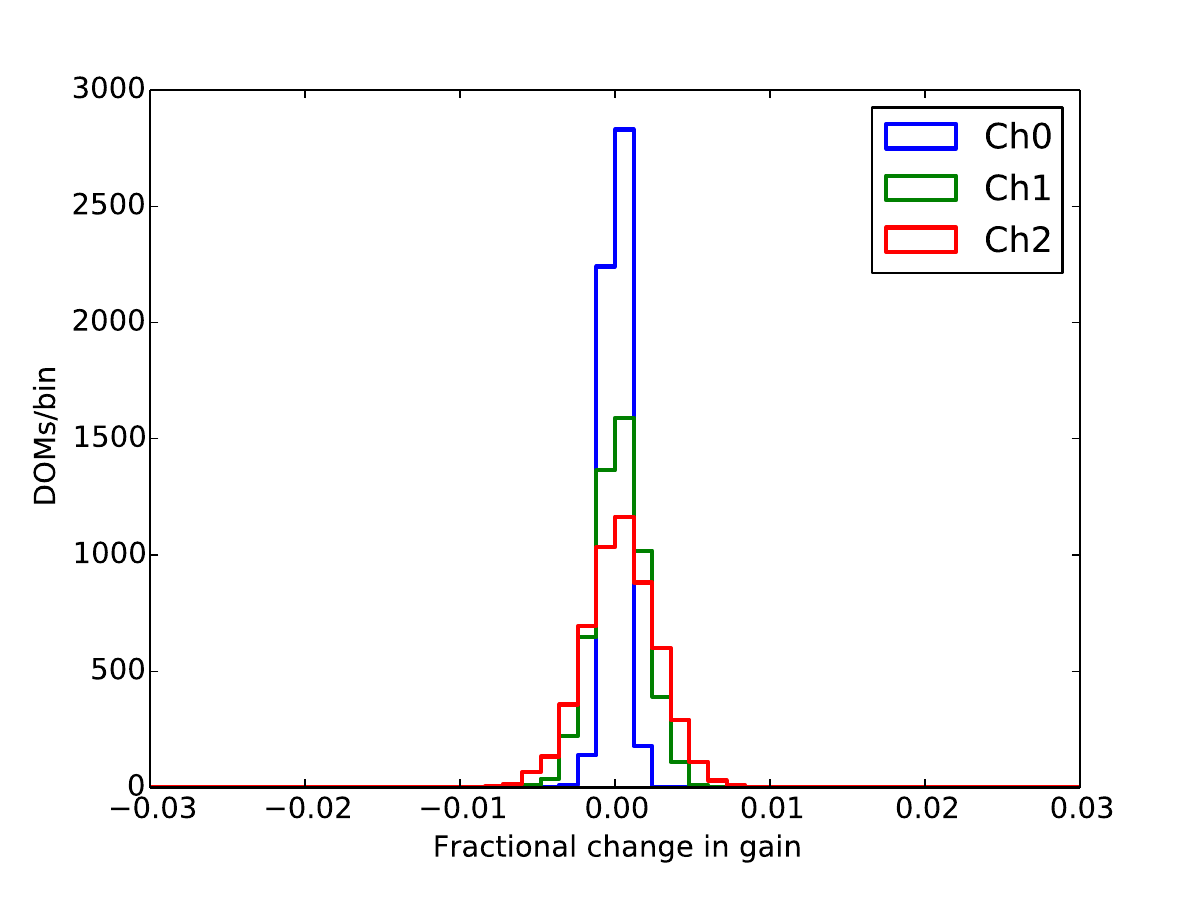}}
  \subfloat[]{\includegraphics[width=0.5\textwidth]{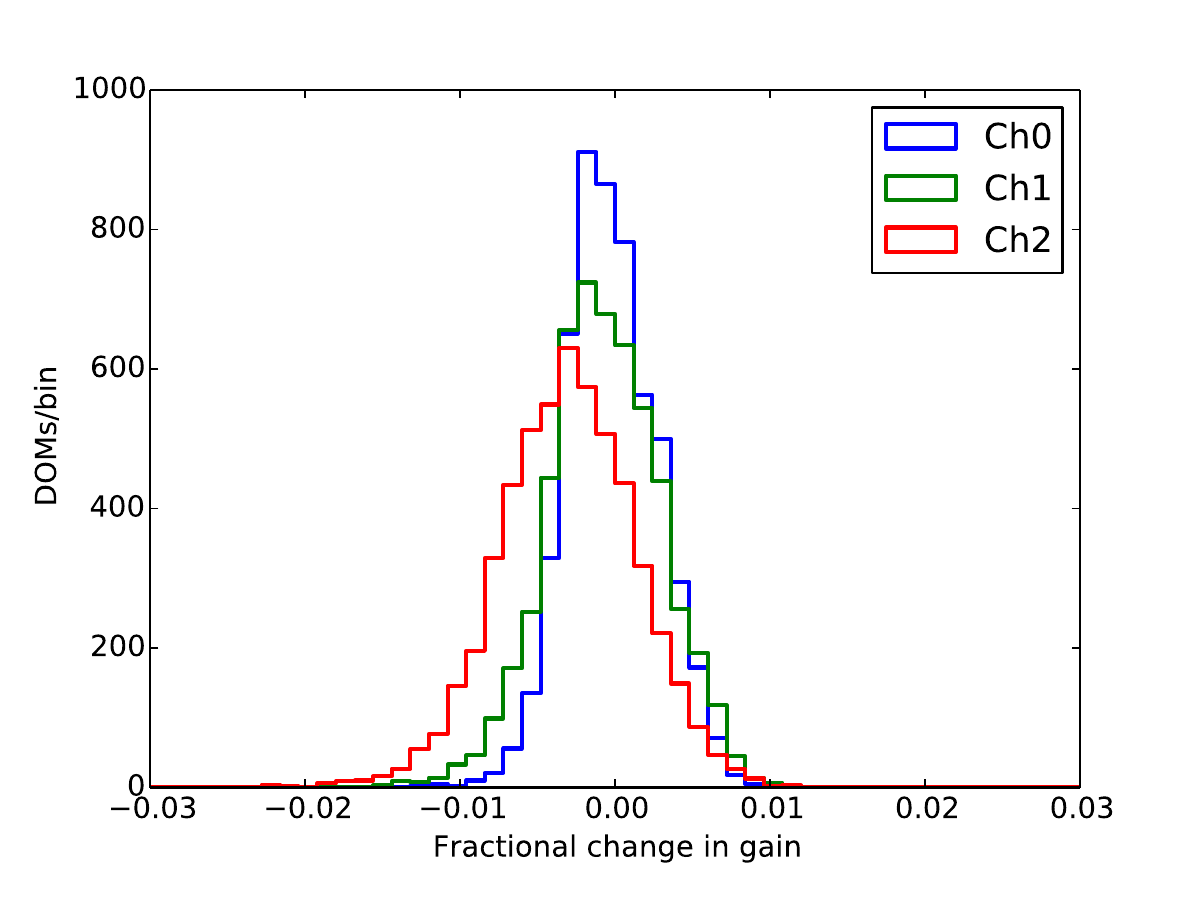}}
  \caption{Fractional ATWD channel amplifier gain shifts, from 2015 to
    2016 (left) and 2011 to 2016 (right).  Channels 0, 1, and 2 are
    high gain, medium gain, and low gain respectively.}
  \label{fig:domcal_ch_gain}
\end{figure}

Variations of the front-end electronic amplifiers or the digitizers
themselves can potentially lead to changes in the overall gain of the DOM electronics.
The stability is demonstrated by comparing the Main Board channel amplifier gains
from sets of calibrations taken from 2011 to 2016
(figure~\ref{fig:domcal_ch_gain}).  From year to year, the amplifier gain 
calibration is repeatable to 0.1\%, 0.2\%, and 0.5\% in the high-gain,
medium-gain, and low-gain channels respectively.  Since detector completion
in 2011, a small systematic shift of $-0.3\%$ is visible in the low-gain
channel, but this is corrected by updating the calibration constants of
each DOM.

\begin{figure}[!h]
 \centering
 \includegraphics[width=0.6\textwidth]{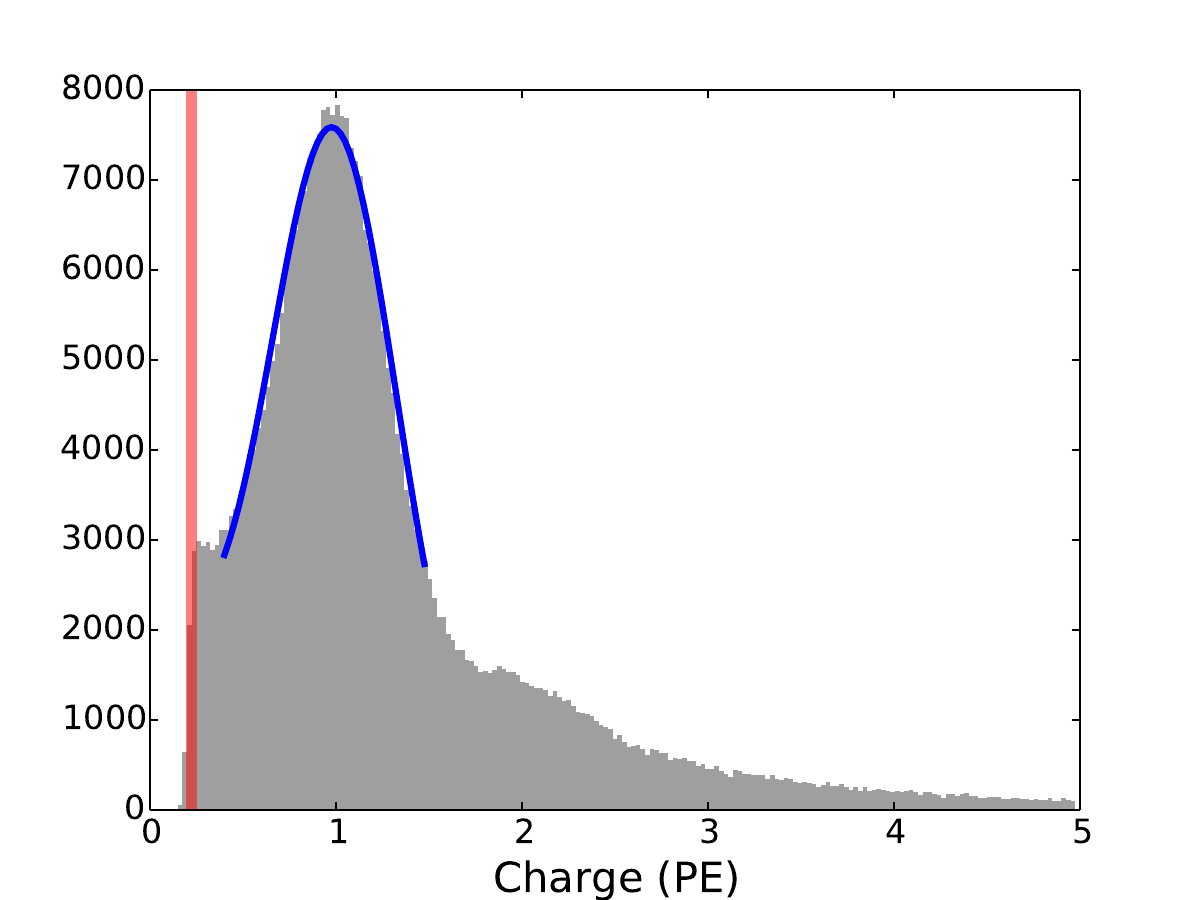}
 \caption{Charge distribution of a typical in-ice DOM. The threshold
   is marked in red, and a Gaussian + exponential fit to
   the SPE region is shown in blue. The mean of the Gaussian is used
   to monitor the gain stability.}
 \label{fig:spe_fit_thresh}
\end{figure}

\begin{figure}[!h]
  \captionsetup[subfigure]{labelformat=empty}
  \centering
  \subfloat[]{\includegraphics[width=0.5\textwidth]{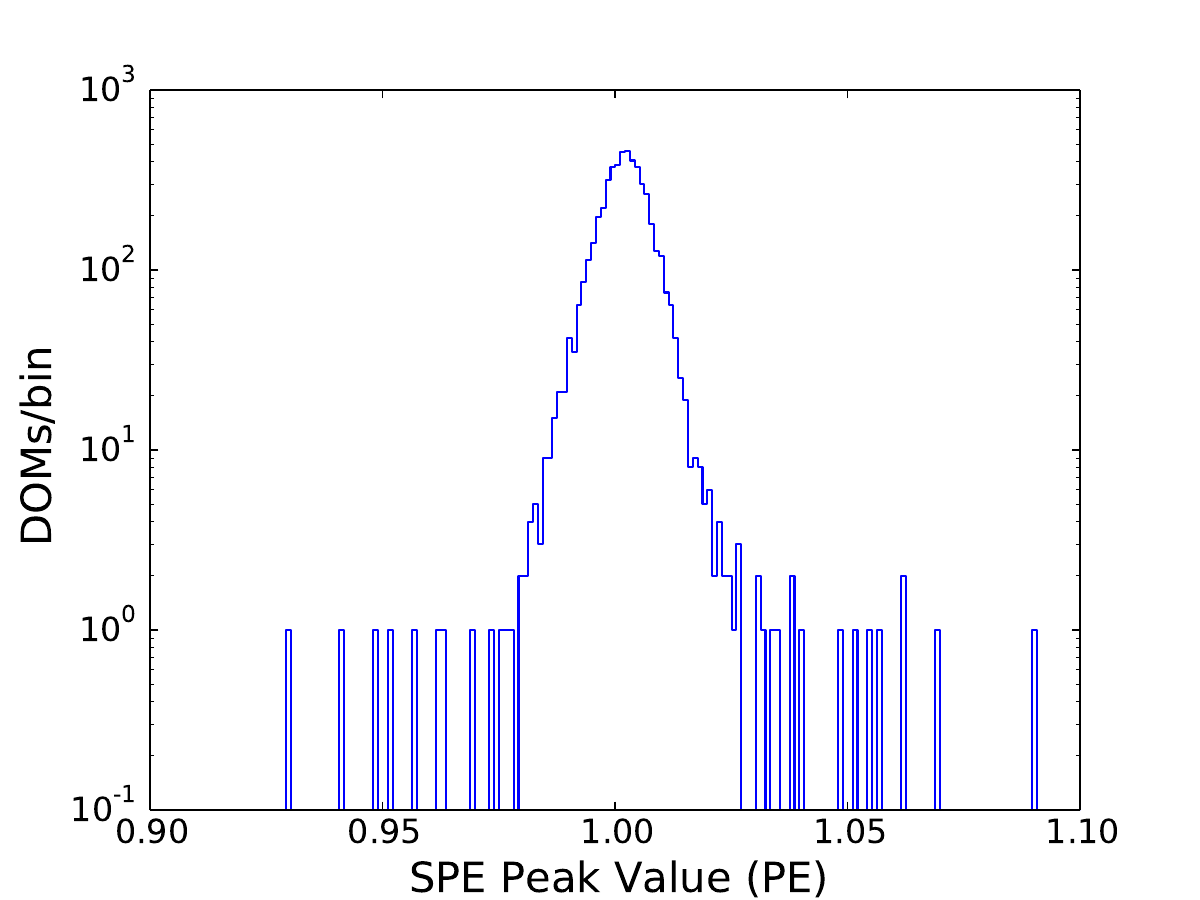}}
  \subfloat[]{\includegraphics[width=0.5\textwidth]{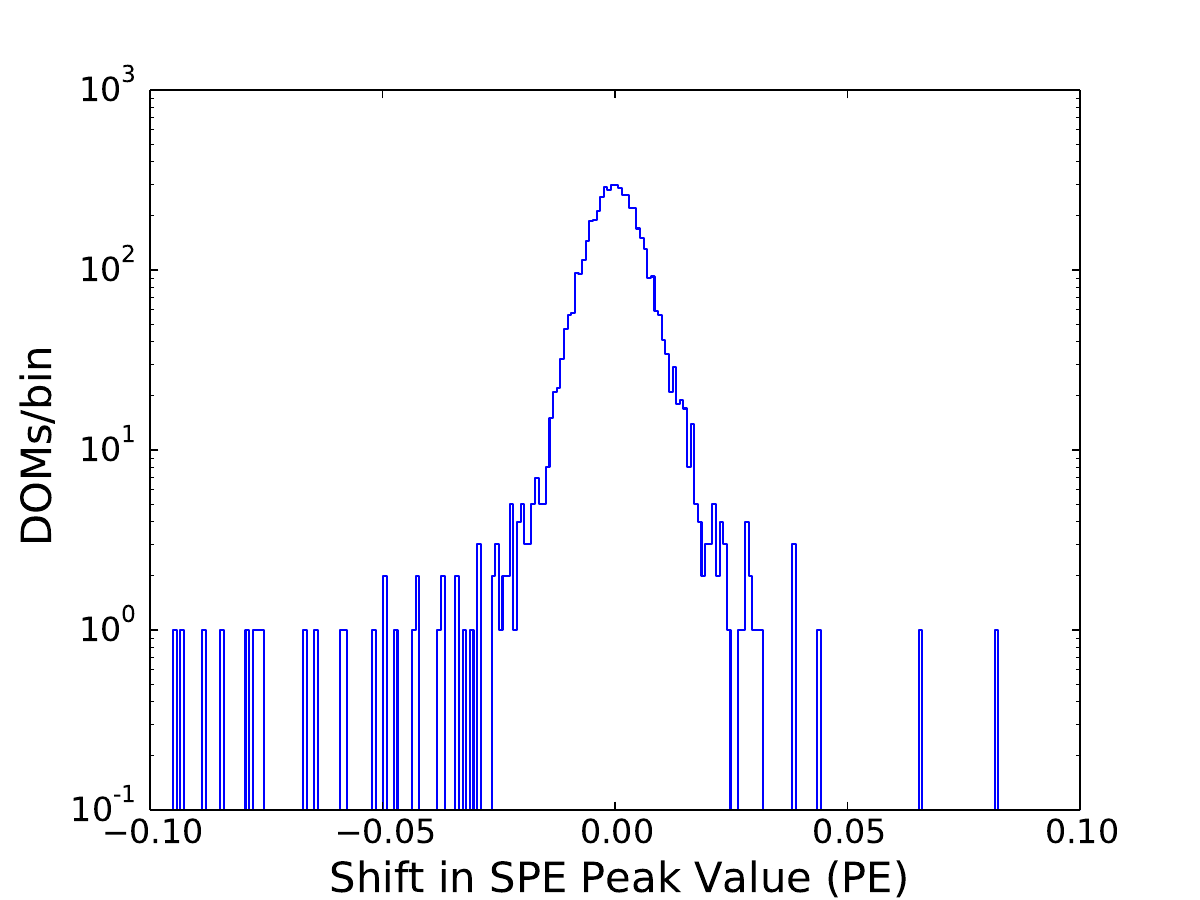}}
  \caption{Distribution of the mean of the Gaussian fit to the SPE
    peak (left) and the shift in this value for each in-ice DOM (right) between May 2015 and April
   2016.}
  \label{fig:gain_spe_stability}
\end{figure}

The stability of the total gain, including both the PMT and
front-end electronics, is monitored during data-taking using the
single photoelectron peak of the charge distribution on each DOM from
cosmic ray muons. A
Gaussian + exponential function is fit to the peak region as in
figure~\ref{fig:spe_fit_thresh}, and the mean of the Gaussian is
tracked throughout the year. The threshold, defined as the bin in the histogram for which 99\% of the area of the histogram is contained in the sum of all higher bins, is also tracked through
the year. The peak position is calibrated to 1~PE
and is stable to within 0.01~PE for 95\% of all DOMs, as shown in
figure~\ref{fig:gain_spe_stability}. Over 99\% of DOMs show no
measurable change in the threshold as long as the discriminator
thresholds are unchanged; these settings are only changed once per year.

\begin{figure}[!h]
 \centering
 \includegraphics[width=0.8\textwidth]{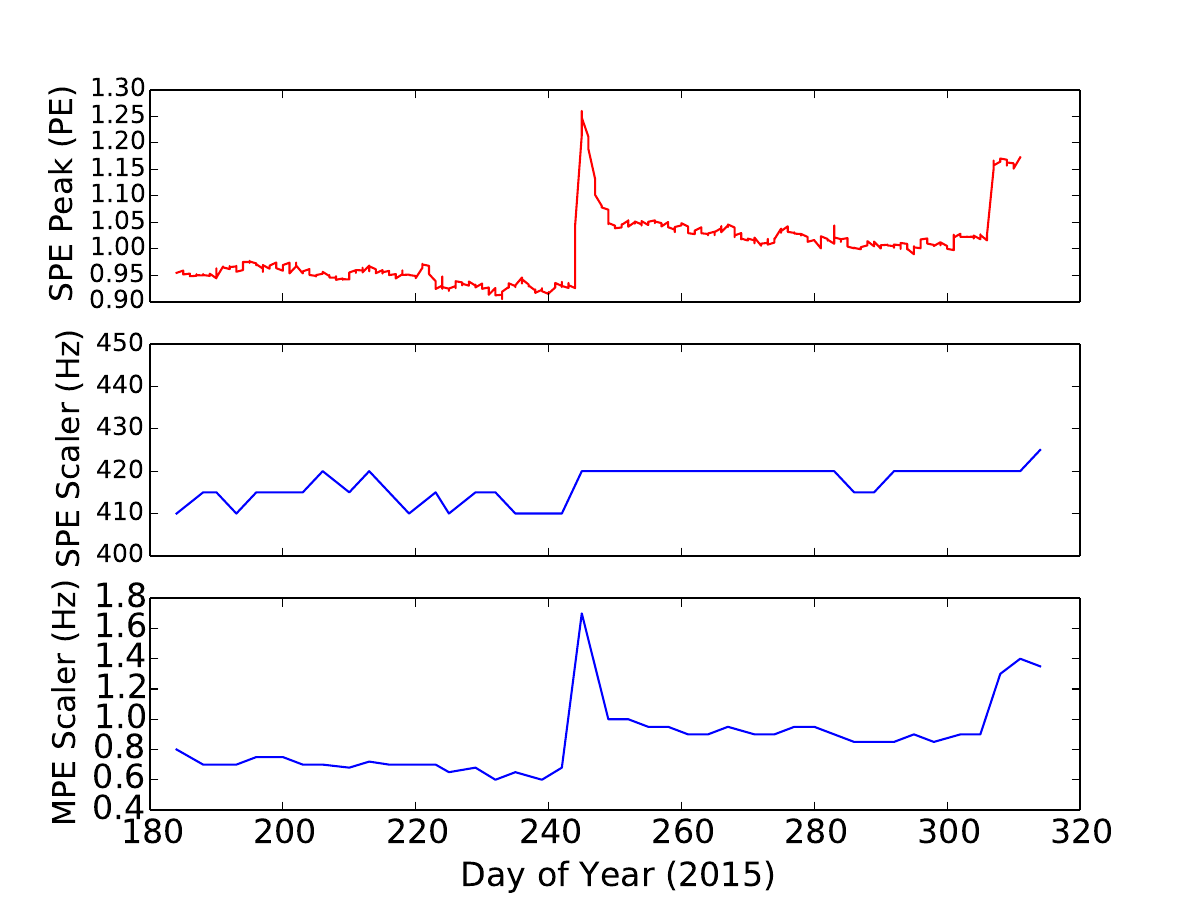}
 \caption{Mean of the Gaussian fit to the SPE peak (top) and the SPE
   scaler rate (middle) and MPE
   scaler rate (bottom) from July 2015 to November 2015 for a DOM
   that shows unpredictable gain shift behavior. The DOM
   configuration was unchanged during this period.}
 \label{fig:gainshift_spe}
\end{figure}

There are about 12~DOMs that show unpredictable, abrupt shifts in the SPE peak
position of 0.05~PE or more. Figure~\ref{fig:gainshift_spe} shows the time history of the
SPE peak position of one of these DOMs over 4~months. The peak shift
corresponds to increases or decreases in the multi-photoelectron (MPE)
scaler rate, where the scaler counts the discriminator crossings.  This
indicates that the SPE peak shift is indeed caused by a change in the DOM
gain. However, the SPE scaler rate is stable to within 2\%, indicating that the
probability to detect single photons is effectively unchanged.

\begin{figure}[!h]
 \centering
 \includegraphics[width=0.7\textwidth]{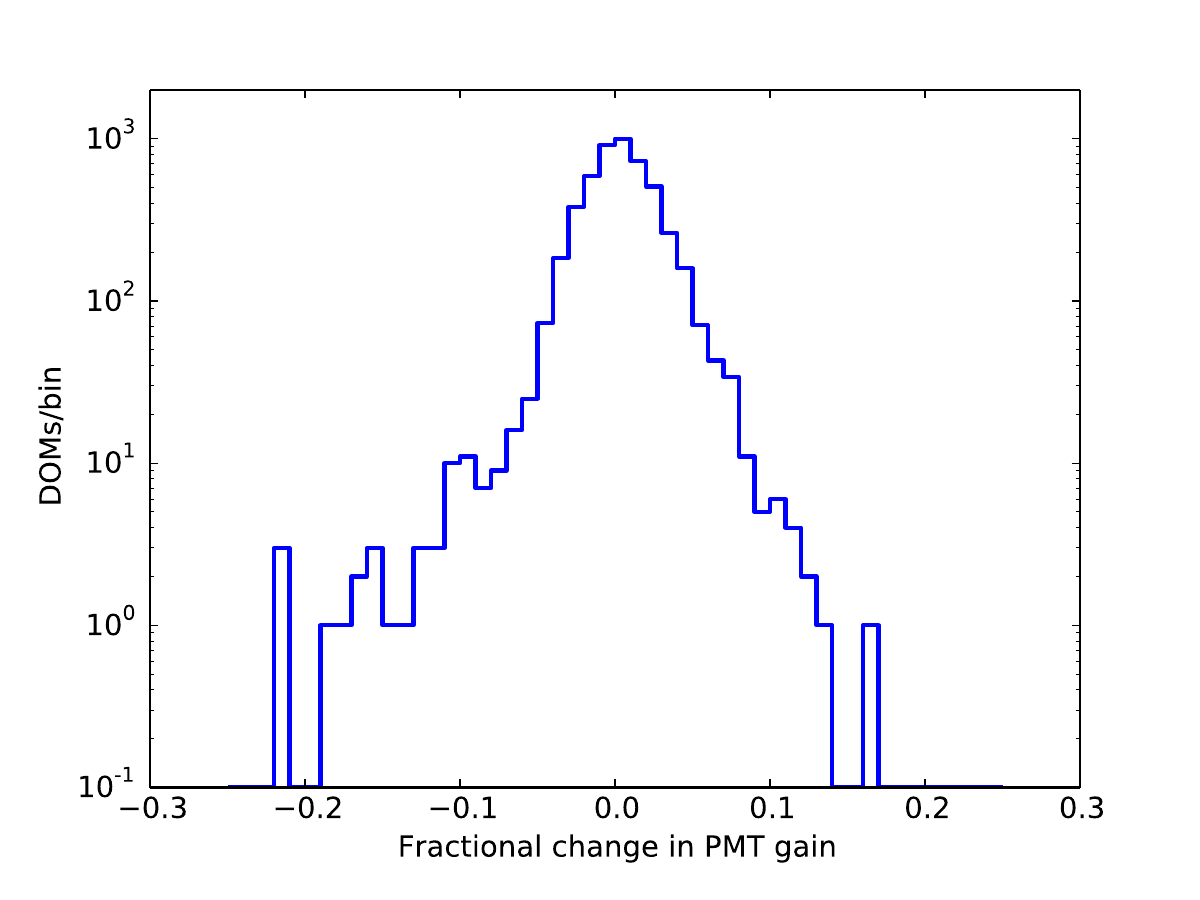}
 \caption{Distribution of the fractional changes in PMT gain from 2011 to 2016 for all in-ice
   DOMs, at the 2011 operating high voltage for each DOM.}
 \label{fig:pmt_gainshift}
\end{figure}

Long-term stability of the PMTs can be tracked by examining any change in
the high voltage vs.~gain calibration over time.  The fractional change in gain for
all in-ice DOMs over a five-year time period is shown in
figure~\ref{fig:pmt_gainshift}.  For most DOMs, the PMT gain is stable to
within $\pm3\%$ over the time period shown, with a median gain shift of $0.3\%$, while approximately $1\%$ 
of DOMs have a gain shift of more than 10\%.  Any shifts are tracked with
regular calibration using the methods previously described.

\subsection{\label{sec:optical_stability}Optical Efficiency Stability}

\begin{figure}[!h]
  \captionsetup[subfigure]{labelformat=empty}
  \centering
  \subfloat[]{\includegraphics[width=0.5\textwidth]{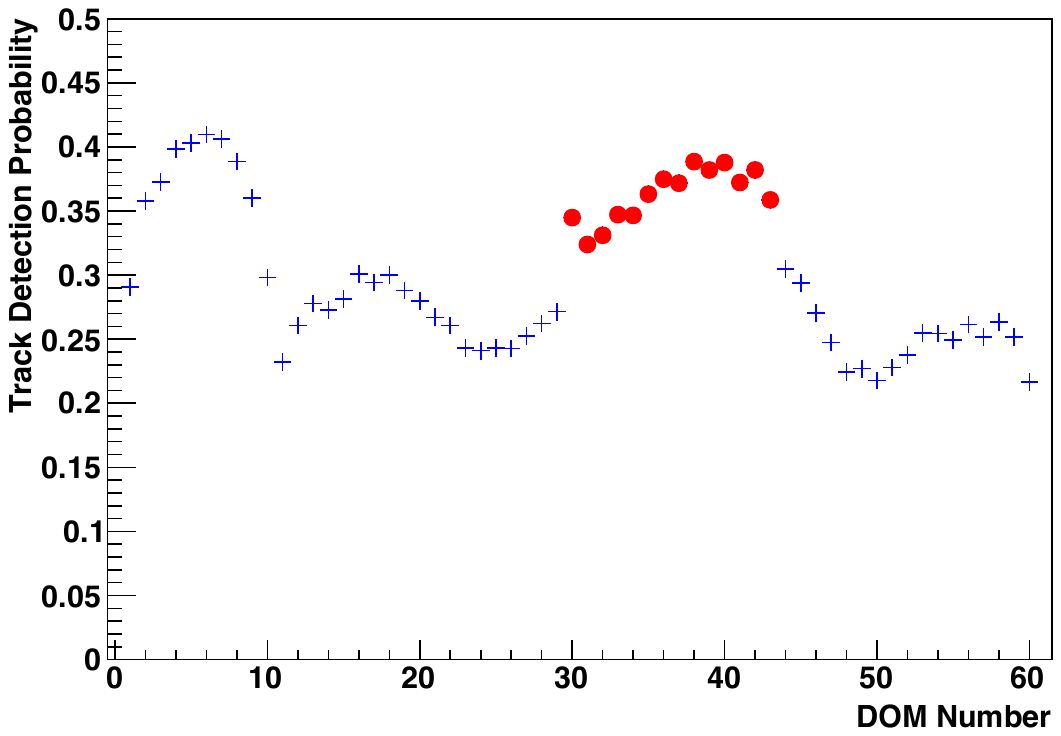}}
  \subfloat[]{\includegraphics[width=0.5\textwidth]{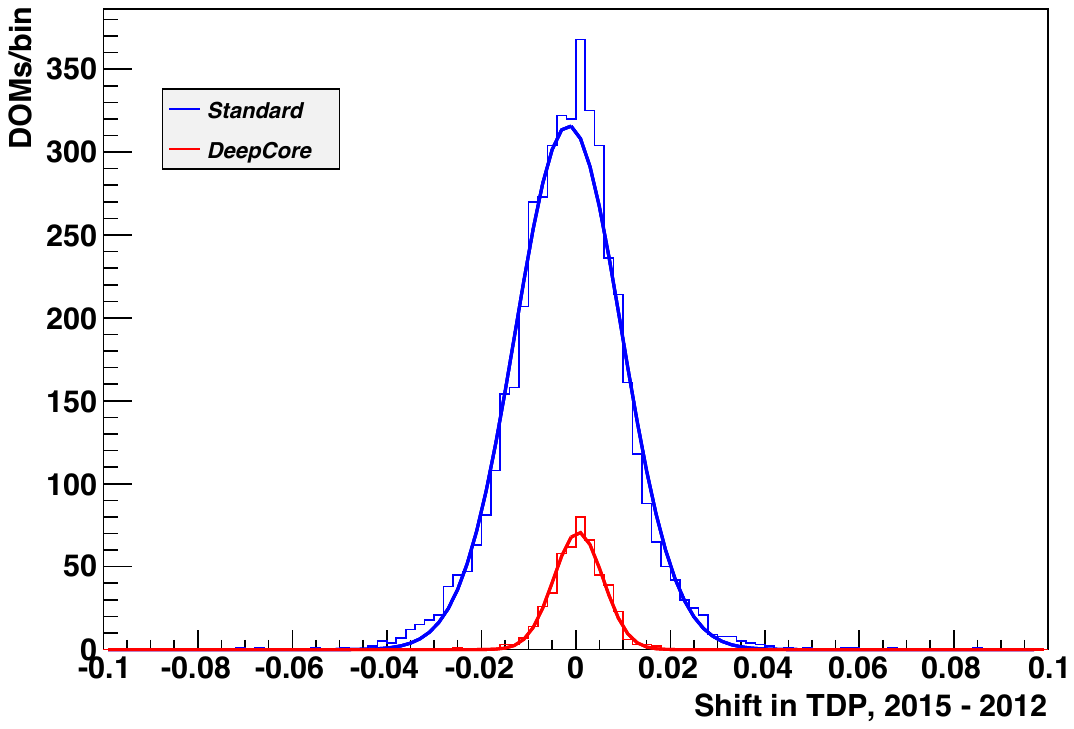}}
  \vspace{-1\baselineskip}  
  \caption{Left: track detection probability (TDP) in String 80, with a
    mixture of standard DOMs (blue crosses) and HQE DOMs (red
    circles). The variation with depth is due to the depth-dependent
    optical properties of the ice. Right: shift in TDP for all in-ice DOMs between 2012 and 2015; standard DOMs are in blue, and
    HQE DOMs are in red. The mean of the Gaussian fit to
    the TDP shift in the standard DOMs is $-0.1$\%, and the mean 
    TDP shift in the HQE DOMs is 0.05\%.}
  \label{fig:tdp}
\end{figure}

The detector response in IceCube is verified with low energy muons as
described in ref.~\cite{IC3:ereco}. The detector response is monitored in each run using the track
detection probability (TDP) calculated from bright muon tracks
with more than 30 hits in IceCube. The muon 
tracks are reconstructed using the likelihood methods described in
ref.~\cite{Ahrens:2003fg}, but charge and time information from the DOM under
study are excluded from the reconstruction. The TDP is
defined for each DOM as the ratio of the number of detected tracks
within $\SI{100}{\meter}$ of the DOM to the total number of tracks within $\SI{100}{\meter}$ of
the DOM. This ratio depends both on the optical properties of the ice
near the DOM and the optical efficiency of the DOM. We do not attempt
to separate
these effects in the TDP, but rather use the TDP to monitor the
overall stability of the detector response. Figure~\ref{fig:tdp} shows the TDP on
String 80, which includes both standard and HQE DOMs; the TDP is
20--25\% higher for HQE DOMs than for neighboring standard
DOMs, whereas the quantum efficiency is 35\% higher. The TDP is stable to within 1\% since 2012, when the baselines
were stabilized by being set in the DAQ configuration. Figure~\ref{fig:tdp} shows
the difference in the TDP for all DOMs between a run in 2012 and a run
in 2015.

The detector response stability is also measured with the {\it in
  situ} light sources in IceCube. Both the in-ice calibration laser
\cite{IC3:SC} and the flasher LEDs show less than 1\% difference in the total
charge collected between 2012 and 2015. 

\subsection{\label{sect:darknoise}Dark Noise}

The vast majority of background hits result from dark noise, i.e. effects
that lead to the emission of an electron from the cathode of the PMT in the
absence of a photon source external to the DOM. 
Dark noise is a complex phenomenon with numerous possible sources,
including thermionic emission, electronic noise, field emission 
within the PMT, Cherenkov light from radioactive decays, and
scintillation / luminescence in the glass of the PMT and pressure sphere.
The average total in-ice hit rate is \SI{560}{\hertz} for DOMs with
standard PMTs and \SI{780}{\hertz} for high quantum efficiency DOMs. The 
contribution from cosmic-ray muons, estimated as the in-ice HLC hit rate,
is small and decreases with depth from 25 Hz to 5 Hz. 

The dark noise can be characterized as a combination of uncorrelated
(Poissonian) noise pulses with a rate between \SI{230}{\hertz} and
\SI{250}{\hertz}, and a correlated component, with a pulse rate from
\SI{280}{\hertz} to \SI{340}{\hertz}.  A comparison at low temperature of
DOM dark noise to that of a bare PMT suggests that the majority of the
noise originates from the glass pressure sphere.  Cherenkov light from
$^{40}\mathrm{K}$ decays contributes to the uncorrelated noise component,
and thus the potassium content of the glass was limited.
Measurements of early samples indicated a $\mathrm{K}_2\mathrm{O}$
concentration of 0.03\% by weight, roughly corresponding to 100 Bq of beta
decays per sphere. 

\begin{figure}
  \centering
  \includegraphics[width=0.7\textwidth]{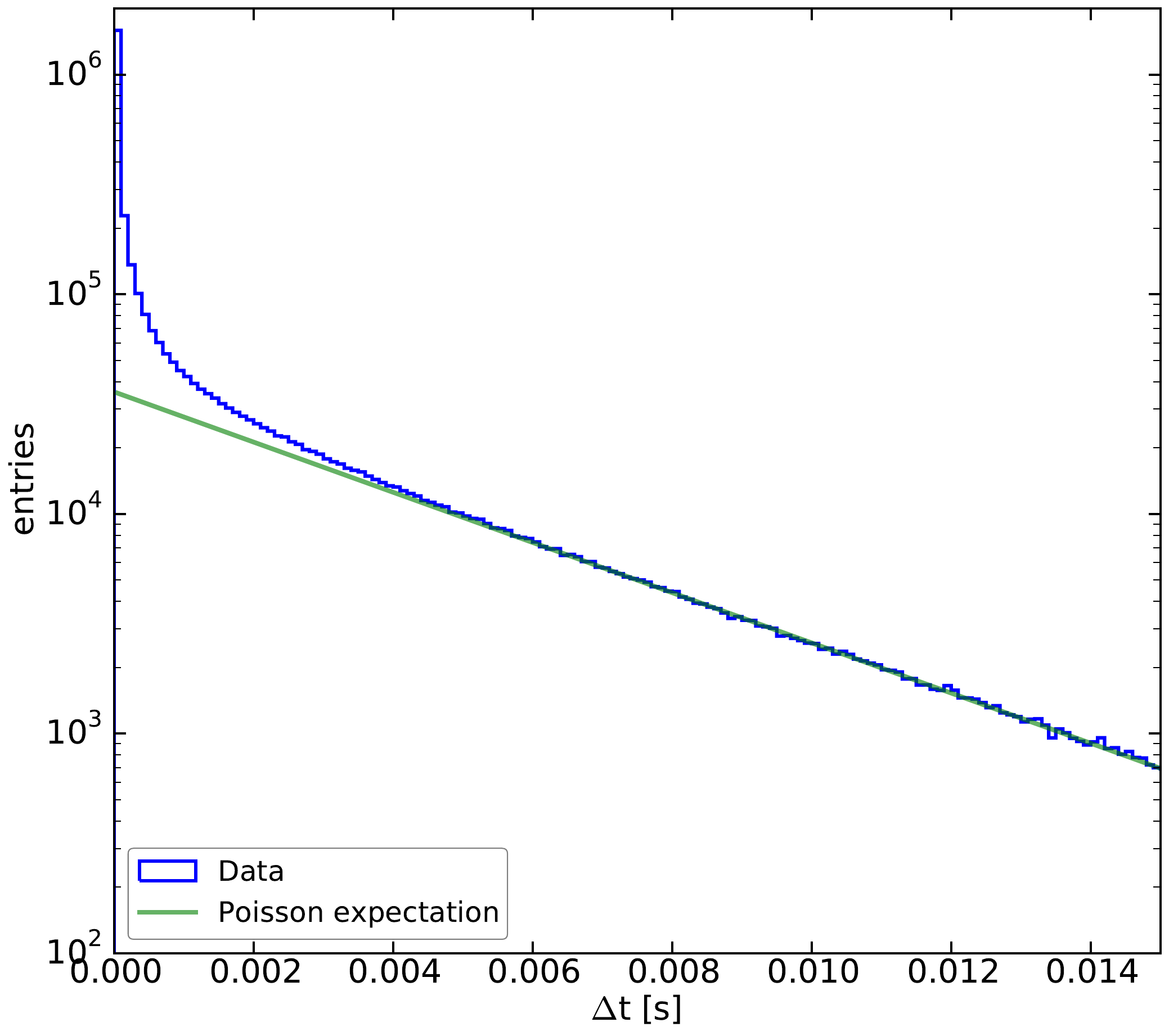}
 \caption{Time interval between successive hits for all next-to-top layer
   DOMs (DeepCore excluded).  The line is an exponential fit to the
   Poissonian regime between 7 and 15 ms.}
 \label{fig:darknoise_deltaT}
\end{figure}

\begin{figure}
  \centering
  \includegraphics[width=0.7\textwidth]{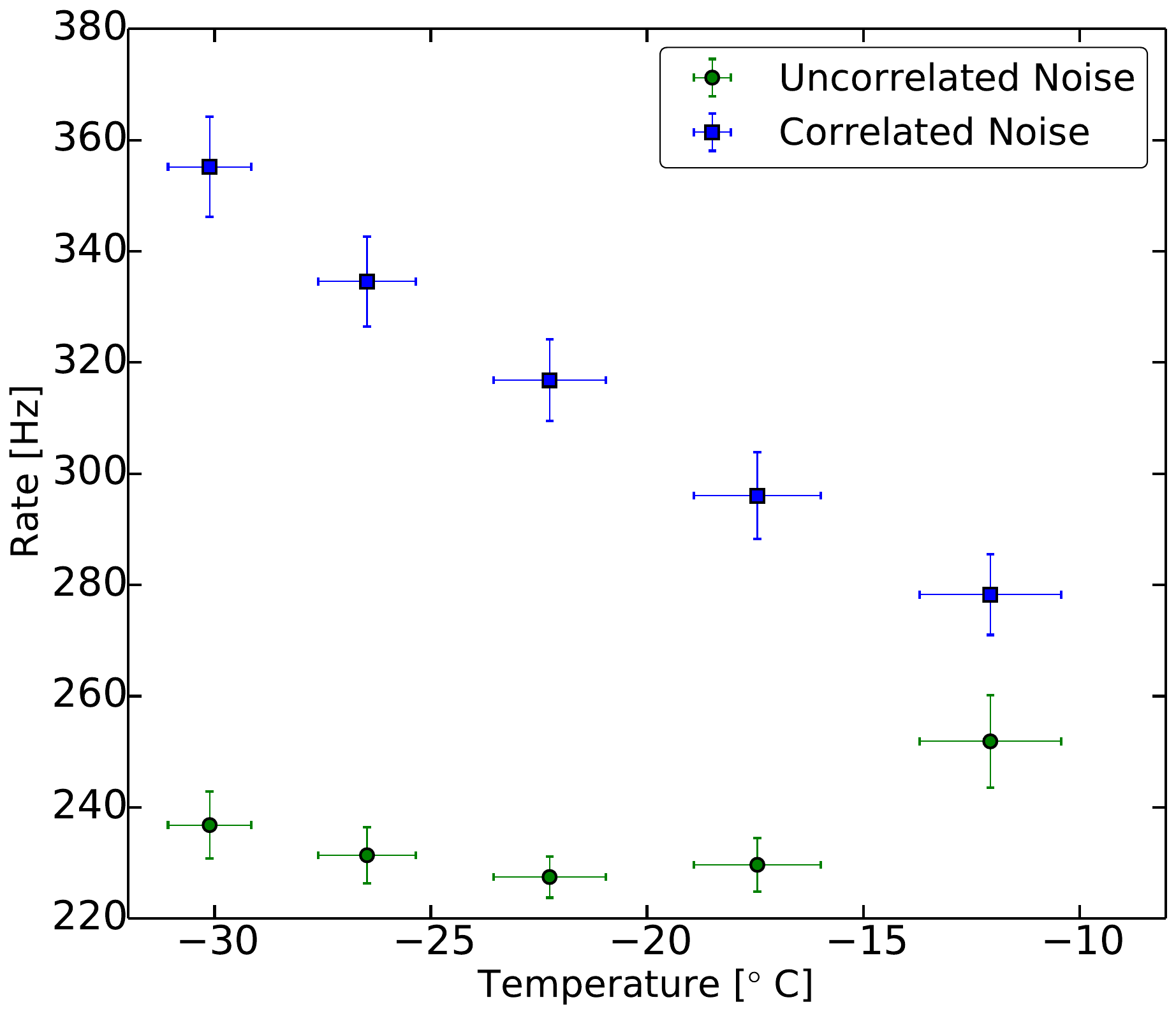}
  \caption{Dark noise rate of DOMs in ice as a function of temperature,
    obtained from untriggered HitSpool data. Each data point represents the
    average of 12 DOM layers from 78 strings (DeepCore excluded).} 
  \label{fig:dom_darknoise_vs_temperature}
\end{figure}

The correlated noise manifests itself as an overabundance of short time
intervals between hits in a single DOM compared to the Poisson expectation
(figure~\ref{fig:darknoise_deltaT}). 
The temperature dependence of the noise rate
(figure~\ref{fig:dom_darknoise_vs_temperature}) was determined by combining a
measured temperature profile of 
the South Pole ice cap \cite{price2002temperature} with a fit of the
Poissonian expectation of the total dark noise rate to every individual
DOM, and was verified in lab measurements.  The temperature of the in-ice
DOMs, measured on the Main Board, increases with depth from
$-31\ ^{\circ}\mathrm{C}$ to $-9\ ^{\circ}\mathrm{C}$
($10\ ^{\circ}\mathrm{C}$ above the ambient ice temperature). 

The short time intervals are clustered in bursts (uninterrupted sequences
of time intervals less than 3 ms) with an average number of
hits per burst increasing from \num{3.3} at \SI{-10}{\celsius} to \num{3.8} at
\SI{-30}{\celsius}. A study with forced-readout data shows that the
phenomenology of the correlated noise component in IceCube is in general in
good agreement with results reported in ref.~\cite{meyer_noise}, but an
unambiguous physical explanation is still to be confirmed.  Stimulation of
glass samples with radioactive sources results in scintillation
\cite{helbing_glass}, suggesting that luminescence of the glass triggered
by the radioactive decay of $^{40}\mathrm{K}$ and other elements in the
pressure sphere is responsible.  Naturally-occurring cerium in the glass is
a candidate for the active scintillator. 

\begin{figure}[!h]
 \centering
  \includegraphics[width=0.8\textwidth]{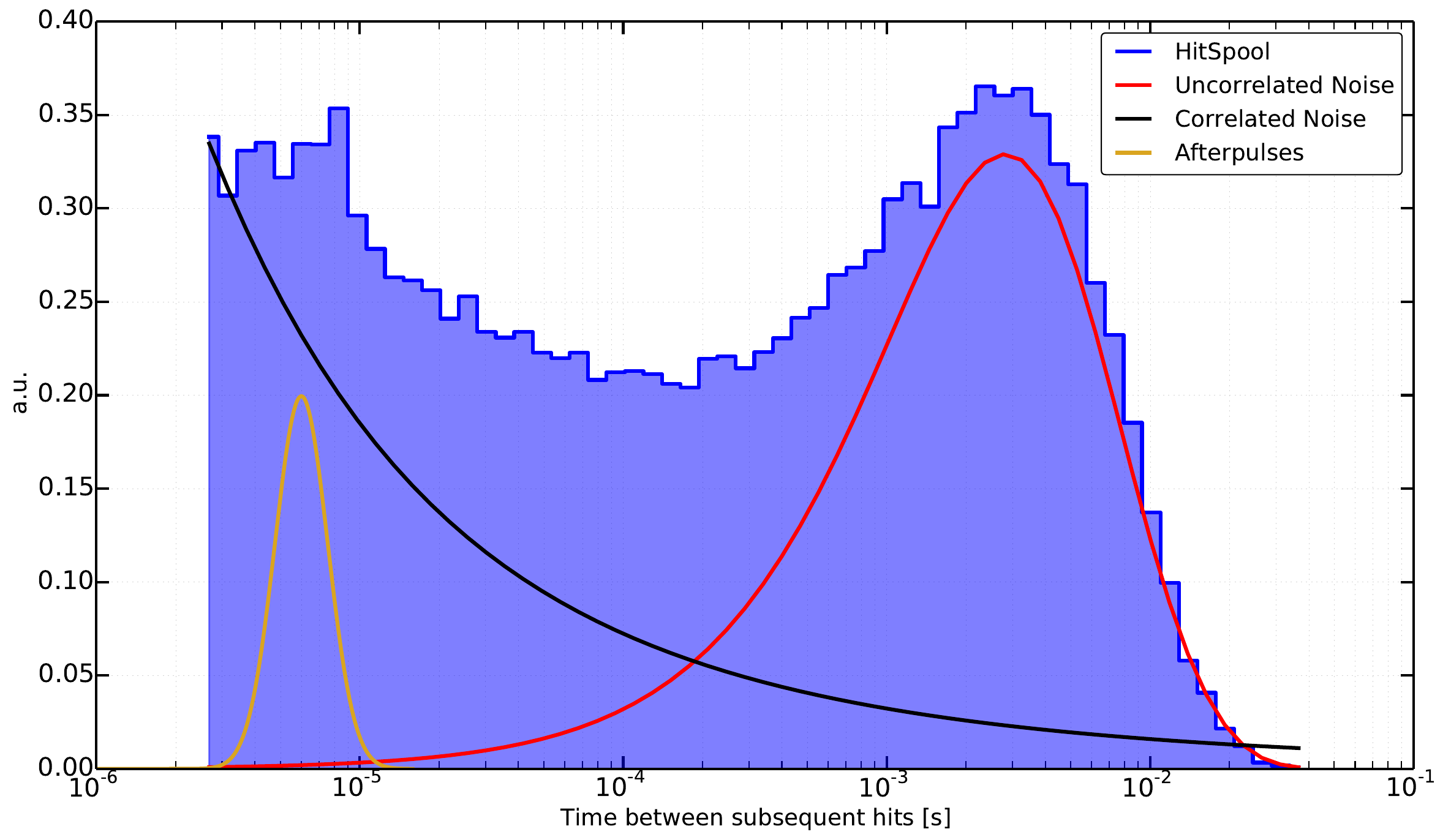}
 \caption{Histogram of time differences between successive hits from HitSpool data of
DOM 15 on string 27 (blue) on a logarithmic scale in order to visualize the
different noise components \cite{heereman2015hitspooling}.  A fourth
subdominant component centered at \SI{100}{\micro\second} is not parameterized and is still under study.}
 \label{fig:darknoise_deltaT_components}
\end{figure}

The various sources of dark noise present in a DOM can be separated using
the time between successive untriggered hits in HitSpool 
data (section~\ref{sec:domhub_hitspool}), as shown in 
figure~\ref{fig:darknoise_deltaT_components}. An overview of the various 
noise components and their parameterizations is given in table \ref{tab:noise}.  Late-time
correlated afterpulses, a common feature of PMTs, is attributed to
ionization of residual gases by electrons that were accelerated between
the dynodes.  Although afterpulses occur at various time delays, this
component is parametrized here with a single average timescale.  A
noise model incorporating these various sources is used for detector
simulation \cite{larson2013simulation}. 

\begin{table}[h!]
\caption{Characteristics of noise components in IceCube DOMs, adapted from
  ref.~\cite{stanisha_noise_14}.}
  \centering
  \footnotesize
\begin{tabularx}{\textwidth}{lXXr}
\toprule
Noise Component& Origin & Distribution & Parameters \\
\midrule
afterpulses & PMT & Gaussian & $\mu = \SI{6}{\micro\second}$, $\sigma = \SI{2}{\micro\second}$\\
uncorrelated & thermal noise, \newline radioactive decay & Poissonian & $\lambda \simeq \SI{250}{\hertz}$\\
correlated & luminescence (?) & log-normal &
$\mu = \num{-6}\ [\log_{10}(\delta t/\mathrm{sec})]$, \\
& & & 
$\sigma = \num{0.85}\ [\log_{10}(\delta t/\mathrm{sec})]$ \\
\bottomrule
\end{tabularx}
\label{tab:noise}
\end{table}

\begin{figure}[!h]
 \centering
 \includegraphics[width=1.0\textwidth]{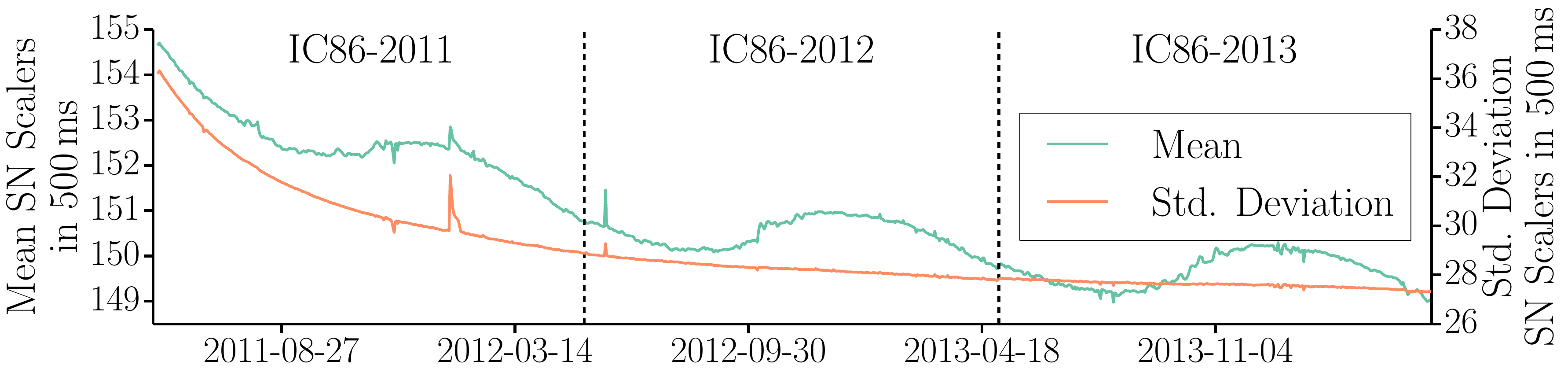}
 \caption{Mean and standard deviation of the supernova scaler distribution for the
   entire detector over the course of the first three years of the
   completed IceCube \cite{briedel_phd}. Spikes are due to calibration
 runs which injected light into the detector, and to fluctuations in
 individual DOMs.} 
 \label{fig:noise_over_time_briedel}
\end{figure}

\begin{figure}[!h]
 \centering
 \includegraphics[width=1.0\textwidth]{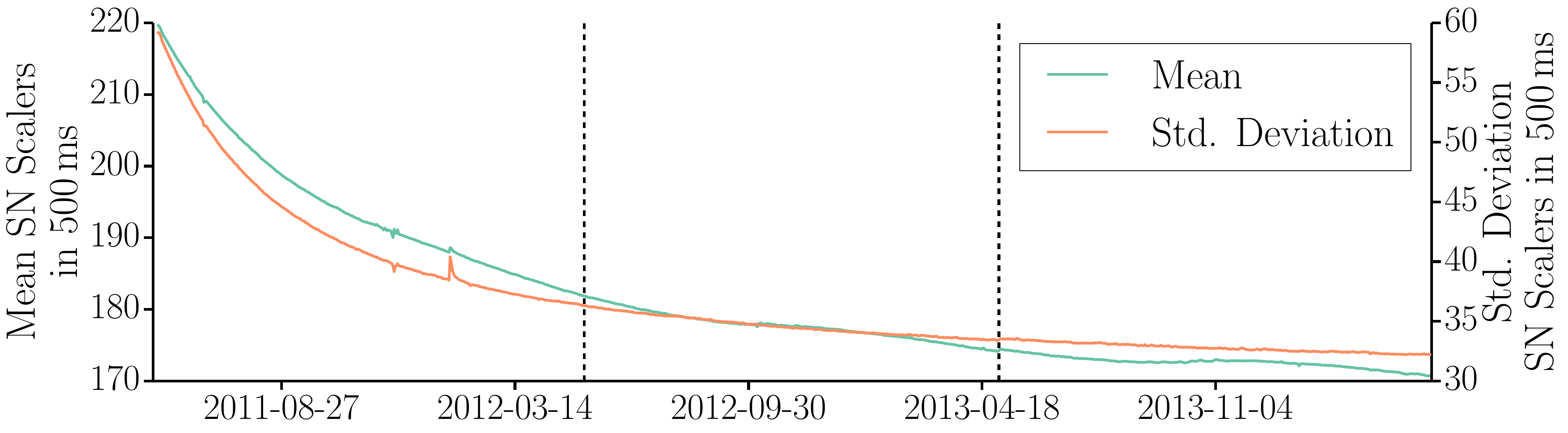}
 \caption{Mean and standard deviation of the supernova scaler rate of strings
   deployed in the last deployment season (austral summer of 2010/2011)
   \cite{briedel_phd}, where changes in the mean rate are
still dominated by the decay component. Spikes are due to calibration
 runs which injected light into the detector, and to fluctuations in
 individual DOMs.} 
 \label{fig:noise_over_time_briedel_lastseasondepoyed}
\end{figure}

The evolution of the dark noise contribution to the total rate over time was investigated using the
supernova scaler stream \cite{IC3:supernova, briedel_phd}, effectively 
the summed dark noise rate of the detector with an artificial deadtime
applied to reduce the effect of correlated noise
(section~\ref{sect:SNDAQ}). A long-term exponential decay in 
the noise rate is visible; this may be caused by 
decreasing triboluminescence \cite{ice_tribo} arising from the initial ``freeze-in''
of newly deployed DOMs, impurities introduced during the drill
process, or a combination of the two effects.  The decay
is especially recognizable in the standard deviation of the scaler
distribution (figure~\ref{fig:noise_over_time_briedel}), which decreases by
25\% over the course of the three years, with the decrease increasing with
depth~\cite{Piegsa09}. Changes in the mean rate are
initially dominated by the decay component and later by the seasonal
variation of atmospheric muons. The noise
rate decay is most pronounced when selecting strings
that were deployed in the final construction season, as shown in
figure~\ref{fig:noise_over_time_briedel_lastseasondepoyed}.

\begin{figure}[!h]
 \centering
 \includegraphics[width=0.95\textwidth]{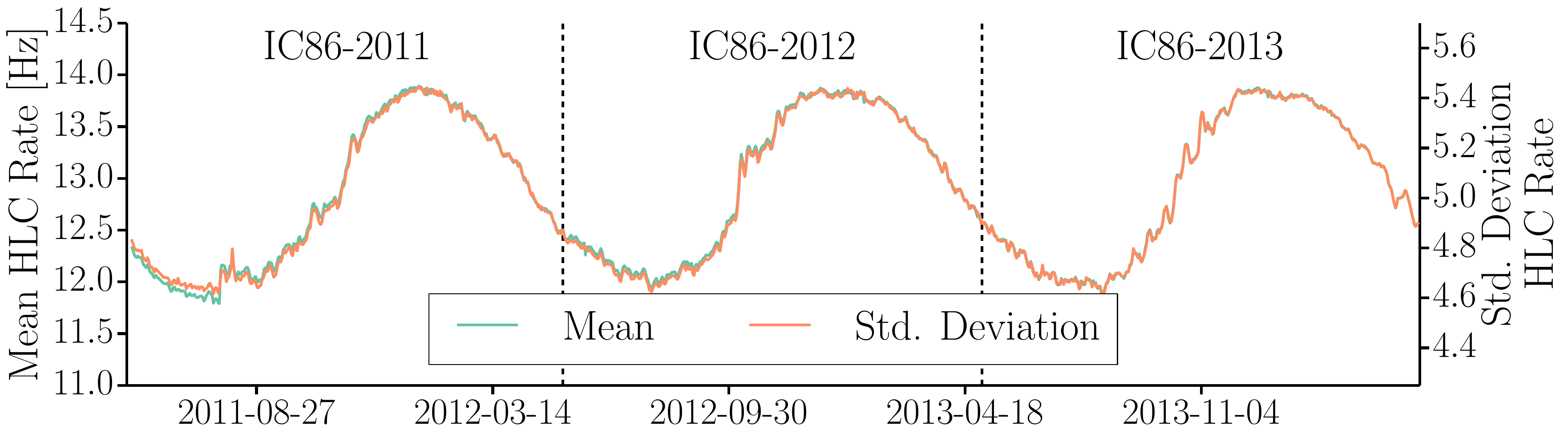}
 \caption{Mean and standard deviation of the local coincidence hit rate distribution
   \cite{briedel_phd}.  Changes in dark noise rate do not
   contribute significantly to changes in the local coincidence hit rate,
  which is dominated by seasonal and weather-related
   changes in the atmospheric muon flux.} 
 \label{fig:hlc_over_time_briedel}
\end{figure}

Since the dark noise components are not correlated between
DOMs, the dark noise rate change is not prominent in local coincidence
hits, which are dominated by atmospheric muons (figure
\ref{fig:hlc_over_time_briedel}).  Thus, the detector trigger rate as well
as many higher-level reconstruction variables are relatively isolated from the
noise rate variations.  Nevertheless, the total hit rates for each DOM are
tracked over time and updated yearly in a database for simulation and
reconstruction purposes.  Seasonal variations of these rates are
below the 1\% level.

%auto-ignore
\section{\label{sec:cable}Cable Systems}

The IceCube detector may be viewed as a digital network of optical
sensors. The IceCube cable systems form the physical backbone of this
network, simultaneously supplying power to the DOMs and bi-directional
digital communications between the DOMs and the readout hardware at the
surface.  Copper was chosen over optical fiber at an early stage in the
project, based on a previous successful demonstration of the technology with
the AMANDA digital-readout optical module string
\cite{AMANDA:string18} and concerns with mechanical robustness of
fiber during freeze-in in the deep ice.

\subsection{Design}

\begin{figure}
  \centering
  \includegraphics[width=\textwidth]{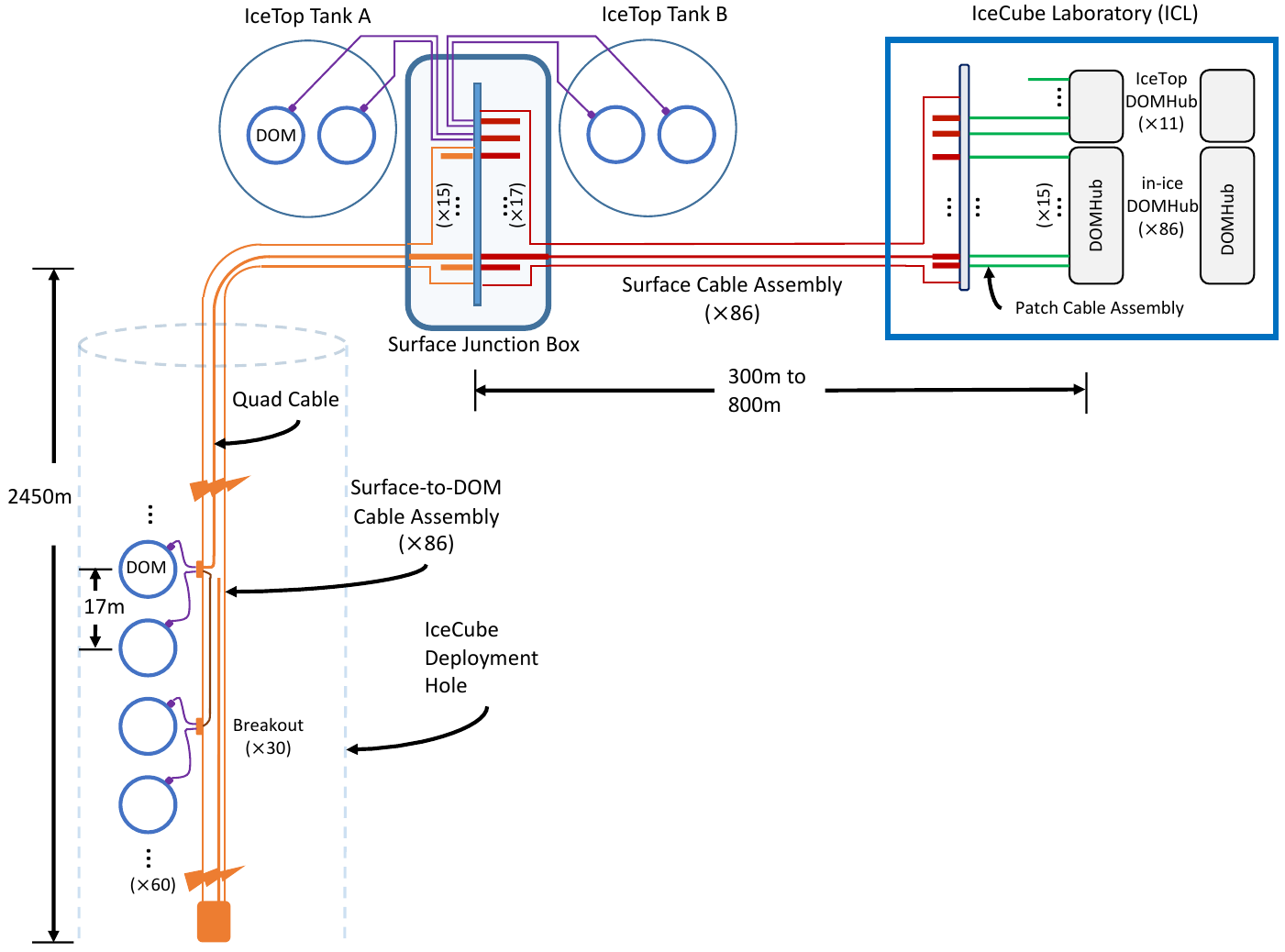}
  \caption{\label{fig:icecube-cables-logical}Schematic of the IceCube cable network architecture.}
\end{figure}

The cable system comprises the following assemblies: the in-ice cable,
IceTop cables, the surface junction box (SJB), the surface cables, and the patch
cables  (figure~\ref{fig:icecube-cables-logical}). The in-ice cable, $\SI{2505}{\meter}$ long, is deployed
into the ice along with 60 DOMs that are attached to connectors at 30
breakouts spaced $\SI{34}{\meter}$ apart.  An adjacent pair of DOMs is connected to a distinct
twisted wire pair. Two wire pairs are combined into four-conductor quad
cables meeting stringent electrical performance requirements; the quad
arrangement provides enhanced cross-talk immunity and improved
mechanical stability during freeze-in compared to a twisted pair.

The 60 DOMs on each cable require a total of 15 quads. An additional 5 quads in the
in-ice cable provide for special instrumentation connections, a spare quad,
and local coincidence connections between adjacent DOMs. The in-ice
cable terminates at the SJB, located just below the snow surface between
the IceTop tanks. The SJB is a stainless steel
enclosure that houses the in-ice cable and surface cable
connections. IceTop cables also terminate and connect to the surface cable
in the SJB. The surface cable is trenched $\SI{1}{\meter}$ deep into the
surface of the snow between the SJB and the ICL. The surface
cables vary from 300~m to approximately 800~m in length depending on hole location. The surface cables are
pulled up two cable towers into the ICL and terminate at patch panels where the individual
quads are separated and connected to patch cables that finally
terminate at the DOMHub, the computer that receives DOM signals in the ICL
(section~\ref{sect:sps}). Each 
DOMHub services a single in-ice string or 8 IceTop stations. IceTop DOMHubs service fewer
DOMs than in-ice DOMHubs due to the higher rates recorded by IceTop DOMs~\cite{ICECUBE:IceTop}.

The IceCube cable system design had to meet a number of stringent
mechanical, electrical and functional requirements, due to the extreme
conditions of installation at the South Pole.  In particular, the in-ice cable
was required to withstand hydrostatic pressure of up to 450~bar during
deployment.  Handling temperatures were specified down to $-40^{\circ}\ \mathrm{C}$ for all 
cable components; deployment took place between late November and
mid-January each year, and temperatures never dropped below
$-40^{\circ}\ \mathrm{C}$ during these months.  The primary electrical requirements for the quad
cables are provided in Table ~\ref{tab:quad_requirements}.

\begin{table}[h]
  \centering
  \caption{Electrical requirements for the cable quads.} 
  \begin{tabularx}{\textwidth}{ l X  X  }
    \toprule
    Characteristic& Specification & Conditions \\
    \midrule

    Attenuation & $\le20~\mathrm{dB}$ & At 1.0 MHz end-to-end \\

    Near-end crosstalk suppression& $\ge50~\mathrm{dB}$ intra-quad & At 2~MHz \\

      &$\ge24~\mathrm{dB}$ quad-quad& At 100 MHz\\

    Far-end crosstalk suppression& $\ge30~\mathrm{dB}$ intra-quad & At 2 MHz \\

      &$\ge24~\mathrm{dB}$ quad-quad& At 20 MHz\\

    Differential impedance & $145\Omega \pm 10\Omega$ & At 1.0 MHz \\

    Loop resistance & $\le160~\Omega$ & \\

    Dielectric breakdown voltage & $\ge2000~\mathrm{VDC}$, $\le1~\mu\mathrm{A}$ leakage & \\
    \bottomrule  
  \end{tabularx}
  \label{tab:quad_requirements}
\end{table}

 The in-ice cable was also required to be less than 50~mm in
 diameter, weigh less than 2~kg / m, have a minimum static bend radius of 40~cm,
 carry a maximum static tensile load of 10~kN, and have a breaking strength
 of 40~kN. During deployment, the maximum tension on the cable was less
 than 8~kN, taking into account the buoyancy of the DOMs.
 
\begin{figure}
  \centering
  \includegraphics[width=0.65\textwidth]{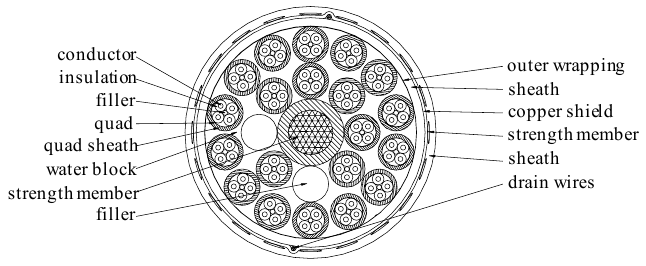}
  \caption{\label{fig:cable_xsection}In-ice cable cross
    section, with a nominal 46 mm diameter and mass of 2 kg/m.} 
\end{figure}

The in-ice cable includes 20 quads, 2 polyethylene fillers to maintain structural
symmetry, Kevlar outer and inner core strength members, a copper foil
shield with drain wires to provide electrical noise isolation, and a water
block that prevented water or ice from damaging the symmetry of the
cable (figure~\ref{fig:cable_xsection}). The surface cable has a similar
construction to the in-ice 
cable with two exceptions: first, the inner Kevlar strength member was replaced
with additional conductors needed to service the IceTop tanks; and second, the
water block was not required.

A competitive proposal process resulted in two main suppliers for the IceCube
cable system.  Ericsson Network Technologies (Hudiksvall, Sweden) was chosen
to produce the raw cable, and SEACON Brantner \& Associates (El Cajon,
California) was chosen to manufacture the cable assemblies. The raw
cable was manufactured and tested to meet all required specifications prior to spooling.  The in-ice cables were then
wound onto custom metal spools, while the shorter surface cables were wound
onto wooden spools. SEACON provided the breakouts
and connectorized the top of the in-ice cable and both ends of the surface
cables. Glass-reinforced-epoxy SEACON XSJJ connectors were used for the in-ice
cable to DOM interface; these connectors are water pressure-rated to 690
bar. The 30 breakouts per in-ice cable were installed by 
slicing open the cable, cutting the correct quads, adding the
XSJJ connectors, waterproofing and overmolding the connectors, and then
resealing the cable. SEACON also
attached 120 Yale Cordage (Saco, Maine) YaleGrips to the in-ice
cable that served as mechanical attachment points for the 60 deployed DOMs
per cable. The separate quads in each cable were terminated with
Military Standard round,
metal-shell connectors. After cable assembly was complete, 
each cable was individually tested for electrical continuity and subjected to high
potential (high voltage) testing. 

\subsection{Installation}

Installation of the IceCube Cable System at the South Pole each season was
broken into five distinct tasks: surface cable trenching and
installation into the ICL and SJB; in-ice cable deployment and
connection to the SJB; IceTop freeze control power and communications installation at the ICL and
IceTop trench; patch cable installation in the ICL; and testing,
troubleshooting and repair after all connections were made. Surface cables
were installed early in the season to support IceTop tank freeze control
operations, which were necessary to ensure clear crack-free ice in the
IceTop tanks~\cite{ICECUBE:IceTop}. A 1-m-deep trench was mechanically
dug between the ICL and each IceTop station. After the cable was placed in
the trench, it was pulled into the ICL and connected to an IceTop
DOMHub. Later, just prior to filling the tanks with water in the field, the
connections were made between the surface cable and the SJB. The IceTop
DOMs with their long 17~m penetrator assemblies were connected to the
appropriate quad located in the center of the surface cable via the SJB. Two additional
cables were connected between the SJB and the tanks that provided power and
communications to the IceTop freeze control units. The in-ice cable
was prepared for installation while the Enhanced Hot Water Drill was coming
up the hole in its reaming phase (section~\ref{sec:hot_water_drilling}).  After the drill was removed,
the end of the in-ice cable was pulled into the drill tower, and DOM deployment
commenced (section~\ref{sec:deployment_inst}). After DOM
deployment was complete, the in-ice cable end was taken off the spool and
dragged to the IceTop station where its connection to the surface cable in the SJB was
made. Finally, patch cables were installed in the ICL between the
individual quads and the DOMHub in the ICL. End-to-end string commissioning
then commenced.

%auto-ignore
\section{\label{sec:drill-deploy}Drilling and Deployment}

\subsection{\label{sec:hot_water_drilling}Hot Water Drilling}

Transforming a cubic kilometer of Antarctic ice into an astrophysical
particle detector required drilling 86 boreholes
approximately 60 cm in diameter to a depth of 2500~m. Hot water drilling
was the only
feasible technology to provide rapid access to the deep ice on this scale.
Instrumentation was deployed into the water-filled boreholes, becoming
frozen in place and optically
coupled with the surrounding ice sheet. The 5~MW Enhanced Hot Water Drill
(EHWD) was designed and built specifically for this task,
capable of producing the required boreholes at a maximum rate of one hole per 48
hours. This section contains an abbreviated description of the drill; a more detailed description
can be found in ref.~\cite{ehwd}.

\begin{figure}[!ht]
 \centering
 \includegraphics[width=0.7\textwidth]{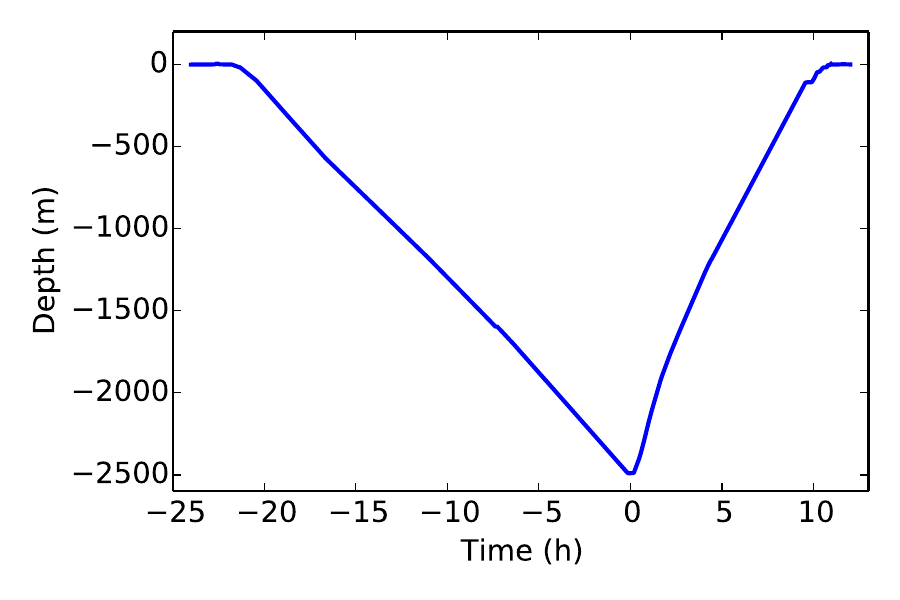}
\caption{Drilling and reaming profile for string 41, from the 2009--2010
  deployment season. Zero time is maximum drill depth, so drilling time is
  negative and reaming time is positive.}
\label{fig:drilldepthtime}
\end{figure}

The procedure involved a drilling phase to create the initial hole,
followed by an upward reaming phase to give the 
hole a targeted diameter profile.  During drilling, the location of
the drill was recorded with an onboard navigational pack consisting of
a Paroscientific pressure sensor
(model 8CB4000-I) to measure depth, two
liquid pendulums to measure tilt, and a fluxgate compass to measure
orientation. The hole diameter was larger than
the DOM (35~cm diameter) to compensate for closure from refreezing to provide sufficient
time to deploy instrumentation, with contingency time for delays.  The
elapsed duration from the end of drilling until the hole closes to
less than the DOM diameter is referred to as the hole lifetime. IceCube drilling was
completed in seven field seasons (approximately 21 months total time).
Peak performance occurred in the 2009--2010 season with 20 holes drilled
(early November to mid-January).  The drill specifications and performance
are shown in Table~\ref{tab:ehwd_system} and
Table~\ref{tab:ehwd_system_peak}. Figure~\ref{fig:drilldepthtime} shows the
drilling and reaming time for a typical hole in the 2009--2010 season.

\vspace{\baselineskip}

\begin{minipage}{\textwidth}
  \centering \captionof{table}{EHWD System Characteristics}
  \begin{tabular}{ l  r }
 \hline
    Specification & Value \\ \hline Total Power (Thermal + Electrical)
    & 5 (4.7 + 0.3) MW \\ Maximum Drill Speed & 2.2 m/min \\ Maximum Ream
    Speed & 10 m/min \\ Water Flow (delivered to main hose) & 760 L/min
    \\ Water Temperature (delivered to main hose) & \SI{88}{\celsius}
    \\ Water Gauge Pressure (at main pumps) & 7600 kPa\\
 \hline
  \end{tabular} 
  \label{tab:ehwd_system}
\end{minipage}
\vspace{\baselineskip}

\begin{minipage}{\textwidth}
  \centering \captionof{table}{Average and peak performance for a 24-hour
    lifetime hole of 2500 m depth. Peak performance corresponds to
    string 32 from the 2009--2010 drilling season, which had the
    fastest drilling + reaming time.}
  \begin{tabular}{ l  r  r }
\hline
    Specification & Avg. Value & Peak Value\\ 
\hline 
Total Fuel\footnote{Total Fuel includes deep drilling/reaming and firn
      drilling}, AN-8 & 21,000 L & 15,000 L \\ Time to Drill/Ream & 30 hr&
    27 hr \\ Hole Production Cycle Time\footnote{Hole Production Cycle Time
      is the elapsed time from start of one hole to start of the next hole}
    & 48 hr & 32 hr \\
\hline
    \end{tabular}
  
  \label{tab:ehwd_system_peak}
\end{minipage}
\vspace{\baselineskip}

Hot water drilling was not practical in the firn layer, the 50~m-thick surface layer of
compressed snow above the ice, because the firn
absorbs hot water without melting a hole.  An independent
firn drill was designed that consisted of a conical drill
head wrapped in copper tubing through which hot water circulated at high speed,
melting the snow by contact. Hot water drilling commenced after the firn
drill completed the first portion of the hole. The firn drill had its own
water supply and heater in order to operate in parallel with hot water drilling
at other holes. The firn drill reliably achieved a rate of 2~m per hour~\cite{ehwd}.

The EHWD system was implemented across two separate sites. The Seasonal
Equipment Site (SES) provided electricity and a stable supply of hot
pressurized water, and the Tower Operations Site (TOS) was where the hole
was drilled.  The two sites were linked by long cables and insulated
hoses. The SES comprised generators, water tanks, pump and heating
buildings, a central control building, mechanical and electrical shops,
spare parts storage, and the Rodriguez well system (Rodwell), which
provides water to compensate for the volume change from ice to water
in each hole~\cite{rodriguez_well}. Hoses and cables connected SES subsystem buildings together, and
wherever necessary custom electrically heated hoses were installed,
providing an effective freeze mitigation strategy. The TOS included the
drill tower and attached operations building as well as the hose and cable
reels.  There were two towers and one set of drill reels.  After drilling,
drill reels were moved to the next hole location, where the second tower
had already been staged; the first tower stayed at its existing location
to support deployment of the instrumentation.  Once deployment had
finished, the first tower could be moved to the next location while
drilling at the second tower was underway.  This leapfrog sequence of the
tower structures reduced hole turnover time and allowed nearly
continuous drilling operations.

Due to the size and complexity of the SES, it remained stationary
throughout each drill season.  At the end of the drill season, the SES was
decommissioned and repositioned for the following drilling season.  The
distance between the SES and TOS had a practical limit, referred to as
reach, which defined the boundary of a seasonal drilling sector.  The
maximum reach
of the EHWD was 450~m, limited by pressure and voltage drop through the
SES--TOS link.

Each drilling season started with a staggered arrival of drill crew members
while the SES and TOS were excavated from accumulated snow drift and commissioned.  Season startup
tasks included SES and TOS warming and hookups, reinstallation of
do-not-freeze equipment (such as motor drives, sensors, and some
rubber items such as gaskets), generator commissioning, safety checkout and system
tuning, initial (``seed'') water delivery to begin the drilling process, and Rodwell development.  This phase typically
took four weeks.

The production drilling sequence was to drill, ream, and move to the next
location. The drill crews worked in three shifts per day of ten people
per shift. Independent firn drilling stayed ahead of deep drilling by at
least two holes, and often the Rodwell and first few holes of
the season had already been firn-drilled the prior season. Hole production rate was 48 hours per hole on average,
and the quickest cycle time was 32 hours. The idle phase
between holes was characterized by minimal flow through the
system and included regular maintenance tasks and deployment of IceCube
instrumentation.  

System shutdown would begin approximately two weeks before the end of the
austral summer season. Shutdown tasks included flushing the system with
propylene glycol, blowing out the plumbing with compressed air, removing
do-not-freeze equipment for warm storage, storing the TOS structures and
other support equipment, and finally, moving the SES into place for the
following season.  Due to a strong safety culture and retention of
experienced personnel, IceCube had only
four reportable drilling-related safety incidents in approximately 52 on-ice 
person-years. 

\subsection{\label{sec:deployment_inst}Deployment Procedures and Instrumentation}

\begin{figure}[!ht]
 \centering
 \includegraphics[width=0.70\textwidth]{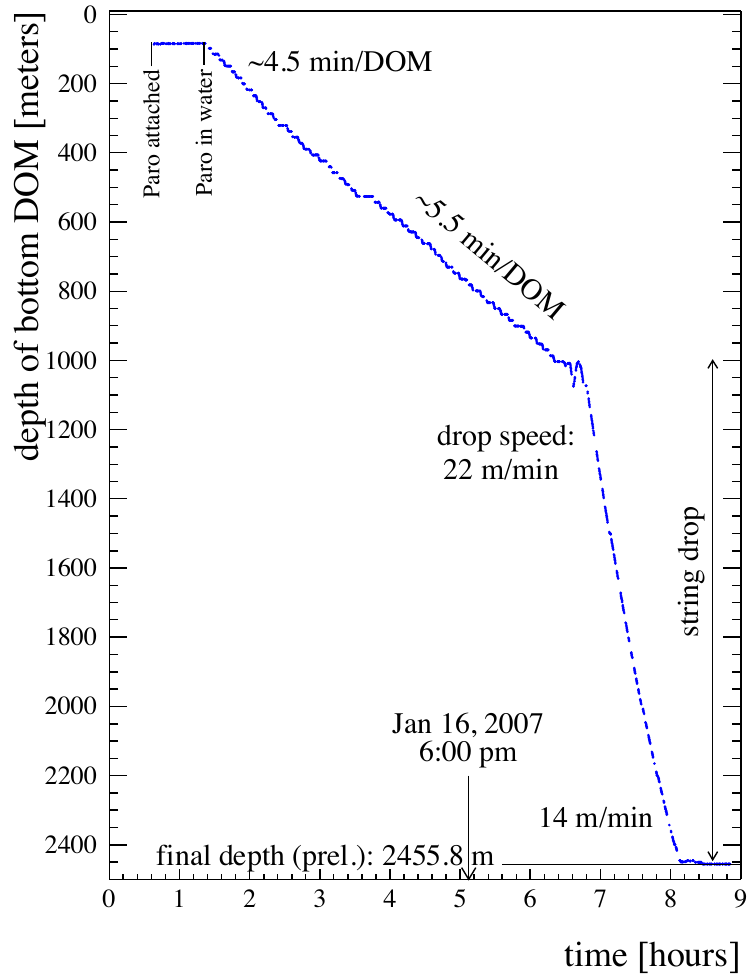}
\caption{Depth of the lowermost DOM vs.~time for String 48, deployed in the
  2006--2007 season. The final depth is the preliminary Stage~1 depth
  measured during deployment (section~\ref{subsec:stage1_geo}),
prior to the Stage~2 corrections derived from LED flasher
measurements (section~\ref{subsec:stage2_geo}).}
\label{fig:deploytime}
\end{figure}

DOM deployment required 60 DOMs staged in the drill tower, various
attachment hardware, pressure sensors, and the in-ice cable on a spool
outside the drill tower. The hole diameter was logged before deployment
using calipers on the drill head. After verification that the hole was 
ready for deployment and a pre-deployment safety check was carried out, the in-ice cable
was pulled over the top of the tower above the open hole. Four 100-pound
weights were connected together and attached via a 2.1~m steel cable to a
DOM that had a 17 meter penetrator assembly (see figure~\ref{fig:domcable}). The weights and
lowermost DOM were attached to the bottom of the in-ice cable via a 15.5~m
steel cable. After lowering of the DOM and weights, the next DOM with a
1.8~m penetrator assembly was attached to the in-ice cable. The two DOMs were
then electrically connected to the in-ice cable at the breakout. A
Paroscientific pressure sensor (model 8CB4000-I) was attached just above
the second DOM, and a Keller AG pressure sensor (model 300DS-340-003700)
was attached in the middle of the 
instrumented section of the string. The
pressure sensors were read out during deployment to confirm that the string
was properly descending into the hole.   DOMs with 17~m and
1.8~m penetrator assemblies were alternately attached to the in-ice 
cable, until all 60 DOMs were attached, after which the remaining 1.5~km of
in-ice cable was lowered into the hole (the ``drop'' phase). The top of the
cable was secured by an anchor trenched into the snow near the hole. After the
cable was secure, its end was taken off of the spool and connected to the
Surface Junction Box (SJB). An example of the time profile of deployment from pressure sensor
attachment to string drop is shown in figure~\ref{fig:deploytime}.

Dust loggers were deployed in selected holes in order to measure the
concentration of dust and volcanic ash as a function of depth at various
locations in the detector, a critical component of the overall measurement
of the optical properties of the
ice~\cite{Aartsen:2013rt,citeulike:2998650}. A dust logger consists of a 404~nm laser line source emitted horizontally in a
fan pattern, paired with an integrated
photomultiplier tube and digital photon counter for light
detection. Disposable dust loggers were deployed in two holes, and reusable dust loggers were
deployed in six holes. Each disposable dust
logger was attached to the main cable between the lowermost DOM and the
weight stack, and an extension cable ran between the dust logger and the
bottom breakout, just above the next-to-lowest DOM. In this mode, the hole
was logged once during deployment, and the logger was not
retrieved. The reusable dust logger was used to
log holes just before deployment, using a separate winch. The reusable
dust logger produced two logs for the 
hole, one in the up direction and one in the down direction, which were used for
reciprocal calibration.

\subsection{Geometry Measurement}

The geometry of the detector was determined using drill and survey data
during deployment (Stage 1), and then corrected and refined using the LED
flashers in ice (Stage 2). The IceCube coordinate system is
defined relative to the South Pole grid coordinate system (Northings and
Eastings) that moves with the ice sheet.  Elevation is defined relative to
Mean Sea Level.  The IceCube coordinate system origin
is located at 46500'E, 52200'N, at an elevation of 2900~ft (883.9
m).  This origin is located close to the center of the instrumented volume of
IceCube, about 2000~m below the surface of the ice. The $y$-axis of
the IceCube coordinate system is aligned with the Prime Meridian (Grid North),
pointing toward Greenwich, UK. The $x$-axis of the IceCube coordinate
system points 90$^{\circ}$ clockwise from the $y$-axis (Grid East). The $z$-axis is
normal to the Earth's surface, pointing upward. 

\subsubsection{\label{subsec:stage1_geo}Stage 1 Geometry Measurement}
The $(x,y)$-coordinates of the string were calculated using the position of
the drill tower. Before deployment, when the drill tower was in position, at
least three of the tower corners were surveyed from at least one control
point.  The coordinates for the center of the hole in the tower floor were
calculated from the corner coordinates. The string was assumed to be
vertical, so the $(x,y)$-coordinates from the tower were used at all
depths on the string. The drill
data show deviations of less than 1~m from vertical that are not
included in the detector geometry but have been validated
with flasher data for select strings, as discussed in section~\ref{sec:trilateration}. 

The depth of the lowest DOM on the string was calculated using pressure
readings from the pressure sensor, converted to depth by correcting for the
compressibility of the water in the hole and the ambient air pressure
measured before the pressure sensor enters the water. The distance from the
tower floor to the water surface was measured with a laser ranger
during deployment. The
vertical DOM spacings were also measured with a laser
ranger aimed down the hole after each DOM attachment. All depths were
converted to $z$-coordinates in the IceCube 
coordinate system.

\subsubsection{\label{subsec:stage2_geo}Stage 2 Geometry Measurement}

\begin{figure}[!ht]
  \captionsetup[subfigure]{labelformat=empty} \centering
  \subfloat[]{\includegraphics[width=0.5\textwidth]{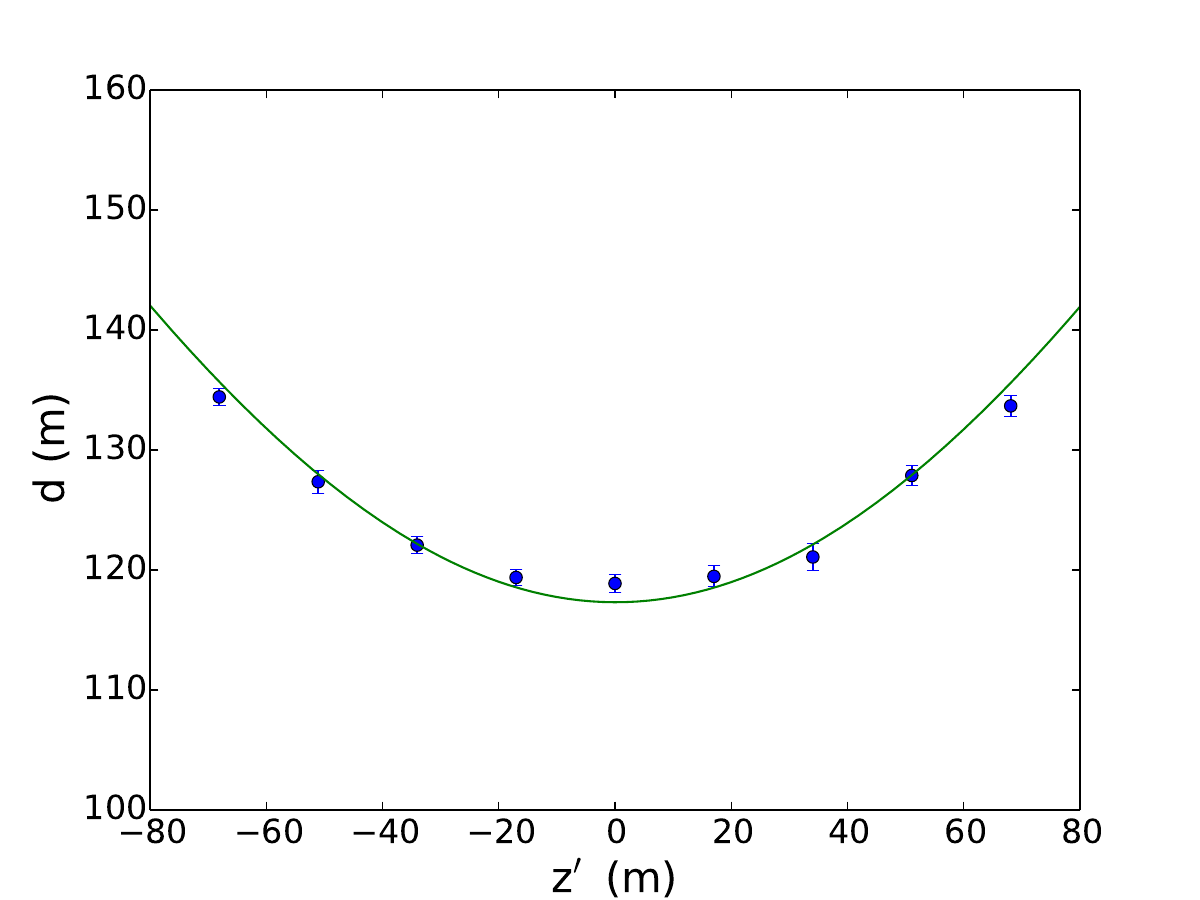}}
  \subfloat[]{\includegraphics[width=0.5\textwidth]{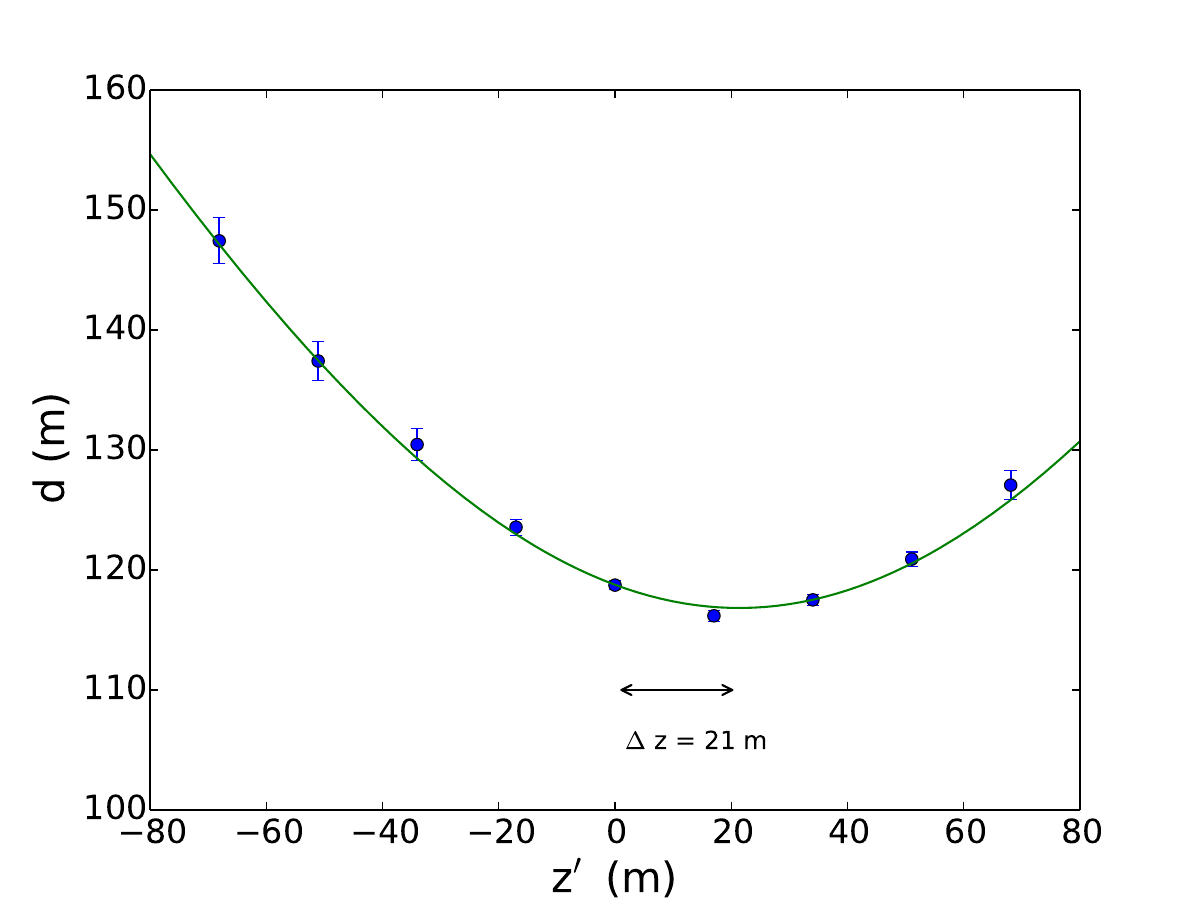}}
  \caption{Left: distance $d$ vs.~relative depth $z'$ as measured by flashers for
    two strings that were deployed at nearly the same depth with no
    anomalies during deployment. Right: the same plot for a different pair;
    in this case, the pressure sensor on the flashing string failed during
    deployment.}
  \label{fig:geohyperbola}
\end{figure}

The LED flashers on the DOMs were used to correct the depths of the strings relative to
the Stage 1 geometry measurements. The correction was calculated by flashing
the horizontal LEDs on a DOM at the surrounding strings and finding the
leading edge of the time distribution of the light recorded by the
receiving DOM, denoted $t_0$. The distance corresponding to the leading
edge time is $d = c_{\mathrm{ice}} \cdot t_0$, and the distances for all receiving
DOMs are plotted as a function of the vertical distance between the flasher
and the receiver, $z' = z_{\mathrm{receiver}} - z_{\mathrm{flasher}}$. The resulting curve
describes a hyperbola, $d = \sqrt{D^2 + (z' -\Delta z)^2}$, where $D$ is
the horizontal distance between the flasher string and the receiver string,
calculated from Stage 1 data, and $\Delta z$ is the relative offset between
the depths of the flashing and receiving string as shown in
figure~\ref{fig:geohyperbola}. The hyperbolic fit was performed
simultaneously on multiple flasher--receiver pairs; in order to avoid
bias, this global fit was performed multiple times on randomly
selected sets of string pairs. An average offset was calculated for
each string and applied as a correction to the Stage~1 $z$-coordinates
of all DOMs on that string. The correction was typically less than 1~m
relative to the Stage 1 data, but could be as large as 20~m (larger than the
17~m DOM spacing) in cases where the pressure sensor failed during string
deployment before it recorded the final depth reading, resulting in a
poor estimate of the Stage~1 depth. The Stage 2 depth correction was less than 3~m for 93\%
of the strings. The uncertainty on the correction varied from string
to string but was typically less than 0.2~m. The uncertainty on the
absolute $z$-coordinate is 1~m, based on timing residual studies from
downgoing muons~\cite{IC3:perf}.

\subsubsection{\label{sec:trilateration}Trilateration Validation of DOM Coordinates}

\begin{figure}[!ht]
  \captionsetup[subfigure]{labelformat=empty} \centering
  \subfloat[]{\includegraphics[width=0.48\textwidth]{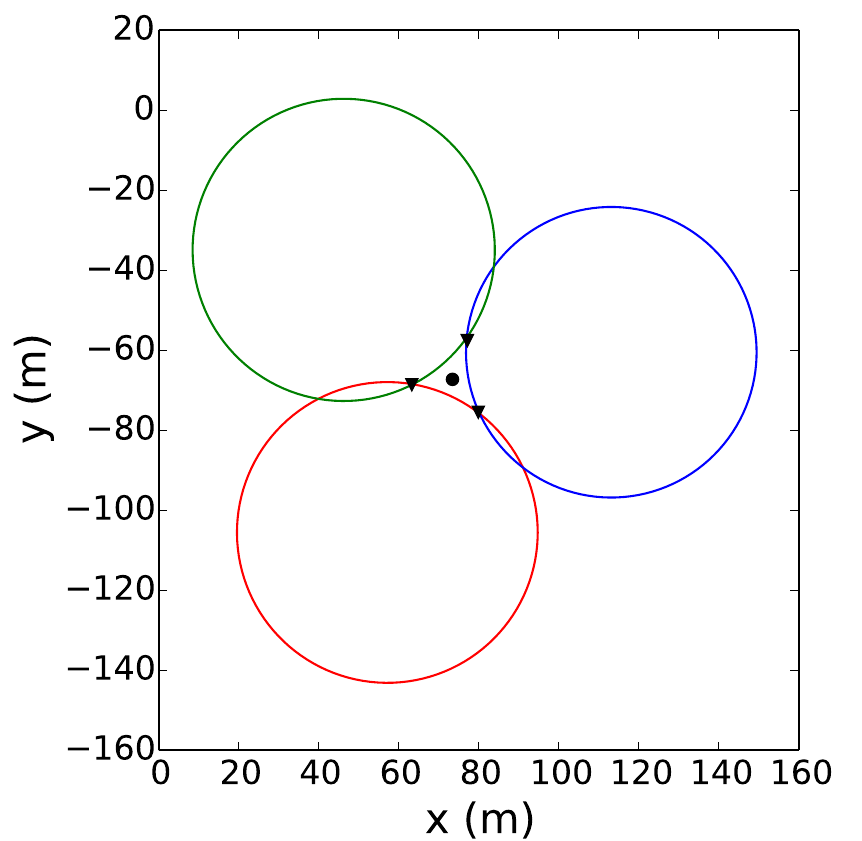}}
  \subfloat[]{\includegraphics[width=0.48\textwidth]{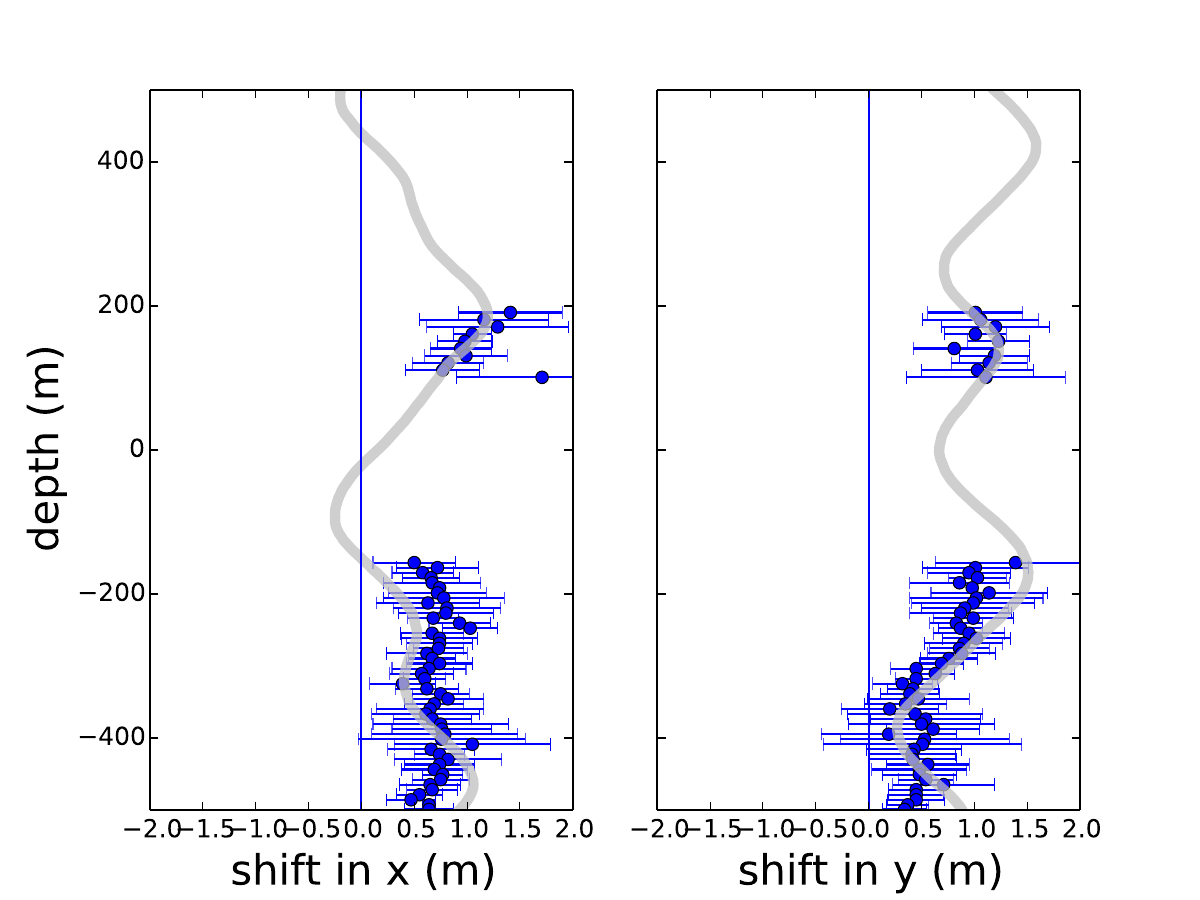}}
  \caption{Left: example of three circles drawn around strings
    receiving flasher data from string~80. The inner intersection
    points of the circles are marked with triangles, and the centroid
    of these points, marked with a black dot, is the fitted flasher
    position.  Right: Results of the trilateration fit to String 85, one of
    the DeepCore strings in the center of the detector. The shifts in the
    $x$- and $y$-coordinates are shown as a function of depth, with the drill
    data shown as a gray line.}
  \label{fig:trilateration}
\end{figure}

The Stage 2 geometry assumes that all strings are vertical. The
location data from the drilling process show that the string is not perfectly vertical,
although deviations from the vertical are typically less than 1~m. The
$(x,y)$-coordinates from the drill location data at varying depth were validated on the DeepCore strings
using a trilateration method. In this method, the 5~DOMs closest to the
flasher on each of the three closest strings surrounding the flasher are
selected, and a circle of radius $r = \sqrt{d^2 - (z')^2}$ is
described around each receiving DOM, where $d$ is the distance between the DOM
and the flasher calculated from the leading edge time of the received
light, and $z'$ is the relative depth of the flashing and receiving
DOMs calculated from the method described above. With 15~circles,
there are 125 possible combinations of three circles that can be used
for trilateration. For each combination, the six intersection points
of the three circles are calculated as shown in
figure~\ref{fig:trilateration}, and the $(x,y)$-position of the flashing DOM 
is taken to be the centroid of the three innermost intersection
points. The final $(x,y)$-position is the average of the values
calculated from each of the 125~combinations of circles. The
error bars on the positions are $1 \sigma$ from a Gaussian fit to all centroid
coordinates; the $x$- and $y$-coordinates are fitted independently. The
measured positions agree with the drill data within the error bars as shown in
figure~\ref{fig:trilateration}. Since the shifts from the nominal
$(x,y)$-coordinates are less than 1~m, and the trilateration measurement is only
practical in a few DeepCore strings where the interstring spacing is 40--70~m, these
corrections were not applied to the IceCube geometry, which still assumes
the vertical string geometry from Stage 2.

\subsubsection{Inclinometers}

The stability of the East Antarctic ice sheet is one key feature making it
suitable for a particle physics installation.  However, ice sheets can
deform mechanically under the stress of their own weight.  To measure the
stability of the ice, IceCube installed 50
inclinometers in the deeper sections of the array: two 
electrolytic inclinometers (Applied Geomechanics, Santa Cruz, CA)
housed in $\sim$20-cm-long aluminum pressure modules were installed during
the 2007--2008 season, and 48 micro-electromechanical (MEMS) tilt sensors
(ADIS16209, Analog Devices) were added to DOMs deployed in seasons
2009--2010 and 2010--2011.  This inclinometer array was intended not only for monitoring
detector geometry but also as a unique glaciology experiment that
permits three-dimensional tracking of deep ice flow and evaluation of complex
full-stress models that cannot be effectively tested under laboratory
conditions \cite{pattyn03}.

\begin{figure}[!ht]
	\centering
    \includegraphics[width=0.7\textwidth]{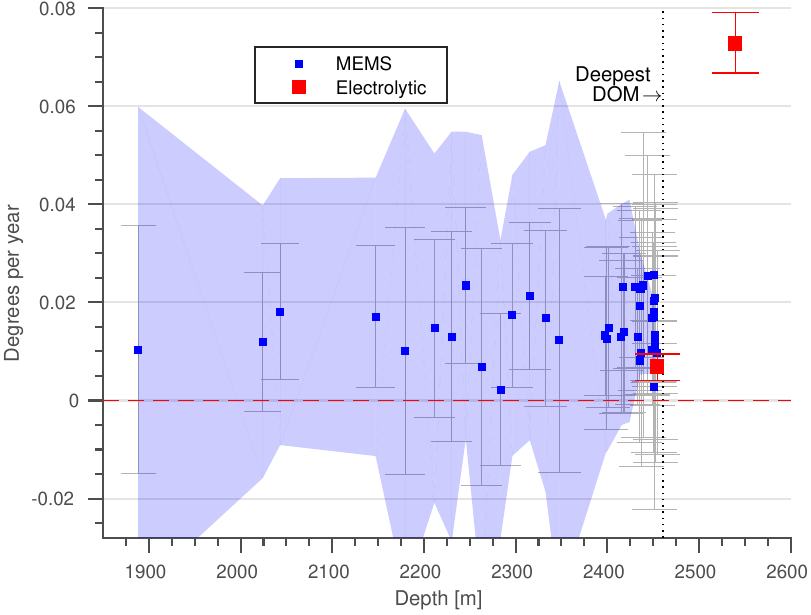}
	\caption{Long term average inclination readings from two electrolytic
      (red) and 42 MEMS (blue) tilt sensors installed in 2007--2011.  Error
      bars show standard deviation from trend; shaded area is MEMS 95\%
      confidence interval in 15 m bins, not including the quoted drift
      from aging tests.  The reading at \qty{2540}{m}
      indicates increasing strain below the IceCube instrumented volume.}
	\label{fig:tilt}
\end{figure}

Figure~\ref{fig:tilt} shows
six years of data from 42 of the DOM-embedded MEMS tilt sensors and
eight years of data from the two electrolytic inclinometers. Measurements are
started after 400~days in ice to avoid  
drift due to initial settling of the inclinometers. The electrolytic
inclinometer at \qty{2455}{m} (86\% ice sheet depth) was 
installed at the bottom of String 68.  For String 45, the drill
descended an additional $\sim$\qty{100}{meters} in order to deploy an
electrolytic inclinometer attached to a 100-pound weight at 2540 m
(90\% ice sheet depth) using an extension cable. Data
points are long-term average DOM inclination in degrees per year, with error
bars indicating the standard deviation from trend.  The MEMS sensors have
higher noise than the electrolytic inclinometers, and aging tests have indicated a long-term
drift of $\sim$\numrange[range-phrase = --]{0.02}{0.03}~m/s$^2$, corresponding to
$\sim$\numrange[range-phrase = --]{0.01}{0.02}~degrees per year
\cite{inclinometer_comm}. The MEMS readings are consistent with
$\lesssim$\qty{0.01}{degrees} of tilt per year with no apparent depth
dependence.  The deep electrolytic sensor 
at \qty{2540}{m} has shown a persistent \numrange[range-phrase =
  --]{0.07}{0.08} degrees of tilt per year since installation (shear
  of $\tan(0.075^\circ) = 0.0013$
per year).

In most glaciers at sufficiently great depth, ice strain undergoes a
transition from compression-dominated to shear-dominated with
small-scale folding at centimeter to meter scales~\cite{montagnat14,jansen16}.  This transition depth depends on
temperature \cite{price2002temperature}, grain size, and impurities, and is
associated with a strong single-maximum polycrystalline ice fabric
\cite{cuffey10}.  Tilt sensors within the IceCube instrumented volume, at
depths \SI{<2450}m, show essentially no movement.  Profiles of the
atmospheric dust embedded in the ice measured with dust loggers \cite{I3:dustlogger} show no
indication of folding and appear undisturbed over the full IceCube depth.

\subsection{Hole Ice Imaging Camera}

An imaging camera system, developed at Stockholm University, Sweden, was deployed on String 80 in the final
construction season in order to monitor the freeze-in process and optical
properties of the drill hole.  The system consists of two video cameras
housed in separate glass pressure spheres \SI{5.8}{m} apart, at a
depth of \SI{2455}{m} between the bottom DOM and the string
weights. The cameras can be rotated to point in multiple directions, with some
limitations due to mechanical constraints.  Each
camera is equipped with four LED lamps and three lasers (red, blue, and
green).  Camera heaters, lights, and movement can be controlled remotely
using a dedicated system in the ICL.  

The camera system observed that the drill hole was completely refrozen after
15 days.  The refrozen hole ice consists of a clear outer
layer and a central core of about 16 cm diameter that has a much
smaller scattering length than the bulk ice \cite{rongen_vlvnt15}.  The
optical properties of this hole ice are still under active investigation.
No long-term changes have been observed with subsequent camera operations.

%auto-ignore
\section{\label{sect:online}Online Systems}

The IceCube online systems comprise both the software and hardware at the
detector site responsible for data acquisition, event selection,
monitoring, and data storage and movement.  As one of the goals of IceCube
operations is to maximize the fraction of time the detector is sensitive to
neutrino interactions (``uptime''), the online systems are modular so that
failures in one particular component do not necessarily prevent the
continuation of basic data acquisition. Additionally, all systems are
monitored with a combination of custom-designed and industry-standard tools
so that detector operators can be alerted in case of abnormal conditions.

\subsection{\label{sect:online:dataflow}Data Flow Overview}

\begin{figure}[!ht]
 \centering
 \includegraphics[width=0.6\textwidth]{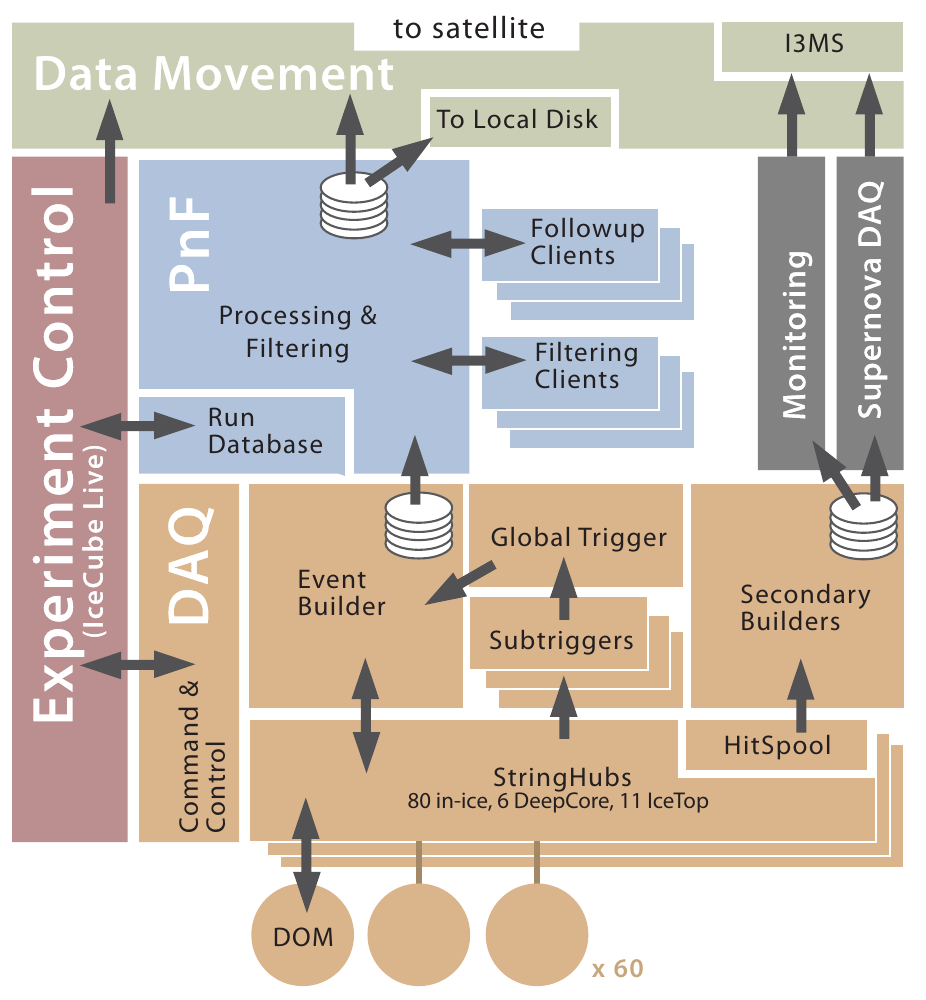}
 \caption{Data flow in the primary IceCube online systems. See details
   on each component in the text.}
 \label{fig:online_dataflow}
\end{figure}

The online data flow consists of a number of steps of data reduction and
selection in the progression from photon detection in the ice to
candidate physics event selection, along with associated secondary
monitoring and data streams.  An overview of the data flow is shown in
figure~\ref{fig:online_dataflow}.

DOM hits are mostly due to dark noise. The first step
in data reduction takes place in the DOM, using the Local Coincidence (LC) condition described in
section~\ref{sec:dom_functional}.  Hits that meet the LC criteria are flagged
as Hard Local Coincidence (HLC hits) and include a full payload of
digitized waveforms, while isolated non-LC hits are flagged as Soft Local
Coincidence (SLC) hits and are compressed more aggressively, with only a
timestamp and minimal amplitude / charge information transmitted
(section~\ref{sect:online:payloads}). 

All DOM hits are read out to dedicated computers on the surface
(section~\ref{sect:sps}) by the data acquisition system (DAQ).  The next level
of data selection is the formation of software triggers by the DAQ
system. HLC hits across the detector are examined 
for temporal and in some cases spatial patterns that suggest a common
causal relationship.  A number of different trigger algorithms run in
parallel, described in section~\ref{sect:online:trigger}.  All hits (both HLC
and SLC) within a window around the trigger are combined into events, the
fundamental output of the DAQ.  The event rate varies
seasonally with the atmospheric muon flux~\cite{ICECUBE:IceTop} from 2.5
kHz to 2.9 kHz, with a median rate of 2.7 kHz, and the total DAQ data rate is
approximately 1~TB/day. 

The DAQ also produces secondary streams that include time calibration,
monitoring, and DOM scaler data.  The scaler data report the
hit rate of each DOM and are used in the supernova data
acquisition system (section~\ref{sect:SNDAQ}).  The time calibration and
monitoring streams are used to monitor the health and quality of the
data-taking runs.

The DAQ event data are then processed further with approximately 25 filters
in order to select a subset of events (about 15\%) to transfer over
satellite to the Northern Hemisphere (section~\ref{sect:online:filter}).  Each
filter, typically designed to select events useful for a particular physics
analysis, is run over all events using a computing cluster in the ICL.
Because of limitations both on total computing power and bounds on the
processing time of each event, only fast directional and energy
reconstructions are used.  This Processing and Filtering (PnF) system is
also responsible for applying up-to-date calibration constants to the DAQ
data. All processed events, even those not selected by the online filters,
are archived locally.

A dedicated system for data movement, JADE, handles the local archival storage to
tape or disk and the handoff of satellite data
(section~\ref{sect:online_jade}).  This includes not only primary data streams
but also monitoring data, calibration runs, and other data streams.
Low-latency communications for experiment control and real-time monitoring
are provided by the IceCube Messaging System (I3MS).  
Experiment control and detector monitoring are handled by the IceCube Live
software system, described in section~\ref{sec:online:icecubelive}.

Data-taking runs are arbitrarily divided into 8-hour periods and assigned
a unique run number; data acquisition need not actually pause during
the run transition.  Detector configuration parameters that affect physics
analyses are changed at most once per year (typically in May), indicating
the start of a new ``physics run''.   

\subsection{\label{sect:sps}South Pole System and South Pole Test System}

The South Pole System (SPS) comprises 19 racks of computing and network
hardware that run the various online systems described in this
section.  The DOM surface cables are connected via passive patch panels to
custom 4U computers, called DOMHubs, one DOMHub per in-ice string and 11 additional
hubs for IceTop.  The remaining servers, including those for higher-level
data acquisition, event filtering, detector monitoring, and core
infrastructure, are currently 2U Dell PowerEdge R720 servers running
Scientific Linux (Table \ref{tab:sps_breakdown}).  The servers are
typically upgraded every three to four years.  The custom hardware components in
the DOMHubs are replaced with spares as failures warrant, and the disks and
single-board computer were upgraded in 2013--14.

\begin{table}[h]
  \centering
\caption{Breakdown of computing equipment at SPS, indicating number of
    machines used for each task.}
  \begin{tabular}{ r  c }
\hline
    Component & \# \\ \hline DOMHubs & 97 \\ Other data
    acquisition & 4 \\
    Monitoring & 3 \\ Event filtering & 24 \\ System infrastructure & 8 \\ Other &
    6 \\
\hline
  \end{tabular}
  \label{tab:sps_breakdown}
\end{table}

Each of the DOMHubs is an industrial computer chassis with custom components for DOM
power, timing, and communication.  A low-power single-board computer
communicates with 8 custom PCI DOM Readout (DOR) cards via an
industry-standard backplane.  An ATX power supply with two
redundant modules powers the DOMHub, while two
48~VDC Acopian power supplies, mounted and connected in series inside the
chassis to double the voltage to 96~V, supply power to the DOMs.  The DOM power is switched and
monitored by the DOR cards and is controlled by software.  Another PCI
card, the DOMHub Service Board (DSB), is responsible for GPS timing fanout
(section~\ref{sect:online:master_clock}).

The SPS computers are connected via switches in each rack that provide
redundant connections to a 10--20 Gbps network backbone.  The DOMHubs are
connected to the rack switches with two bonded 1 Gbps links.  Typical network I/O during
data-taking for the DAQ Event Builder (section~\ref{sect:online:evbuilder}) is about 240~Mbps in each direction.
The PnF Central Server sees 200 Mbps in and 640 Mbps out; the output
stream is significantly higher than the input as PnF distributes the
events to filtering clients and generates multiple output streams
(section~\ref{sect:online:filter}).  

Redundancy and continuous monitoring of SPS is one of the keys to a high
detector livetime (section~\ref{sec:operational_performance}).
Nagios monitoring software detects and flags problems, 
including issues with DOM power and communication on the DOMHubs.  Severe
problems impacting data-taking result in a page to the IceCube winterover 
personnel via the station's Land Mobile Radio (LMR) system.  A dedicated
powered spare server can replace any failed DAQ node, and spare PnF filtering
clients can be started to increase throughput in case of a data filtering
backlog.  SPS hardware is also connected to uninterruptible
power supplies (UPS) in order to continue data-taking through station power
outages of up to 15 minutes.

Total power usage of the detector and online systems, including computing servers, is
approximately 53 kW.  The majority is consumed by the DOMs, with an
average power consumption of 5.7~W each, including power supply efficiency
and transmission losses, for a total of 30.6 kW.  The DOMHubs require 128~W
each, not including the DOM power, for a total of 12.4 kW.  The
computing servers consume approximately 200--300~W each, depending on
configuration.  Most of the of the remaining power is used by the PnF
Filter Clients: 20 servers of 300W each for a total of 6 kW.

A scaled-down version of SPS, the South Pole Test System (SPTS) located in
Madison, Wisconsin, U.S.A., allows testing and validation of both hardware
and software in the Northern Hemisphere before rollout to SPS.  Servers and DOMHubs
identical to those at SPS, along with a small number of DOMs in chest
freezers, are used in the test system.  Although the number of DOMs
available is much smaller than in the real detector, recent software
improvements allow the ``replay'' of pre-recorded raw SPS hit data
on SPTS systems, providing a data stream to higher-level DAQ and PnF
components identical to SPS.  Another test system includes a full-length
in-ice cable and is used primarily for validation of DOM communications and
timing.

\subsection{Data Readout and Timing}

While the low-level communications and timing systems of the DOM are
described in detail in ref.~\cite{ICECUBE:DAQ}, we review those here in
the broader context of the online systems.

\subsubsection{\label{sect:online:comms}Communications}

Digital communication between the DOR card and DOM occurs via copper
twisted pairs, with two DOMs per pair on the in-ice cable, and one IceTop
DOM per pair for increased bandwidth (IceTop hit rates can exceed 3 kHz).
The physical layer signaling uses on-off keying with bipolar pulses.  The
protocol is a custom 
packet-based scheme.  Each packet is assigned a sequence number, and all
received packets are acknowledged if the sequence number is correct.  Each
packet also contains a cyclic redundancy checksum to detect transmission errors.
Out-of-sequence packets received are ignored, and non-acknowledged packets
are retransmitted. The total bandwidth of the communication channel
is 720 kbps per twisted pair.

Messaging is managed from the surface, in that the DOR requests data from
each DOM in turn; only one DOM per pair can transmit at a time.  Communication is
paused once per second to perform a timing calibration (RAPCal; section~\ref{sect:dom:rapcal}); this enables time transfer of DOM clock to DOR
clock for every DOM.  

\subsubsection{\label{sect:online:master_clock}Master Clock System}

\begin{figure}[!ht]
 \centering
 \includegraphics[width=0.8\textwidth]{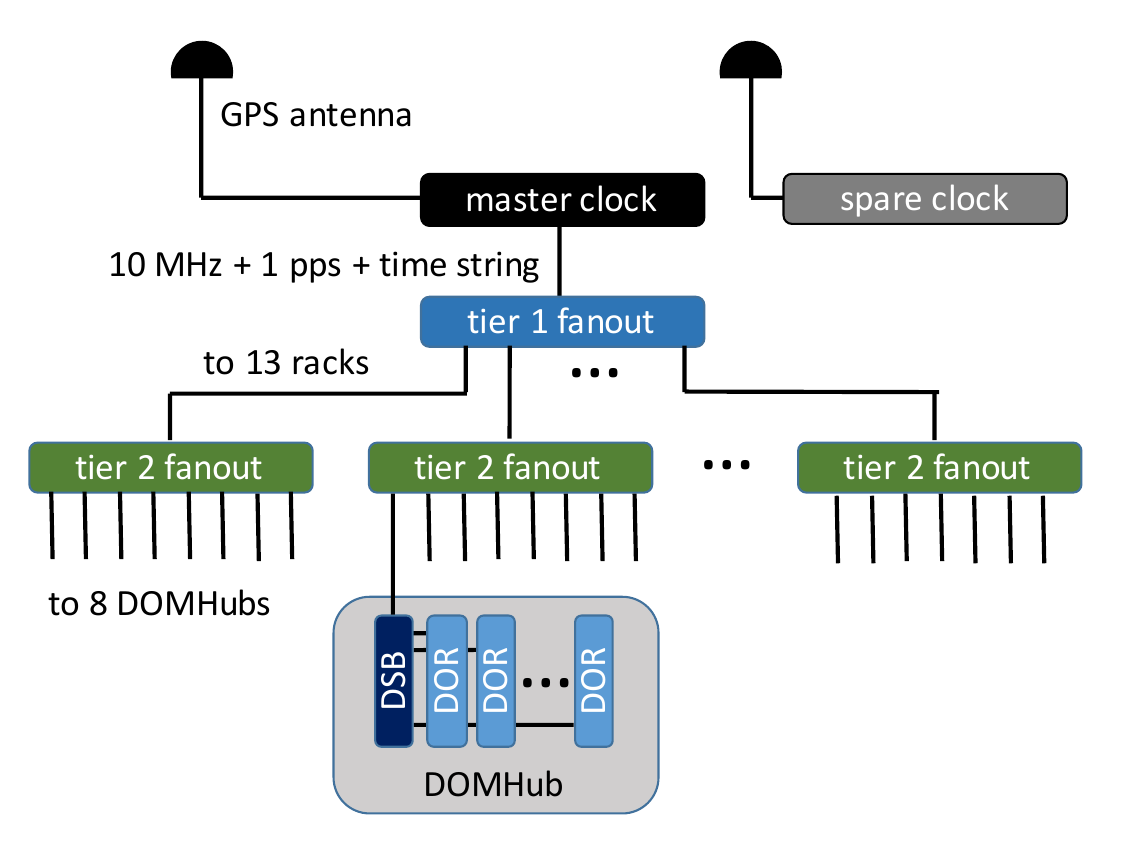}
 \caption{Master Clock fanout system, from GPS receiver to DOR cards in
   each DOMHub.}
 \label{fig:clock_fanout}
\end{figure}

The DOR clocks themselves are synchronized to UTC via an active fanout
system from a single Symmetricom ET6000 GPS receiver with a
temperature-stabilized 10 MHz oscillator, also known as the Master
Clock. The fanout tree is shown in
figure~\ref{fig:clock_fanout}. The 10 MHz output, a 1 Hz output, and a
serial time string indicating
the UTC date and time are distributed to the DOMHubs via a series of
fanouts, using shielded, delay-matched twisted-pair cables.  Within the
DOMHub, the DSB card continues the fanout via short delay-matched patch
cables to each DOR card.  The local 20 MHz clocks of each DOR card are
phase-locked to the distributed 10 MHz signal.  

To avoid the Master Clock being a single-point failure for the detector, a
hot spare receiver, using its own GPS antenna and powered through a
separate UPS, is continuously active and satellite-locked in case of
problems with the primary.

\subsubsection{DOR Card and Driver}

Each DOR card is connected to up to 8 DOMs, with 8 DOR cards in a
DOMHub. The DOR card controls the 96~VDC power supply to the DOMs and
modulates the communications signaling on top of this DC level; the DOMs
can accept input voltages from 40 V to 120 V.
Dedicated circuitry monitors the current draw and voltage levels on each
twisted pair. A ``firmware fuse'' can disable power if the current draw
deviates from programmable maximum or minimum levels, and this mechanism is
supplemented with standard physical fuses.  

The software interface to the DOR card, and thus to the DOMs, is provided
with a custom Linux device driver.  Access to DOR card functions, including
DOM power control, communication statistics, RAPCal, and current / voltage
monitoring, is facilitated using the Linux \texttt{/proc} filesystem
interface.  Data transfer from the cards to the single-board computer is
achieved via DMA over the PCI bus.  The driver
provides a device file for each DOM for read/write access by higher-level software.

\subsubsection{\label{sect:online:payloads}DOM Hit Payloads}

The content of the DOM hit payloads transmitted to the surface depends on whether local
coincidence was satisfied, i.e. whether the hit was flagged as HLC or SLC.
The DOM Main Board ID and the timestamp of the hit in DOM clock counts are
always transmitted, along with trigger and LC flags.

For HLC hits, the digitized ATWD and fADC waveforms are transmitted.
Waveforms from lower-gain ATWD channels are only included if the signal
amplitude in the higher-gain channel exceeds 75\% of the digitizer range.  The waveforms are
compressed losslessly in the DOM using a delta-compression algorithm that
encodes the difference between subsequent samples.  The difference values
are packed into words of length 1, 2, 3, 6, or 11 bits depending on
magnitude, and special values in the bitstream are used to transition
between different word lengths.

For both HLC and SLC hits, a chargestamp is included that provides an
estimate of the amplitude/charge even if, as in the SLC case, the full
waveform is not transmitted.  For in-ice DOMs, the chargestamp consists of
three samples of the fADC waveform centered around the peak value, along
with the peak sample number.  For IceTop DOMs, the chargestamp is the sum
of all samples of the ATWD waveform, after pedestal subtraction.

\subsection{Data Acquisition Software}

IceCube's data acquisition (DAQ) system is a set of software components
running on the DOMHubs and dedicated servers in the ICL.  These components are shown in
figure~\ref{fig:online_dataflow} and include StringHub, Trigger, Event
Builder, Secondary Builder, and a Command and Control server.  The DAQ is
responsible for detecting patterns of hits in the detector likely to be
caused by particle interactions and storing these collections of hits as
events.

Hits are read continuously from the DOMs by the
StringHub components running on each DOMHub, and a minimal representation of each HLC hit is
forwarded to the Trigger components (either the in-ice or IceTop Trigger.)
The Trigger components apply a
configurable set of algorithms to the hit stream and form windows around interesting temporal
and/or spatial patterns.  These time windows are collected by the
Global Trigger and used to form non-overlapping trigger requests by merging
subtriggers as needed, ensuring that the same hit doesn't appear in
multiple events.  The merged trigger requests are used by the Event Builder
component as templates 
to gather the complete hit data from each StringHub and assemble the final
events.

\subsubsection{StringHub and HitSpool}
\label{sec:domhub_hitspool}

The StringHub software component that runs on each DOMHub is responsible
for reading all available data from each of its connected DOMs each second
and passing that data onto the downstream consumers.  It also saves all
hits to a local ``HitSpool'' on-disk cache and queues them in an
in-memory cache to service future requests from the Event Builder for full
waveform data.

The StringHub component is divided into two logical pieces: the front
end is called Omicron, and the back end is the Sender. Omicron controls all
of the connected DOMs, forwarding any 
non-physics data (calibration, monitoring) to its downstream consumers and
sorting the hits from all 
DOMs into a single time-ordered stream before passing them to the Sender.  

Omicron is also responsible for translating DOM hit times into
UTC-compatible ``DAQ time'', which counts the number of 0.1-ns periods
since the UTC start of the year (including leap seconds).  The translation
uses the RAPCal procedure as described in section~\ref{sect:dom:rapcal},
performed for each DOM every second.  

The Sender caches SLC and HLC hits in memory, then forwards a
condensed version of each HLC hit to the appropriate local Trigger. Each
condensed HLC hit record contains the hit time, a DOM identifier, and the
trigger mode.  After the
Trigger components have determined interesting time intervals, 
the Event Builder requests each interval from the Sender which returns a list of
all hits within the interval and prunes all older hits from the in-memory hit
cache after each interval.

One core requirement of the DAQ is that each component operates on a
time-ordered stream of data.  The DAQ uses its ``Splicer'' to accomplish
this.  The Splicer is an object that gathers all input streams
during the setup phase at the beginning of a data-taking run; no inputs can
be added once started.  Each stream 
pushes new data onto a ``tail'', and the Splicer merges the data from all
streams into a single sorted output stream.  When a stream is closed, it
issues an end-of-stream marker that causes the Splicer to
ignore all further data.  Details of the Splicer algorithm can be found in
ref.~\cite{vlvnt13_trigger}.  

As hits move from Omicron to the Sender, they are written to the
HitSpool disk cache.  These files are
written in a circular order so that the newest hits overwrite the oldest
data.  The files are catalogued in a SQLite database to
aid in fast retrieval of raw hit data.

One limitation of the current design is that it only reads data when
the full DAQ system is running, so the detector is essentially ``off''
during certain hardware failures or the periodic full restarts of the
system that occur every 32 hours.  A future enhancement 
will split the StringHub into several independent pieces to eliminate these
brief pauses.  The front end (Omicron) will be moved to a daemon
that continuously writes data (including secondary, non-physics data and
other metadata) to the disk cache.  Part of the back end (Sender) 
will become a simple HitSpool client that reads data from the disk cache
and sends it to the downstream consumers, while another simple component
will listen for requested hit readout time intervals from the Event Builder
and return lists of hits taken from the HitSpool.

\subsubsection{\label{sect:online:trigger}Triggers}

The DAQ trigger algorithms look for clusters of HLC hits in space and time
that could indicate light due to a particle interaction in the detector, as
opposed to uncorrelated dark noise.   An algorithm searches for a given
multiplicity of HLC hits, possibly with an additional geometric
requirement, within a trigger time window.  The time scale of the trigger window is
set by the light travel time in ice and the geometry requirement
involved. Longer readout windows are appended before and after the trigger
windows to save early and late hits with the events.

Triggers are generally restricted to a subset of DOMs, such as all in-ice DOMs,
IceTop DOMs, or DeepCore DOMs.  The algorithms run in parallel over all
hits in the DOM set, and then overlapping triggers are merged.  The various
trigger algorithms are described below, and a summary of the algorithm
parameter settings is found in Table \ref{tab:triggers}.  Trigger settings
are changed at most once per year.

The fundamental trigger for IceCube, IceTop, and DeepCore is the Simple
Multiplicity Trigger (SMT).  The SMT requires $N$ or more HLC hits within a
sliding time window of several $\mu\mathrm{s}$, without any locality
conditions.  Once the multiplicity condition is met, the trigger is 
extended until there is a time period of the length of the initial trigger
window without any HLC hits from the relevant DOM set.  The
multiplicity value $N$ is tuned to the energy threshold of the sub-detector,
which fundamentally is set by the string or tank spacing.

\begin{figure}[ht]
  \centering \subfloat[]{
    \includegraphics[scale=0.45]{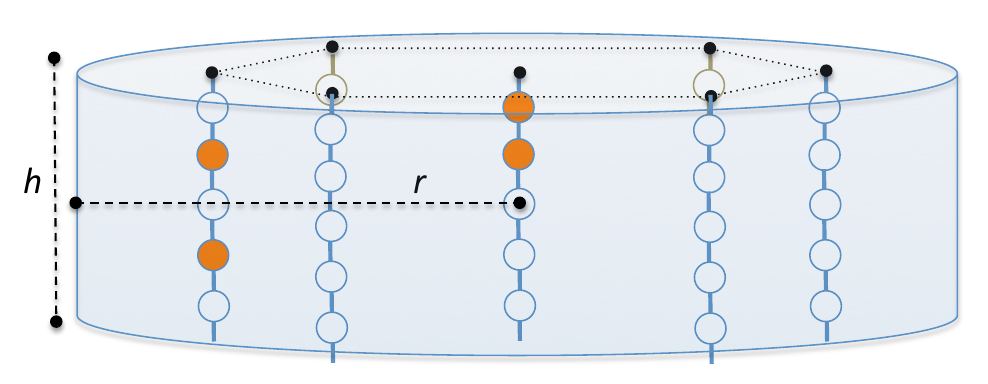}
    \label{fig:trig_cylinder}
  }
  \quad
  \subfloat[]{
    \includegraphics[scale=0.5]{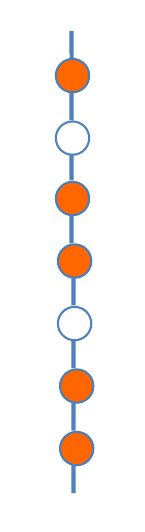}
    \label{fig:trig_string}
  }
  \caption{Schematic representation of triggers using spatial coincidences.  Shaded circles
    represent HLC-hit DOMs.  Left: Volume Trigger.  Right: String Trigger. }
\end{figure}

Other triggers use a lower multiplicity threshold by adding constraints on
the HLC hit topology.  The time windows for these triggers are based upon
the size of the locality volume. The Volume Trigger defines a cylinder of fixed size around
each hit DOM and requires a given multiplicity within this cylinder
(figure~\ref{fig:trig_cylinder}); this allows IceCube to trigger on localized
low-energy events that do not satisfy the SMT condition.  The Volume Trigger
has an additional simple multiplicity parameter that fires the trigger when
a certain number of hits is reached, regardless of any spatial
restrictions; this prevents the trigger 
algorithm from slowing down when the detector has triggered already from
the primary SMT. The String Trigger requires a certain number of hits
within a span of DOMs along a single string 
(figure~\ref{fig:trig_string}); this allows one to trigger on low-energy
muons that pass vertically through the detector.

IceCube can detect hypothetical subrelativistic heavy
particles such as magnetic monopoles that may catalyze nucleon decays along
their trajectory \cite{Aartsen:2014awd}.  However, because these
particles may travel at velocities less than $0.01c$, the time
windows used in the standard triggers are too short.  A dedicated Slow
Particle (SLOP) trigger has thus been developed to search for slow
track-like particle signatures.

\begin{figure}[!ht]
 \centering
 \includegraphics[width=0.6\textwidth]{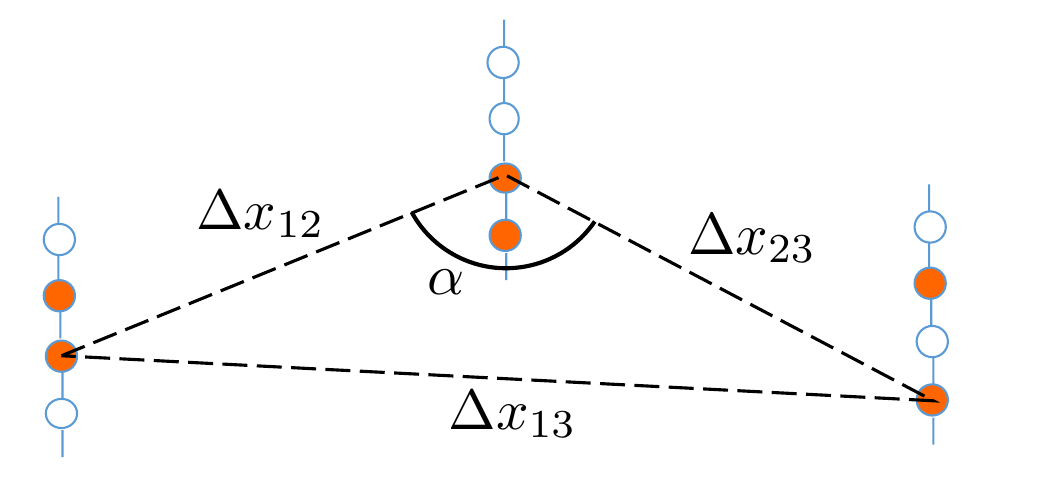}
 \caption{Geometry of a SLOP trigger triplet of HLC pairs.}
 \label{fig:slop}
\end{figure}

The SLOP trigger operates in several stages.  The HLC hits, which by design
occur at least in pairs along a string, are cleaned by removing pairs that
are proximate in time ($\Delta t < T_{\mathrm{prox}}$); $T_{\mathrm{prox}}$
is tuned to remove most hits from particles traveling near $c$, such as muons.
For all parameters, the trigger algorithm considers the time and 
position of the first hit within each HLC pair.  Next, triplets of HLC
pairs within a time window $T_{\mathrm{max}}$ 
are formed.  The geometry of each triplet formed (figure~\ref{fig:slop})
must satisfy track-like 
conditions: the largest inner angle $\alpha$ of the triangle formed by the
HLC pairs must be greater than $\alpha_{\mathrm{min}}$, and the
``velocities'' along the triangle sides must be consistent.  Specifically,
the normalized inverted velocity difference $v_\mathrm{rel}$, defined as

\begin{equation}
  v_\mathrm{rel}=\frac{|\Delta
  v_\mathrm{inverse}|}{\overline{v}_\mathrm{inverse}} = 
  3\cdot\frac{|\frac{1}{v_{23}}-\frac{1}{v_{12}}|}
  {\frac{1}{v_{12}}+\frac{1}{v_{23}}+\frac{1}{v_{13}}}
\end{equation}

\noindent where $\ v_{ij} = \Delta x_{ij}/\Delta t_{ij}$, must be less than
or equal to a predefined maximum value
$v_{\mathrm{rel}}^{\mathrm{max}}$.  Finally, the total number of track-like triplets
must be greater than or equal to $N_{\mathrm{triplet}}$, set to 5, and all
of these track-like triplets must overlap in time.  

Other special-purpose triggers exist to collect minimum bias data of
various sorts.  The Fixed-Rate Trigger (FRT) reads out 10~ms of hit data from
the full detector at fixed intervals.  This is especially useful for studies
of DOM noise.  The Calibration Trigger selects a particular type of hit
such as special IceTop non-HLC hits that have full waveform readout, and promotes
them to a trigger. The Calibration Trigger can also be configured to
include all events due to LED flashers in cases where flasher 
operations require disabling standard triggers. Finally, a Minimum Bias
trigger can select one of every $N$ HLC hits and promote this hit to a trigger, adding
readout windows as usual; currently an IceTop Minimum Bias trigger with a
prescale factor $N$ of 10000 is active.

\begin{table}
  \centering \footnotesize
\caption{Trigger parameters (as of May 2016) and typical trigger
  rates of each algorithm.  Most rates vary seasonally with the atmospheric
  muon flux.  The merged event rate varies from 2.5 to
  2.9 kHz.}  
\begin{tabular}{lrrrrr}
  \hline Trigger & DOM set & $N$ HLC hits & Window & Topology & Rate\\
  & & & ($\mu$s) & & (Hz) \\
  \hline
  SMT & in-ice & 8 & 5 & --- & 2100\\
  SMT & DeepCore & 3 & 2.5 & --- & 250\\
  SMT & IceTop & 6 & 5 & --- & 25\\
  Volume & in-ice & 4 & 1 & cylinder (r=175m, h=75m) & 3700\\
  Volume & IceTop infill & 4 & 0.2 & cylinder (r=60m, h=10m) & 4\\
  String & in-ice & 5 & 1.5 & 7 adjacent vertical DOMs & 2200\\
  SLOP & in-ice & $N_{\mathrm{triplet}} = 5$ & $T_{\mathrm{prox}} = 2.5$, &
  $\alpha_{\mathrm{min}} = 140^\circ,\ v_{\mathrm{rel}}^{\mathrm{max}}
  = 0.5$ & 12\\
  & & & $T_{\mathrm{min}} = 0$, & &\\
  & & & $T_{\mathrm{max}} = 500$ & &\\
  FRT & all & --- & --- & --- & 0.003\\
  \hline
\end{tabular}
\label{tab:triggers}
\end{table}

Many events will satisfy more than one of the trigger conditions, sometimes
multiple times.  In order to avoid overlapping events, possibly containing
the same DOM hits, the triggers and their associated readout windows are
merged, while retaining information about the separate triggers.  The
merged trigger is referred to as the Global Trigger.

\begin{figure}[!ht]
 \centering
 \includegraphics[width=0.8\textwidth]{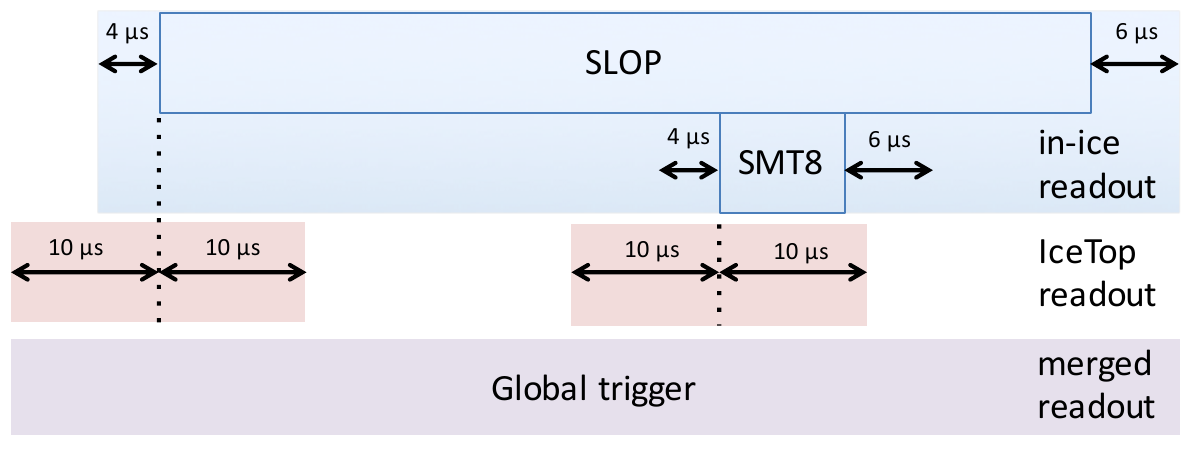}
 \caption{In-ice, IceTop, and merged readout windows for a long event
   satisfying SLOP and SMT8 triggers.}
 \label{fig:trigger_readout}
\end{figure}

Each trigger has defined readout windows around the trigger window; all
hits from the full detector, including those DOM sets not involved in the trigger,
are requested from the StringHub components and built into events.  For the
DOM set involved in an in-ice trigger, the readout windows are appended at each end of the trigger
window, while for other DOM sets, the readout windows are centered around
the trigger start time.  Readout windows around IceTop triggers are global
and include hits from all other DOM sets before and after the trigger
window. The union of overlapping readout windows defines 
an event (figure~\ref{fig:trigger_readout}).  Long events such as SLOP or FRT
triggers typically contain several causally independent ``physics'' events;
these typically are re-split before reconstruction and analysis.

\subsubsection{\label{sect:online:evbuilder}Event Builder}

The Event Builder receives requests from the Global Trigger, extracts the
individual readout windows, and sends them to the appropriate subset of the
StringHubs.  The StringHubs each send back a list of all hits within the
window. When all StringHubs have returned a list of hits, these are bundled with the trigger data into an event.

Events are written to a temporary file.  When the temporary file
reaches a preset configurable size, it is renamed to a standard unique name.  When the PnF
system sees a new file, it accepts it for processing and filtering
(section~\ref{sect:online:filter}).  The total latency from detection of
photons at the DOMs to DAQ events written to disk is approximately five
seconds. 

\subsubsection{\label{sect:online:daqdomconfig}DAQ and DOM Configuration}

The configuration of the DAQ is managed by two sets of XML files: a cluster
configuration file and a hierarchical tree of run configuration files.

The cluster configuration file contains system-level settings used to
launch the DAQ, such as component host servers, startup paths, command-line
options, etc.  Components (other than StringHub) can easily be moved to
different hosts for troubleshooting, load balancing, and maintenance.

Run configuration files list the trigger and DOM configuration files to be
used for taking data.  The trigger configuration file specifies
configuration parameters for all 
trigger components (in-ice, IceTop, and global) used in a run.  These
include the list of algorithms run by each trigger component, along with
readout window sizes and any other configurable parameters (multiplicity
threshold, trigger period, prescale factor, etc.).

DOM configuration files (one per hub) list all DOMs that contribute to
the data-taking run.  All configuration parameters for each DOM are
specified, including PMT high voltages, ATWD operating parameters,
discriminator thresholds, local coincidence settings, baselines and others.

Run configuration files (including trigger and DOM files) are versioned and
frozen once used for data-taking.  All relevant configuration parameters
are also stored in a database for use in analysis.

An additional geometry XML file contains the $(x,y,z)$ and (string,
position) coordinates of the DOMs, needed by the Trigger components.  The
DOMs entries are indexed by their unique Main Board ID.  This ensures that cabling changes
on a DOMHub do not result in changes in data-taking or errors in the geometry.

\subsubsection{Component Control}

The DAQ components are managed by a single ``command-and-control'' daemon,
CnCServer, that manages and monitors components and acts as the main
external interface to the DAQ.  It uses a standard component interface to query and
control the components, and a separate interface for components to expose
internal data used for monitoring the health of the detector or for
debugging purposes.

CnCServer dynamically discovers the detector components during a launch
phase, and instructs them to connect to each other as needed.  Using the
run configuration files, it then distributes each component configuration
appropriately.  The components are then started to begin a data-taking run.
When a run is in progress, CnCServer regularly checks that components are
still active and that data are flowing between components.

\subsubsection{\label{sect:SNDAQ}Supernova Data Acquisition System}

The IceCube DAQ has a parallel triggering and analysis pathway designed
specifically for the detection of the many $O(10)$ MeV neutrinos from a
Galactic core-collapse supernova.  In the case of such an event, these
neutrinos will produce interactions in
the detector that, individually, are too dim to trigger the standard DAQ,
but because of their high number, can cause a coherent rise in the
individual hit rates of the DOMs~\cite{IC3:supernova}.

Each DOM monitors its hit rate and sends a stream of binned counts, using a
bin width of 1.6384~ms ($2^{16}$ clock cycles at 40 MHz).  An artificial
deadtime of $250\ {\mu}\mathrm{s}$ is applied after each hit to reduce the
impact of correlated hits (section~\ref{sect:darknoise}).  Each
StringHub collects the rate stream of each DOM, supplies UTC-based timestamps,
and forwards the streams to the Secondary Builder.

The supernova DAQ (SNDAQ) system receives the Secondary Builder stream,
rebins the individual DOM rates, and monitors the sum of rates over several
timescales for a significant rise.  This analysis is described in
detail in ref.~\cite{IC3:supernova}.  One complication is that light
deposition from cosmic-ray muons distorts the significance
calculation.  To correct for this, the trigger rate of the standard DAQ is
continuously sent to SNDAQ, and any significant alert is corrected
\cite{IC3:icrc15_sndaq}.  At a high significance threshold, the capture of
all untriggered data around the alert time is initiated using the HitSpool
system (section~\ref{sect:hitspool}), and the Supernova Neutrino Early Warning
System (SNEWS) \cite{SNEWS} is notified.  SNDAQ latency of approximately 7
minutes is dominated by the sliding window algorithm used to determine
average DOM rates. 

\subsubsection{\label{sect:hitspool}HitSpool Request System}

In the event of a significant transient event, subsystems such as SNDAQ can
request all untriggered DOM hits from the detector in 
a particular time interval by sending requests to a HitSpool Request daemon. Presently,
the HitSpool Request System has three clients; 
their basic characteristics are described in
Table~\ref{tab:hsclients}.  The central daemon passes the request on to 
every DOMHub, where hits in the requested time
interval are gathered and forwarded to a ``sender'' component.  The hits
are then bundled and transferred to the Northern Hemisphere for further analysis.

The time windows of SNDAQ HitSpool data requests are based on the
statistical significance of the alert and are shown in
Table~\ref{tab:hsclients}. The online High Energy Starting Event (HESE) 
analysis system requests HitSpool data from a symmetrical time window of
1~s around events with a total deposited charge of greater than 1500~PE.
The recently implemented HitSpool client for solar flare analyses is
triggered externally by significant Fermi Large Area Telescope (LAT) events~\cite{fermilat:flare}
and requests HitSpool 
data from a symmetrical time window of one hour around the trigger
time. Unlike the other two clients, these data sets are not transferred
over the satellite due to their size but are stored locally on disk, with
transmission over satellite only pursued in extraordinary cases.

\begin{table}
  \caption{HitSpool data-requesting services and request characteristics.
    The SNDAQ quantity $\xi$ is related to the statistical significance of
    the rate fluctuation.}
  \centering
  \footnotesize
\begin{tabularx}{\textwidth}{lcXXXX}
  \toprule Client & Trigger Threshold & Time\newline Window & Request
  Length & Raw \newline Data Size & Frequency \\
  \midrule
  SNDAQ & $7.65 \le \xi < 10$  & $[-30\,\mathrm{s},+60\,\mathrm{s}]$ &
  $90 \,\mathrm{s}$& $15 \,\mathrm{GB}$&
  $0.5/\mathrm{week}$ \\
   & $\xi \ge 10$ &  $[\pm250\,\mathrm{s}]$ & $500\,\mathrm{s}$ & $85
  \,\mathrm{GB}$ & $0.0006 / \mathrm{week}$ \\
  HESE & $1500 \,\mathrm{PE} $ &
  $[\pm0.5\,\mathrm{s}]$& $1\,\mathrm{s}$ & $175\,\mathrm{MB}$ &
  $4/\mathrm{day}$ \\
   & 6000 PE & & & & $1/\,\mathrm{month}$ \\ 
  Solar Flare & Fermi-LAT & $[\pm30\,\mathrm{min}]$ & $1\,\mathrm{h}$&
  $~600\,\mathrm{GB}$& $ 7 / \mathrm{year}$ \\
  & significant event & & & & 
  \\ \bottomrule
\end{tabularx}
\label{tab:hsclients}
\end{table}

\subsection{\label{sect:online:filter}Online Processing and Filtering}

\subsubsection{Overview}

The online Processing and Filtering (PnF) system handles
all triggered events collected by the DAQ
and reduces the data volume to a level that can be accommodated in our
satellite bandwidth allocation (about 100 GB/day).  This treatment
includes application of calibration constants, event
characterization and selection, extraction of data quality monitoring
information, generation of realtime alerts for events of astrophysical
interest, and creation of data files and metadata information for long-term
archiving.  The PnF system is a custom software
suite that utilizes about 20 standard, multi-processor servers located in
the SPS computing cluster.  

The first step in the analysis of triggered events is the calibration of
the digitized DOM waveforms, as described in section~\ref{sec:waveformcal}.
The geometry, calibration, and detector status (GCD) information needed to
process the data is stored in a database.  Next, each DOM's waveform is
deconvolved using the known DOM response to single photons to 
extract the light arrival time and amplitude information~\cite{IC3:ereco}.
This series of time and amplitude light arrival information for each DOM is
the basis for event reconstruction and characterization.  PnF encodes this
information in a compact data format known as the Super Data
Storage and Transfer format (SuperDST); this format uses only 9\% of the storage
size of the full waveform information.  The encoding does introduce a
small level of discretization error to the data, measured to be 1.1~ns in time and
0.04~PE in charge, smaller than the calibration uncertainties on these
values.  Any DOM readout whose SuperDST information is found not to be a
good representation of the original waveform, or sees large amounts of
light, also has the full waveform data saved in addition to the
SuperDST record.

Each event is then characterized with a series of reconstruction
algorithms that attempt to match the observed patterns of recorded light in
the SuperDST with known patterns of light from track and shower event
hypotheses~\cite{IC3:ereco}.  The reconstructed vertex position, direction,
energy, and the goodness-of-fit are used to select interesting events by various
filter selections.  The filter criteria are set by the collaboration
each year and are tuned to select events of interest to specific
analyses.  Each year there are about 25 filter selections in
operation; as of 2016, approximately 15\% of all triggered events are
selected by one or more filters.  Some of these filters are designed to search for
neutrino events of wide astrophysical interest to the scientific community
and trigger alerts that are distributed to followup observatories
worldwide~\cite{Abbasi:2011ja,Aartsen:2015trq}.

The PnF system also extracts and aggregates data quality and monitoring
information from the data during processing.  This information includes
stability and quality information from the DOM waveform and calibration
process, rates of DOM readouts, and rates and 
stability information for all detector triggers and filters.  This
information is aggregated for each data segment and reported to the IceCube
Live monitoring system (section~\ref{sec:online:icecubelive}).

Finally, the PnF system generates several types of files for satellite
transmission and for long term archival. These
include:   
\begin{enumerate}
\item Filtered data files containing events selected by the online filter
  selections.  These events generally only include the SuperDST version of
  the DOM information and results from the online event reconstructions.
  The files are queued for satellite transmission to the IceCube data
  center in the Northern Hemisphere by the data handling system.
\item SuperDST data files containing the SuperDST version of DOM readout
  information for all triggered events as well as summary information from
  the online filtering process.  This file set is intended as the long-term
  archive version of IceCube data.
\item Raw data files containing all uncalibrated waveforms from all DOMs for
  every event.  
\end{enumerate}

During normal operations, the DAQ produces a raw data output of
$\sim$1 TB per day, resulting in a raw data file 
archive of the same size.  The SuperDST and filtered data archive, after
data compression, are $\sim$170 GB/day and $\sim$90 GB/day, respectively.

\subsubsection{System Design}

The PnF system uses a modular design based on
a central master server node and a scalable number of data processing client
nodes.  The central master server handles data distribution and
aggregation tasks (requesting data blocks from the DAQ, collating event
SuperDST, reconstruction, filter, and monitoring information and writing
data files), while the clients handle the per-event processing
tasks (event calibration, reconstruction, analysis, and filtering).

The system is built upon the IceCube analysis software framework,
IceTray~\cite{DeYoung:2005zz}, allowing standard IceCube algorithms to be
used online without modifications.
The system uses the Common Object Request Broker Architecture
(CORBA) system as a means for controlling, supervising and interconnecting
the modular portions of the system.  Specialized CORBA classes allow
data to stream from one component to another using IceTray formats.
Clients can also be easily added and removed as needed to meet the
processing load.

\subsubsection{Components}

\begin{figure}[!ht]
 \centering
 \includegraphics[width=0.8\textwidth]{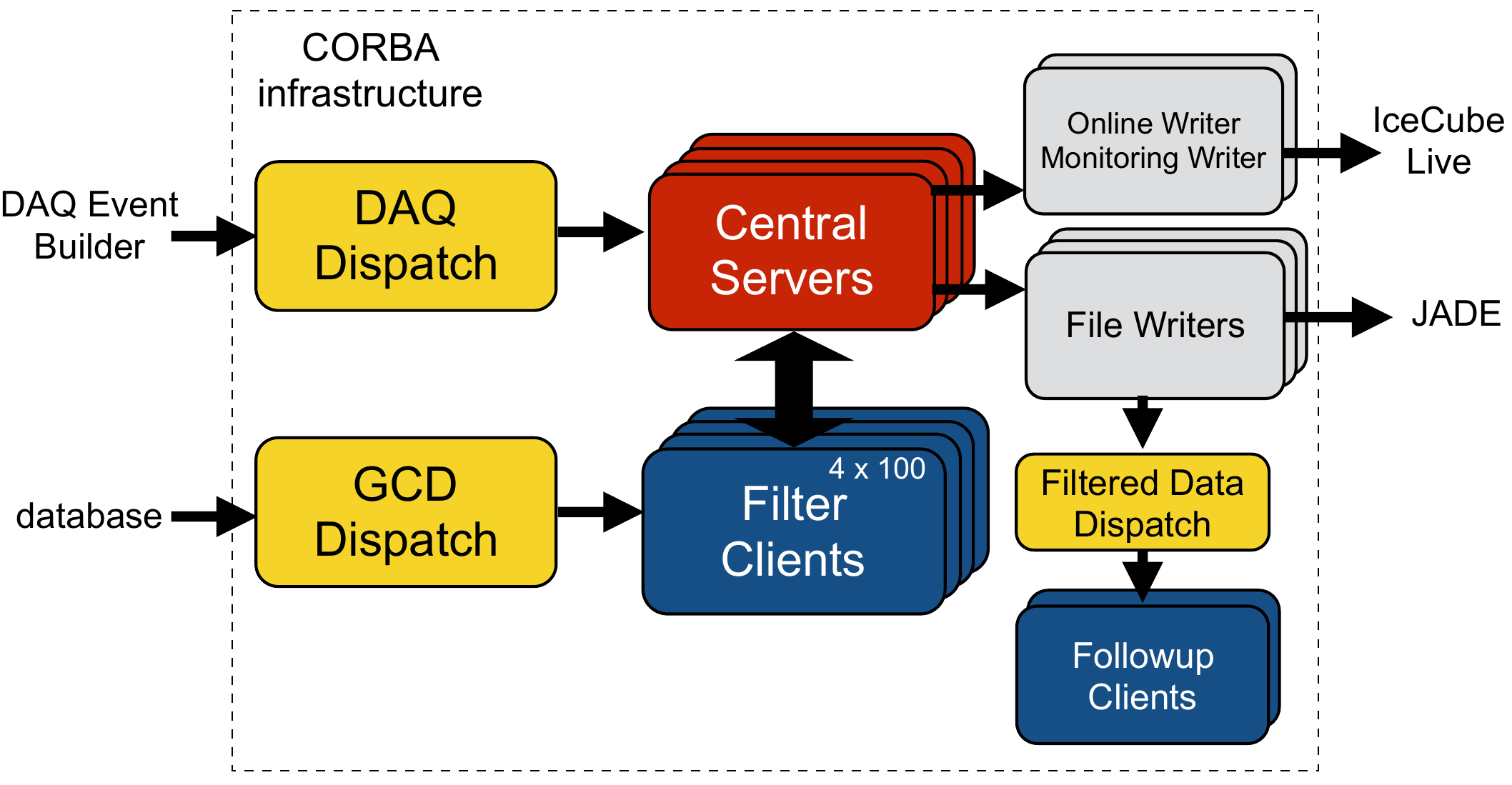}
 \caption{Internal components of the PnF
   system.  Arrows show the flow of data within the system.}
 \label{fig:online_pnf_internals}
\end{figure}

The flow of triggered event data in the PnF
system is shown in figure~\ref{fig:online_pnf_internals}.  Standard
components include: 
\begin{enumerate}
\item DAQ Dispatch, a process to pick up event data from the DAQ data cache
  and forward to the Central Server components. 
\item Central Servers, which
  receive data from the 
  DAQ Dispatch event source, distribute events to and record results
  returning from the PnF client farm, and send events to Writer components.
  Typically there are four servers in operation.
\item Filter Clients, where the core calibration, reconstruction and
  filtering processes are applied to each triggered event.  Up to 500 of
  these clients operate in parallel to filter 
  events in real time.
\item GCD Dispatch, a database caching system to prevent the
  Filter Client processes from overwhelming the GCD database at run transitions.
\item File Writers, responsible for creation of files and metadata for
  the data archive.  There is one writer component for each file type created.
\item Online Writer, responsible for extracting event reconstruction and
  filter information from the data for events of astrophysical interest and
  sending this information out in real time via the IceCube Live alert
  system.
\item Monitoring Writer, responsible for aggregating per-event monitoring
  information, creating histograms, and forwarding to the IceCube Live
  monitoring system.
\item Filtered Data Dispatch and FollowUp Clients, responsible for
  looking for bursts of neutrino events on timescales from 100 seconds up
  to 3 weeks in duration.  Any significant burst of neutrinos found generates alerts
  sent to partner observatories worldwide.
\end{enumerate}

\subsubsection{Performance}

\begin{figure}[!ht]
 \centering
 \includegraphics[width=0.85\textwidth]{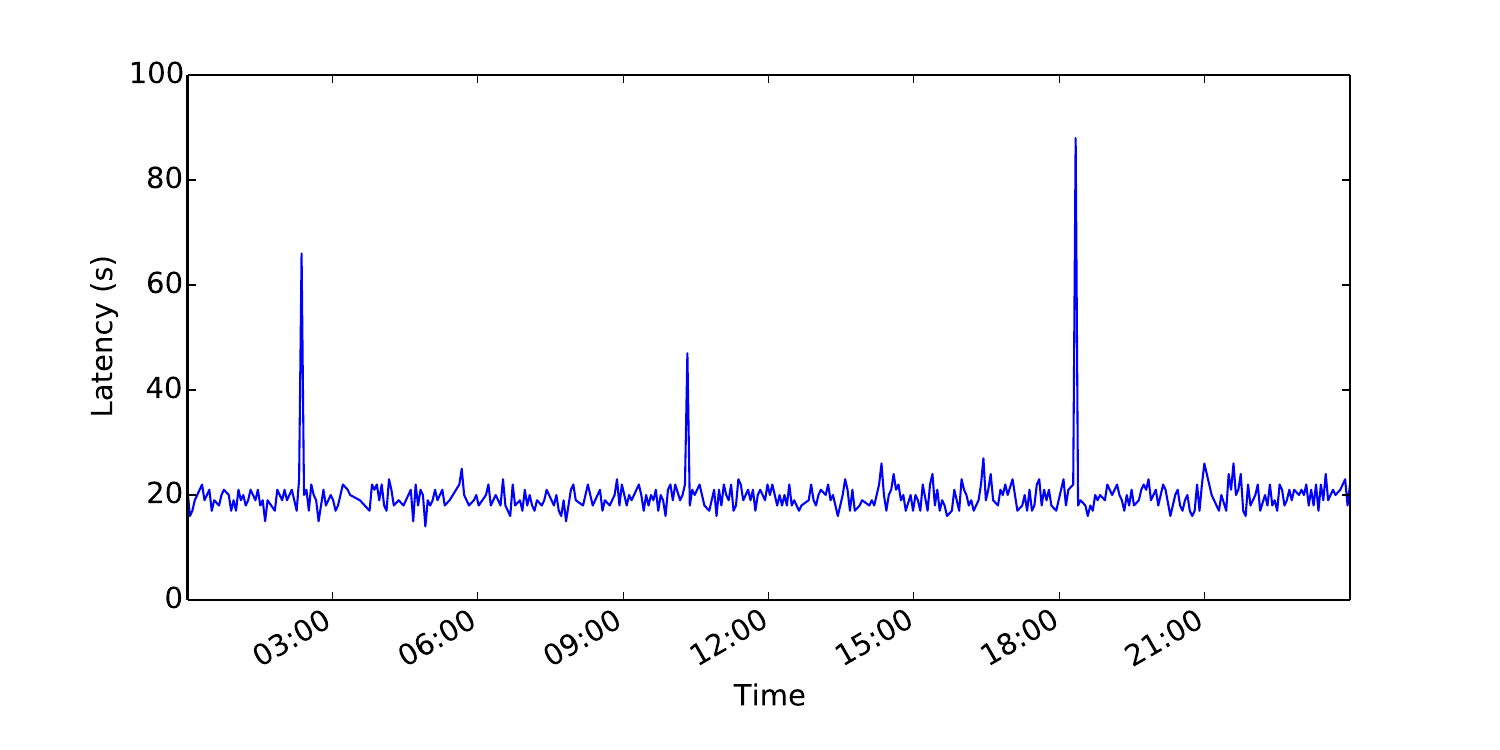}
 \caption{Typical PnF system latency for a
   24-hour period.  The latency is defined as the time between DAQ trigger
   time and time when the online filtering processing is complete.  The
   spikes in latency correspond to DAQ run transitions, when
   geometry, calibration, and detector status information is updated and
   distributed to the filtering clients.}
 \label{fig:online_pnf_latency}
\end{figure}

The PnF system is designed to filter triggered events as quickly as
possible after collection by the data acquisition 
system.  A key performance metric is processing system latency, defined as the duration
of time between the DAQ trigger and the completion of event
processing and filtering.  A representative latency history for the system is
shown in figure~\ref{fig:online_pnf_latency}, showing typical system
latencies of about 20 seconds.

The filter selections used have been relatively stable over several years
of operation of the completed IceCube detector, with most seeing only minor
updates at the 
start of each season.  The majority of physics analyses derive from a small
set of core filters, including:

\begin{enumerate}
\item A muon track filter that searches for high-quality track events from all
  directions.  Up-going events for all triggered energies are selected,
  while only high-energy 
  down-going tracks are selected to avoid the large background of
  down-going atmospheric muons at lower energies.  These selected events
  are frequently used as the input to point source and transient neutrino searches.
\item A shower event filter that searches for events producing large energy
  depositions in or near the instrumented volume.  These selected events are
  frequently used as the input to searches for high-energy shower events arising from
  atmospheric and astrophysical neutrinos.
\item A high-charge filter that searches for any event depositing a
  large amount of energy leading to a recorded charge of $\geq1000$
  photoelectrons in the 
  instrumented volume.  While having a large overlap with the muon track
  and shower filters at high energies, this filter targets the highest
  energy neutrino events of all types. The selected events are used as
  inputs to searches for high-energy astrophysical and cosmogenic
  neutrinos as well as for relativistic magnetic monopoles.
\item Cosmic ray filters that search for extended air-shower events in
  IceTop.  The selected events are used as inputs to analyses 
  targeting the flux, spectrum, and composition of the primary cosmic rays
  observed in the Earth's atmosphere.
\item A DeepCore contained event filter that searches for contained, lower-energy
  neutrino events (in the range of 10--100 GeV) from atmospheric neutrino interactions
  that are contained within the more densely instrumented DeepCore region.
  The selected events are used as inputs to analyses that measure
  neutrino oscillation effects and search for indirect signatures of dark matter.
\end{enumerate}

\noindent Other filters are employed for more specialized searches, as well as for minimum bias selections.

\subsection{\label{sect:online_jade}Data Handling}

The bulk of South Pole Station data traffic is handled by geosynchronous
satellite links.  Due to the location, only
geosynchronous satellites with steeply inclined orbits reach far enough
above the horizon to establish a link.  For a given satellite, this link
provides four to six hours of communications once per sidereal day.
Multiple satellites are currently utilized by the
U.S.~Antarctic Program, providing a window of about 12 hours of connectivity with
bandwidth of 250 Mbps for uni-directional data transfer and bandwidth
of 5 Mbps for bi-directional internet connectivity.  For the remainder of the day, Iridium
communication satellites allow limited voice and data connectivity and provide up to 2.4
kbps of bandwidth per modem.

IceCube incorporates Iridium modems into two separate systems, the legacy IceCube
Teleport System (ITS) and the IceCube Messaging System (I3MS).  ITS uses
the Iridium Short Burst Data mode to send short 
messages of 1.8 kB or smaller with a typical latency (transmission time) of 30 seconds.
Messages may either originate or terminate at the ITS Iridium modem at the
South Pole.  Messages also contain a recipient ID indicating the intended
host to receive the message, allowing a many-to-many communications
infrastructure between systems running at the South Pole and systems in the
Northern Hemisphere.  ITS was retired in 2016.

The newer IceCube Messaging System (I3MS), deployed in 2015, incorporates
multiple Iridium modems and uses the Iridium RUDICS data mode, providing a
2.4 kbit/s bidirectional serial stream per modem and a minimum latency of
about 1.5 seconds.  I3MS runs as a daemon on both ends of the link, accepts
messages via the ZeroMQ distributed messaging protocol, and transports
those messages across the link based on message priority and fair sharing
of bandwidth among all users. I3MS message recipients listen for messages
using ZeroMQ publish-subscribe (PUB-SUB), allowing a given message to be
sent to multiple recipients.  I3MS also provides low-bandwidth secure shell
(ssh) connectivity to the South Pole, allowing off-site operators
access to SPS in the case of detector issues.

Data handling is provided by three servers running the Java Archival and
Data Exchange (JADE) software. JADE is a
recent Java-based reimplementation and expansion of earlier software, the
South Pole Archival and Data Exchange (SPADE).  JADE has 
four primary tasks: data pickup, archiving, satellite transmission, and
real-time transmission. The three servers operate independently of one
another, and each is capable of separately handling the nominal
data volume; thus, data handling can continue seamlessly in case of
hardware failure or maintenance. 

JADE is configured with a number of input data streams, 
each consisting of a data server, a dropbox directory, and a filename pattern.  The
data stream dropbox directories are checked on a regular basis for new
files.  File completion is indicated by the producer creating a matching
semaphore file.  For each file, a
checksum calculated on the data server is compared to a checksum calculated
on the JADE server.  After validation, the original data file is removed
from the pickup location. 

Files are then routed according to the configuration of their
data stream, either transmitted via satellite link or 
archived locally.  Archival data
were formerly written to Linear Tape Open (LTO) tapes; the tape system was
retired in 2015, and archival data are now written to disks.
All of the archival data are buffered on the server until the storage medium
is full. In case of media failure, the buffered files can be 
immediately written to new archival media with a single command.  
Two copies of each archival data stream are saved, and the disks are
transported from the South Pole to the IceCube data center each
year.  Archival data are not regularly reprocessed but are retained
indefinitely in case of an issue with the filtered data streams or a
significant improvement in the low-level calibration.

Data streams intended for satellite transfer are queued separately.  
JADE combines multiple smaller files or splits large files to create $\sim1$
GB bundles, allowing satellite link operators to manage the daily data
transmission.  A configurable number of bundles is then transferred to the
satellite relay server.  If satellite transmission is temporarily
interrupted, the excess bundles are staged on the JADE server. 

Small files ($<$50 KB) with high priority are sent via
the I3MS Iridium link.  In cases where the real-time link is not available, I3MS
will queue the messages to be sent when the link becomes available. All
I3MS messages are also sent to JADE to send via the geosynchronous satellite link to
ensure delivery if the Iridium link should be unavailable for an extended
period of time.

\subsection{\label{sec:online:icecubelive}IceCube Live and Remote Monitoring}

IceCube operations are controlled and monitored centrally by IceCube Live.
IceCube Live consists of two major components: LiveControl,
responsible for controlling data-taking operations and collecting
monitoring data, and the IceCube Live website, responsible for processing
and storing monitoring data as well as presenting this data in webpages and
plots that characterize the state of the detector.

\subsubsection{LiveControl}

LiveControl is responsible for controlling the state of DAQ and PnF, starting and
stopping data-taking runs, and recording the parameters of these runs.
Human operators typically control the detector and check basic 
detector status using a command-line interface to the LiveControl
daemon. Standard operation is to 
request a run start, supplying a DAQ run configuration 
file.  LiveControl then records the run number, configuration, start time,
etc. and sends a request 
for DAQ to begin data-taking.  After data-taking commences successfully,
LiveControl waits a specified amount of time, generally eight hours, then
stops the current run and automatically starts a new run using the same
configuration.

This cycle continues until stopped by a user request or a
run fails.  In case of failure, LiveControl attempts to restart data-taking
by starting a new run.  Occasionally a hardware failure occurs, and a new
run cannot be started with the supplied configuration because requested
DOMs are unpowered or temporarily unable to communicate.  In this case,
LiveControl cycles through predefined partial-detector 
configurations in an attempt to exclude problematic DOMs.  This results in
taking data with fewer than the full number of available strings, but it
greatly reduces the chance of a prolonged complete outage where no IceCube
data are recorded.

A secondary function of LiveControl is the collection, processing, and
forwarding of monitoring data from DAQ, PnF, and other
components.  The associated JavaScript Object Notation (JSON) data are
forwarded to LiveControl using the ZeroMQ protocol and queued internally 
for processing.  Certain monitoring quantities indicate serious problems with
the detector, e.g. the PnF latency is too high.  LiveControl
maintains a database of critical monitoring quantities and raises an alert
if the value is out of the specified range or 
hasn't been received in a specified amount of time.  The alert usually
includes an email to parties responsible for the affected subsystem and,
for serious problems, triggers an automated page to winterover operators.
Alerts generated by controlled components such as DAQ or PnF may also
generate emails or pages.
All monitoring data are forwarded to the IceCube Live website for further
processing and display.

\subsubsection{IceCube Live Website}

Two operational copies of the IceCube Live website exist: one inside the
IceCube network at the South Pole, and one in the Northern Hemisphere.
Monitoring data reach the northern website based on relative priority and using
both geosynchronous and Iridium data transport, summarized in table
\ref{i3messages}.

\begin{table}[!ht]
  \centering
  \caption{Typical data volume and latencies for IceCube Live monitoring
    messages.} 
  \label{i3messages}  
  \begin{tabularx}{0.85\textwidth}{|c|c|c|c|X|}
    \hline Priority & Transport System & Messages/day~ & Data/day & Latency\\
    \hline 1 & I3MS (Iridium) & 10,000 & 1 MB & 1 min. \\
    \hline 2 & I3MS (Iridium) & 150,000 & 5 MB & 1--5 min. \\
    \hline 3 & JADE (Geosynchronous) & 300,000 & 100 MB & 1 day \\
    \hline
  \end{tabularx}
\end{table}

\begin{figure}[!ht]
 \centering
 \includegraphics[width=1.0\textwidth]{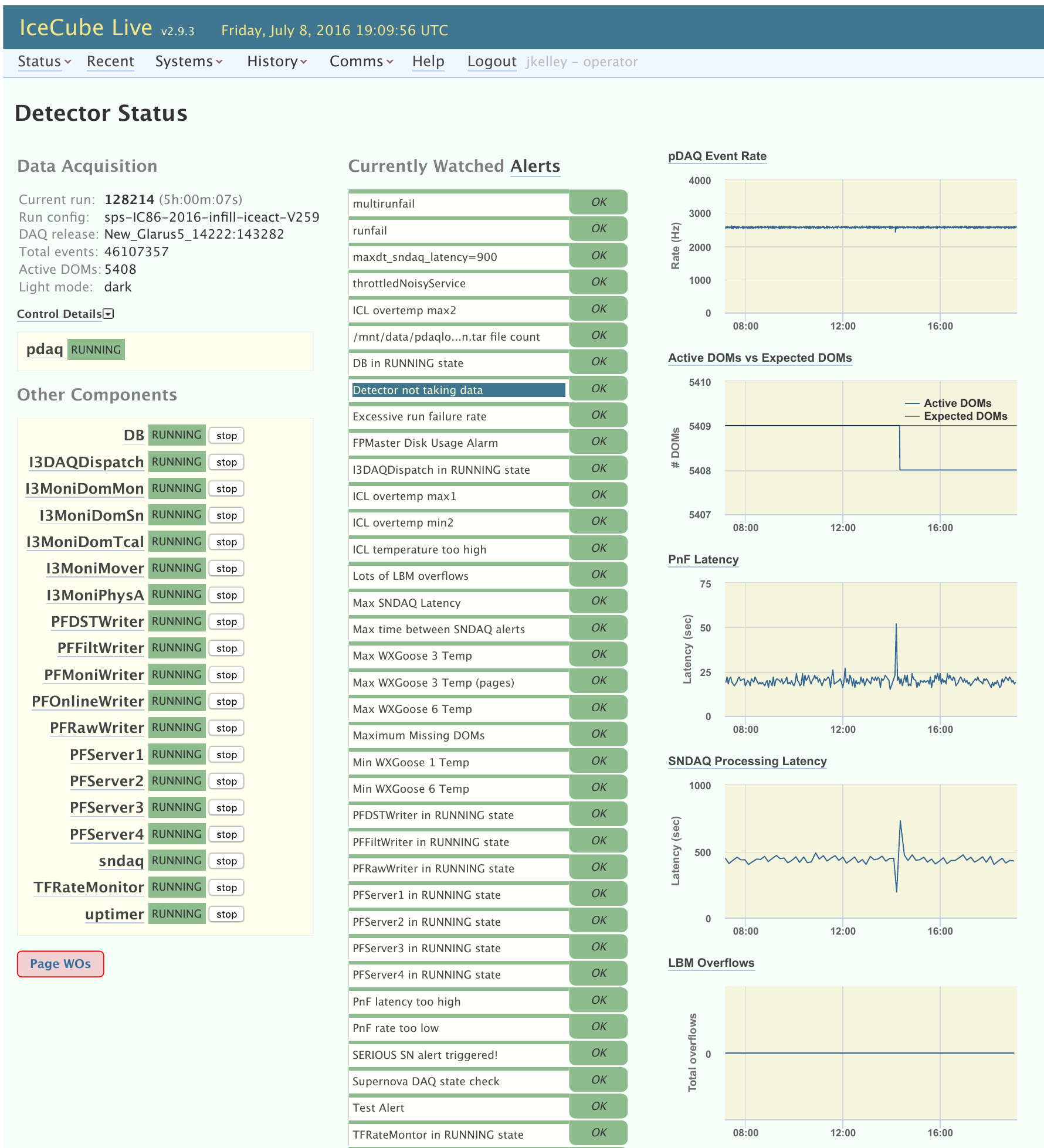}
 \caption{The IceCube Live website front page, showing near-realtime detector
   status.  One DOM has dropped out of data-taking in the run shown.  This
   ``operator'' view includes controls to start/stop components and to page
   the winterovers.} 
 \label{fig:live_screenshot}
\end{figure}

Messages reaching the website are processed by a database server daemon
(DBServer).  Messages also may contain directives requesting DBServer to send email, by
specifying email recipients and content, or requesting that the monitoring
message be published using the ZeroMQ PUB-SUB framework, allowing the message to be
passed to an external process.  The IceCube Live website itself uses the
Django web framework and contains pages that create graphical views of
monitoring data stored in the database.  These pages include a front page
displaying active alerts and plots of event rates and processing latencies
from the previous few hours (figure~\ref{fig:live_screenshot}), and a page
for each run that displays start  
time, stop time, and other essential data.  The run page contains low-level
diagnostic data that include e.g. DOM charge histograms, digitizer baselines,
DOM occupancy, etc., and are used to diagnose any problems that occurred
during the run and to determine if the run can be used in physics
analysis.  Historical monitoring data are available from runs back to
2011.  

Finally, the IceCube Live website in the Northern Hemisphere can transmit
messages to LiveControl at SPS using the Iridium systems.  This capability
is used to link the commercial Slack chat service to a chat web page in
IceCube Live, allowing the IceCube winterover operators to communicate with
experts in the Northern Hemisphere during periods with no geosynchronous satellite
connectivity.  This connection also provides a limited capability to
control the detector, allowing operators in the Northern Hemisphere to
start/stop components or remotely issue HitSpool requests.

\subsection{\label{sec:operational_performance}Operational Performance}

Detector operational uptime is highly valued, in order to remain
sensitive to rare astrophysical transient events.  Many redundancies and
failsafes have been implemented, allowing an average detector uptime of greater
than 99\%. The detector uptime measures the fraction of total time
that some portion of the detector is taking data; sources of downtime
include the transitions between data-taking runs, power outages, and
DAQ interruptions due to software or hardware issues. All DOMHubs and servers are equipped with redundant power
supplies; these are in turn connected to redundant UPSes. This backup battery
power allows the detector to remain fully operational for about 15 minutes in the
event of a power outage.

Industry-standard Nagios monitoring software tracks the SPS hardware status and provides the link
between LiveControl and the paging system.  In the event of a hardware
or software failure that interrupts data-taking, a partial-detector configuration can
often be started automatically in parallel with the paging alert.
Within minutes, winterover operators start an optimal detector configuration
excluding only affected string(s), and then proceed to diagnose and fix the
issue.  Software issues can often be addressed via network access from the
South Pole Station, while hardware failures may require a visit to the ICL
for repairs.

\begin{figure}[!ht]
 \centering
 \includegraphics[width=1.0\textwidth]{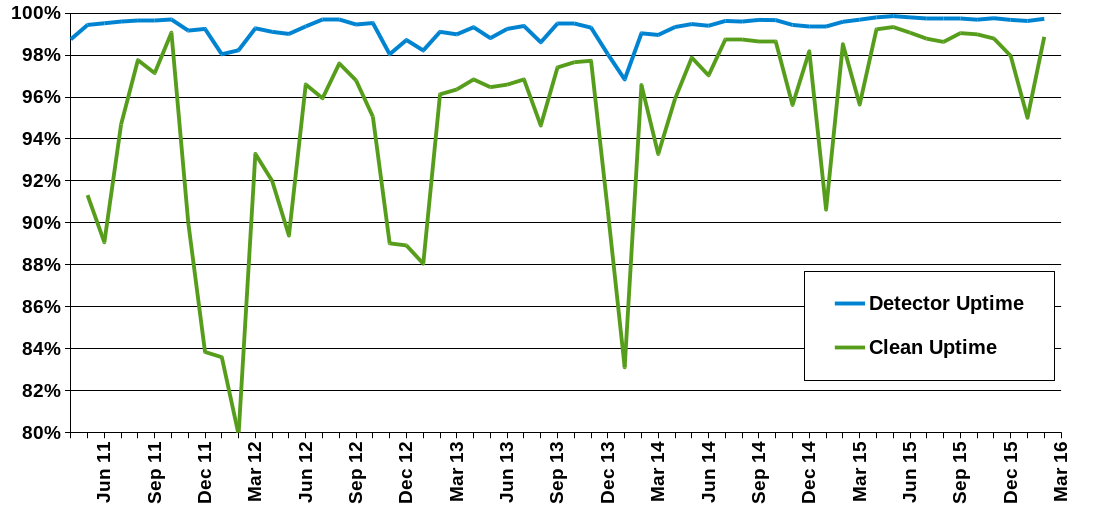}
 \caption{Detector uptime each month since the start of full-detector
   operation. ``Clean uptime'' indicates periods of analysis-ready,
   86-string data.} 
 \label{fig:clean-uptime}
\end{figure}

The recovery of data from all but the last few minutes of runs that fail,
and the recent implementation of continuous data-taking, have improved
detector stability and decreased detector downtime. These features
contribute to an average ``clean uptime'' in recent years of 97--98\% of
full-detector, analysis-ready data, exceeding our target of 95\%.
Historical total detector uptime and clean uptime are shown in
figure~\ref{fig:clean-uptime}.   

\begin{figure}[!ht]
	\centering
    \includegraphics[width=0.8\textwidth]{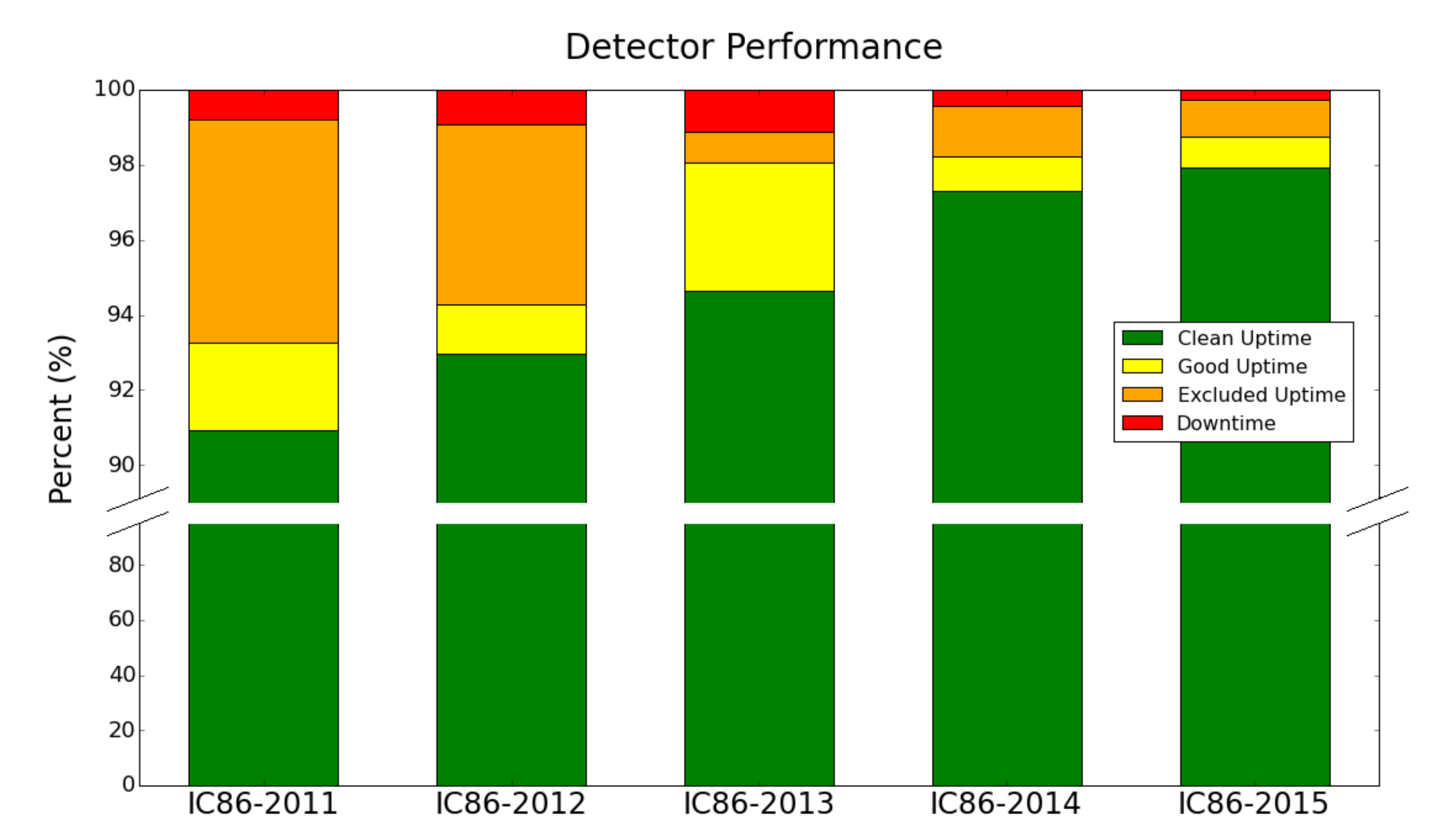}
	\caption{Detector performance breakdown by
      run period.  Clean uptime indicates pristine, full-detector
      data; ``good uptime'' includes partial-detector runs that are
      otherwise healthy; and 
    ``excluded uptime'' indicates runs that are declared unusable for
      analysis.}
    \label{fig:period-performance}
\end{figure}

About 0.2\% of the loss in clean uptime is due to the failed portions of
runs that are not usable for analysis.  There is around 1\% of clean
uptime loss due to runs not using the full-detector configuration. This
occurs when certain components of the detector are excluded from the run
configuration during required repairs and maintenance; these
partial-detector runs are still useful for analyses that have less strict
requirements on the active detector volume or in the case of an exceptional
astrophysical transient event. There is approximately a 1\%
loss of clean uptime due to maintenance, commissioning, and verification
runs, and short runs that are less than 10 minutes in duration.  A
breakdown of total detector time by run period is shown in
figure~\ref{fig:period-performance}.  

%auto-ignore
\section{Outlook}

IceCube's online systems continually evolve in order to enhance the
capabilities and performance of the detector and meet new science needs.  
For example, the 2016 physics run start
included a new low-energy IceTop trigger, a new filter designed to search for
moderate-velocity magnetic monopoles, and a transition to a more flexible
realtime alert system that transmits candidate neutrino events at a rate of
3 mHz via I3MS to a followup analysis server in the Northern Hemisphere.
Specialized filter streams for rare events likely to be astrophysical in
origin have also been added within the past two years.  

In addition to online software improvements and regular computing upgrades, a
number of calibration and maintenance hardware activities are 
planned. For example, in order to mitigate the effects of continued snow
accumulation on the IceTop tanks --- increased energy threshold and reduced
sensitivity to the electromagnetic air shower component --- a small number
of scintillator-based  detector panels have been installed at existing
IceTop stations, and development of additional panels is underway.

The stability of the primary triggers and filters and a high detector
uptime enable continued detection of additional astrophysical neutrino
candidates.  Searches for neutrino point sources are ongoing, and a robust
realtime followup campaign facilitates multi-wavelength observations of
reconstructed neutrino source directions, with the eventual goal of
definitive identification of the astrophysical hadron accelerators.

\subsection{IceCube Gen2 and PINGU}

As neutrino astronomy matures, designs for next-generation detectors are underway.   Neutrino
detectors under construction or planned for the Northern Hemisphere include
KM3NeT in the Mediterranean \cite{km3net} and GVD in Lake Baikal \cite{gvd}.  

IceCube Gen2 is an experiment under design for the South Pole
consisting of a high-energy in-ice array, a surface air shower array with
additional veto capabilities, a low-energy in-ice infill array (the Precision IceCube Next
Generation Upgrade, or PINGU), and potentially a shallow sub-surface 
array of radio antennas \cite{gen2_whitepaper}.  The high-energy array features an 
increased string spacing that allows instrumentation of up to 10 times larger 
effective target volume relative to IceCube; the scientific focus will be the
detection and characterization of astrophysical neutrino sources at the PeV
energy scale.  The PINGU sub-array \cite{pingu_loi} will densely instrument
6 MTon of ice in the center of DeepCore, enabling precision neutrino
oscillation measurements down to the GeV energy range, determination of the
neutrino mass ordering, and dark matter searches at energies above a
few GeV.  Updated calibration devices will be deployed in the new
boreholes in order to measure the optical
properties of the ice more precisely and improve event reconstruction. An extended surface
array, potentially several times larger in diameter than the high-energy
array, will be used both for cosmic ray studies and to veto downgoing
atmospheric muons and neutrinos.  Radio-frequency detection of ultra-high-energy
neutrino-induced showers in the ice is a relatively recent
technique which shows considerable promise to achieve effective
target volumes of about 100 times IceCube at \qty{E18}{eV} where neutrinos
originating from scattering of ultra-high-energy cosmic rays on the 
cosmic microwave background are expected \cite{ara2}.  The 
combination of optical and radio-frequency technologies offers the possibility
of a broadband neutrino observatory with diverse astrophysics and particle
physics science capabilities. 

The baseline sensor design for the array is a modernized version of the
IceCube DOM~\cite{pingu_loi}.  Mechanical components such as the glass sphere and
penetrator, as well as the high-quantum-efficiency PMT, remain unchanged,
while the triggering and digitization electronics are being redesigned.
Alternative sensor designs are also under study that increase photocathode
area, photon collection, angular coverage, and/or directional resolution.
Slimmer, more cylindrical profiles may 
allow smaller hole diameters, decreasing drilling time and fuel usage.
The recent development of the capability to deliver cargo and fuel to the
station via overland traverse rather than aircraft will reduce fuel
costs. This 
is part of an overall effort to reduce logistical support requirements
compared to IceCube construction.  

The IceCube detector has achieved ``first light'' in neutrino astronomy and
has the capability to continue operating at least until the end of the next decade, supporting 
a diverse neutrino physics and astrophysics program and providing unique datasets
to the scientific community.  The IceCube Gen2 facility will continue this legacy 
and contribute to further discoveries in neutrino astronomy and multi-messenger astrophysics.

%auto-ignore
\acknowledgments

We acknowledge the support from the following agencies: U.S. National
Science Foundation --- Office of Polar Programs, U.S. National Science
Foundation --- Physics Division, University of Wisconsin Alumni Research
Foundation, the Grid Laboratory Of Wisconsin (GLOW) grid infrastructure at
the University of Wisconsin -- Madison, the Open Science Grid (OSG) grid
infrastructure; U.S. Department of Energy, and National Energy Research
Scientific Computing Center, the Louisiana Optical Network Initiative
(LONI) grid computing resources; Natural Sciences and Engineering Research
Council of Canada, WestGrid and Compute/Calcul Canada; Swedish Research
Council, Swedish Polar Research Secretariat, Swedish National
Infrastructure for Computing (SNIC), and Knut and Alice Wallenberg
Foundation, Sweden; German Ministry for Education and Research (BMBF),
Deutsche Forschungsgemeinschaft (DFG), Helmholtz Alliance for Astroparticle
Physics (HAP), Research Department of Plasmas with Complex Interactions
(Bochum), Germany; Fund for Scientific Research (FNRS-FWO), FWO Odysseus
programme, Flanders Institute to encourage scientific and technological
research in industry (IWT), Belgian Federal Science Policy Office (Belspo);
University of Oxford, United Kingdom; Marsden Fund, New Zealand; Australian
Research Council; Japan Society for Promotion of Science (JSPS); the Swiss
National Science Foundation (SNSF), Switzerland; National Research
Foundation of Korea (NRF); Villum Fonden, Danish National Research
Foundation (DNRF), Denmark.

\clearpage
%auto-ignore
% List of acronyms used in the paper

\section*{IceCube Acronym List}

\begin{longtable}{p{.20\textwidth} p{.80\textwidth}}

  \textbf{ADC} & analog-to-digital converter \\

  \textbf{ATWD} & Analog Transient Waveform Digitizer; front-end signal
  digitizers in the DOM \\

  \textbf{cDOM} & color (LED) DOM \\

  \textbf{CORBA} & Common Object Request Broker Architecture; software
  process communications middleware used in PnF \\
  
  \textbf{DAQ} & data acquisition \\

  \textbf{DESY} & Deutsches Elektronen-Synchrotron national research
  center, Germany \\
  
  \textbf{DFL} & dark freezer lab used for DOM testing \\

  \textbf{DMA} & direct memory access \\
  
  \textbf{DOM} & digital optical module; IceCube's primary light sensor \\

  \textbf{DOMCal} & DOM calibration software \\

  \textbf{DOMHub} & DOM readout computer in the ICL \\
  
  \textbf{DOR} & DOM readout card used to power and communicate with
  the DOMs \\

  \textbf{DSB} & DOMHub service board providing GPS clock signals to
  the DOR cards \\
  
  \textbf{EHWD} & Enhanced Hot Water Drill used to drill IceCube holes \\
  
  \textbf{ESD} & electrostatic discharge \\
  
  \textbf{fADC} & fast ADC; front-end signal digitizer in the DOM \\
  
  \textbf{FAT} & final acceptance testing of the DOMs \\

  \textbf{FHWM} & full width half maximum \\

  \textbf{FPGA} & field-programmable gate array \\
  
  \textbf{FRT} & fixed-rate trigger; a software trigger algorithm used
  in the DAQ \\

  \textbf{GCD} & geometry, calibration, and detector status information
  needed to process IceCube data \\
  
  \textbf{GPS} & Global Positioning System; primary time source for IceCube
  \\

  \textbf{HESE} & high-energy starting event; a bright neutrino event
  candidate first visible inside the detector \\
  
  \textbf{HLC} & hard local coincidence, as in ``HLC hits'';
  DOM hits satisfying this condition are read out with full
  waveform information \\

  \textbf{HQE} & high quantum efficiency (PMT) \\

  \textbf{I3MS} & IceCube Messaging System; enables 24/7
  connectivity to the South Pole \\
  
  \textbf{ICL} & IceCube Laboratory; central building for IceCube systems
  at the South Pole \\

  \textbf{ITS} & IceCube Teleport System; legacy communications system used
  to provide 24/7 connectivity to the South Pole \\

  \textbf{JADE} & Java Archival and Data Exchange; data-handling software
  system \\

  \textbf{JSON} & JavaScript Object Notation; a lightweight
  data-interchange format \\

  \textbf{LBM} & lookback memory, the SDRAM hit buffer on each DOM \\
  
  \textbf{LC} & Local coincidence of DOMs on a string or between tanks in 
  an IceTop station; see also HLC and SLC \\
  
  \textbf{LED} & light-emitting diode \\

  \textbf{LTO} & linear tape open storage media format \\
  
  \textbf{MPE} & multiple photoelectron, as in ``MPE discriminator'' \\

  \textbf{PCI} & Peripheral Component Interconnect; a computer bus \\
  
  \textbf{PE} & photoelectron; also a unit of amplified charge resulting
  from one converted photon incident on a PMT \\

  \textbf{PINGU} & Precision IceCube Next Generation Upgrade \\
  
  \textbf{PMT} & photomultiplier tube \\

  \textbf{PnF} & processing and filtering online software system \\
  
  \textbf{PSL} & Physical Sciences Laboratory, Stoughton, Wisconsin, U.S.A \\

  \textbf{RAPCal} & reciprocal active pulsing calibration; time calibration
  scheme used to synchronize the DOMs \\

  \textbf{RFI} & radio-frequency interference \\

  \textbf{SJB} & surface junction box; passive connection of the in-ice
  cable, surface cable, and IceTop tanks \\

  \textbf{SES} & Seasonal Equipment Site used to provide power and hot
  water for drilling operations \\
  
  \textbf{SLC} & soft local coincidence, as in ``SLC hits''; SLC hits do
  not satisfy an LC condition and are read out with a timestamp and minimal
  amplitude / charge information \\

  \textbf{SLOP} & slow particle, as in ``SLOP trigger''; a software trigger
  algorithm used in the DAQ \\
  
  \textbf{SMT} & simple multiplicity trigger; a software trigger algorithm
  used in the DAQ \\

  \textbf{SNDAQ} & supernova data acquisition software \\

  \textbf{SNEWS} & Supernova Neutrino Early Warning System \\

  \textbf{SPADE} & South Pole Archival and Data Exchange; legacy
  data-handling software system \\
  
  \textbf{SPE} & single photoelectron, as in ``SPE discriminator'' \\

  \textbf{SPS} & South Pole System, the IceCube computing and networking
  systems at the detector site \\

  \textbf{SPTS} & South Pole Test System, a scaled-down version of SPS for
  testing in the Northern Hemisphere \\

  \textbf{SuperDST} & Super Data Storage and Transfer compressed event
  format \\ 
  
  \textbf{TDP} & track detection probability \\

  \textbf{TOS} & Tower Operations Site used at each hole for drilling
  and string deployment \\

  \textbf{UPS} & uninterruptible power supply \\
  
  \textbf{UTC} & Coordinated Universal Time, the time standard used in
  IceCube \\
  
\end{longtable}

\clearpage
\bibliographystyle{JHEP}
\bibliography{i3iosart}

\end{document}